\documentclass[aip,amsmath,amssymb,reprint]{revtex4-2}

\usepackage{graphicx}
\usepackage{dcolumn}
\usepackage{bm}
\usepackage[utf8]{inputenc}
\usepackage[T1]{fontenc}
\usepackage{mathptmx}
\usepackage{physics}
\usepackage{color}

\begin{document}

\preprint{AIP/123-QED}

\title[Artificial neurons based on antiferromagnetic auto-oscillators as a platform for neuromorphic computing]{Artificial neurons based on antiferromagnetic auto-oscillators as a platform for neuromorphic computing}

\author{H. Bradley}%
\affiliation{Department of Physics, Oakland University.}
\author{S. Louis}
\affiliation{Department of Electrical and Computer Engineering, Oakland University.}
\email{slouis@oakland.edu}
\author{C. Trevillian}%
\affiliation{Department of Physics, Oakland University.}
\author{L. Quach}%
\affiliation{Oakland University William Beaumont School of Medicine}
\author{E. Bankowski}%
\affiliation{United States Army DEVCOM Ground Vehicle Systems Center}
\author{A. Slavin}%
\affiliation{Department of Physics, Oakland University.}
\author{V. Tyberkevych}
\affiliation{Department of Physics, Oakland University.}

\date{\today}

\begin{abstract}

Spiking artificial neurons emulate the voltage spikes of biological neurons, and constitute the building blocks of a new class of energy efficient, neuromorphic computing systems. 
Antiferromagnetic materials can, in theory, be used to construct spiking artificial neurons.
When configured as a neuron, the magnetizations in antiferromagnetic materials have an effective inertia that gives them intrinsic characteristics that closely resemble biological neurons, in contrast with conventional artificial spiking neurons.
It is shown here that antiferromagnetic neurons have a spike duration on the order of a picosecond, a power consumption of about 10$^{-3}$ pJ per synaptic operation, and built-in features that directly resemble biological neurons, including response latency, refraction, and inhibition. 
It is also demonstrated that antiferromagnetic neurons interconnected into physical neural networks can perform unidirectional data processing even for passive symmetrical interconnects.
Flexibility of antiferromagnetic neurons is illustrated by simulations of simple neuromorphic circuits realizing Boolean logic gates and controllable memory loops. 

\end{abstract}

\maketitle

\section{\label{seclevel1}Introduction}

Artificial Intelligence (AI) and machine learning seek to replicate the cognitive functions of the human brain.
AI is currently employed in many applications, for example, in image recognition, data processing, natural language processing, computer vision, and decision making. 
In the near future, AI will be of increasing importance for self-driving cars, medical diagnostics, and for many other areas.
Interestingly, computers using deep learning and other machine learning algorithms can identify handwritten numbers with an accuracy that surpasses the ability of the human brain \cite{baldominos2019survey, lecun2015deep}. 

Machine learning algorithms typically use an artificial neural network (ANN) running on a conventional, silicon semiconductor based computer platform.
Unfortunately, training ANNs can be quite computationally intensive and consequently consume a substantial amount of energy; for example, AlphaGo --- the first AI program to defeat a professional human player in the game of Go --- required more than 1 MW for operation\cite{silver2016mastering, egri2020game}.
In another example, the energy cost to train an ANN for natural language processing is estimated to exceed ten million dollars, with significant negative environmental consequences\cite{sharir2020cost, strubell2019energy}. 
Computational intensity also limits the utility of AI for applications which require low power; for example, in mobile phones, at remote data sensors, or other edge-computing applications that bring computational resources closer to data sources.
Moreover, for applications that require processing and learning from vast quantities of data in real time, traditional silicon based computing platforms greatly limit the utility of AI. 

\begin{figure}
\includegraphics{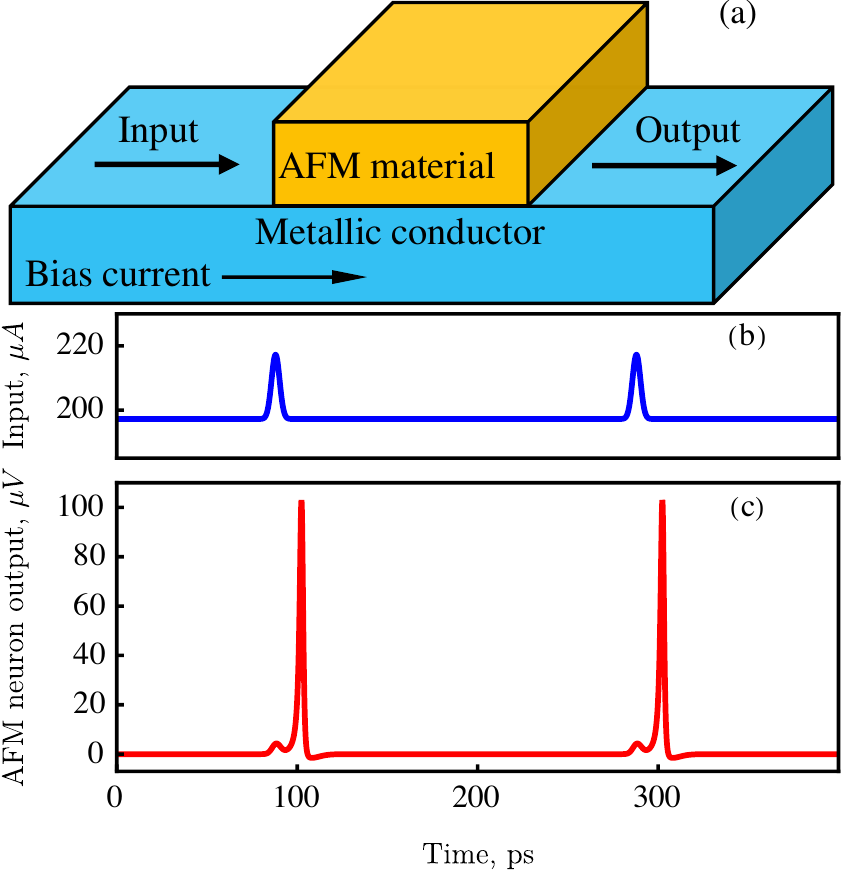}
\caption{\label{neuron} AFM neuron. 
(a) A cartoon of the AFM neuron, which consists of an AFM material and a metallic conductor. Input and output of the neuron are as labeled. 
(b) Simulated input current supplied to the AFM neuron. Bias current is $\sim 198~\mu$A; with input impulses just before 100 ps and 300 ps.
(c) Simulated output spikes from an AFM neuron in response to the input, at 100 and 300 ps.}
\end{figure}

In contrast, the human brain can perform these tasks in real time with a power requirement that is substantially less than 20 watts\cite{levy2021communication}.
It is anticipated that the speed and power efficiency of AI algorithms will be greatly enhanced by the use of specialized brain-inspired, neuromorphic computing hardware\cite{Grollier_Querlioz_Camsari_EverschorSitte_Fukami_Stiles_2020,Capra_Bussolino_Marchisio_Shafique_Masera_Martina_2020,Belkhir_Elmeligi_2018,Masanet_Shehabi_Lei_Smith_Koomey_2020}. 
It has been proposed that neuromorphic computers that use spikes in a manner similar to biological neurons can greatly improve the computational speed and power consumption of spiking neural networks (SNN), and will be substantially lower than that of conventional ANNs\cite{kendall2020building}. 
For example, Intel recently developed a 2 billion transistor Loihi chip that has an architecture based on spiking neurons\cite{Davies_Srinivasa_Lin_Chinya_Cao_Choday_Dimou_Joshi_Imam_Jain_et2018, davies2021advancing}. 
Before that, IBM designed and fabricated TrueNorth, which has 1 million digital neurons and a power consumption of less than 65 mW\cite{Akopyan_Sawada_Cassidy_AlvarezIcaza_Arthur_Merolla_Imam_Nakamura_Datta_Nam2015, Merolla_Arthur_AlvarezIcaza_Cassidy_Sawada_Akopyan_Jackson_Imam_Guo_Nakamura2014, Esser_Merolla_Arthur_Cassidy_Appuswamy_Andreopoulos_Berg_McKinstry_Melano_Barch2016}.
Other prominent computing platforms that feature spiking neurons include SpiNNaker (University of Manchester)\cite{Furber_Galluppi_Temple_Plana_2014}, Braindrop (Stanford)\cite{neckar2018braindrop, Hemsoth_2017}, and BrainChip\cite{Smith_2021}.
These systems all exhibit improved efficiency by employing a computer architecture that is primarily event driven, much like the human brain.
They employ spiking neurons that consist of multiple silicon based transistors per neuron, which have a behavior that differs substantially from biological neurons\cite{zhu2020comprehensive}.

Artificial neurons composed of antiferromagnetic (AFM) material, which we call ``AFM neurons'', have characteristics that resemble biological neurons\cite{Khymyn_Lisenkov_Voorheis_Sulymenko_Prokopenko_Tiberkevich_Akerman_Slavin_2018, Sulymenko_Prokopenko_Lisenkov_kerman_Tyberkevych_Slavin_Khymyn_2018, Sulymenko_Prokopenko_2019}. 
A schematic diagram of a nanometer-sized, single element artificial AFM neuron is shown in Fig. \ref{neuron}(a). 
In this figure, the AFM material is shown in yellow, and is in contact with a metallic electrical conductor, shown in blue.
The metallic conductor acts as a terminal for the AFM neuron, with an input and an output.
When an electrical current, with a DC bias current and an input current impulse, passes through the conductor (Fig. \ref{neuron}(b)), the AFM material will respond by generating a short-duration voltage spike, as shown in Fig. \ref{neuron}(c). 
This output spike is similar to an action potential that is generated by a biological neuron.
It is evident from this figure that a characteristic output spike for an AFM neuron is quite fast, with a duration that is $\approx 3$~ps. 
This high speed, which is faster than other artificial neurons, is a direct result of the intrinsic properties of the AFM material. 

As it will be shown in this paper, there are several distinct advantages of AFM neurons over the currently employed silicon based alternatives. 
Firstly, AFM neurons exhibit biologically realistic characteristics including refraction, latency, bursting, and inhibition, all of which can be controlled dynamically. 
These are intrinsic physical characteristics of AFM neurons.
Secondly, the duration of spikes are in the picosecond timescale, allowing for an operational speed that is substantially faster than conventional computers.
Thirdly, the power to operate this device is several orders of magnitude lower than the state of the art.
Fourthly, as this artificial neuron consists of a single element, it has the potential to greatly simplify the design and fabrication stage of manufacture.
Moreover, the spatial dimension is on the nanometer scale so that they can, in principle, be included with integrated circuits that use complementary metal-oxide-semiconductor (CMOS) technology\cite{zahedinejad2018cmos}.
Lastly, AFM neurons operate at room temperature, and when interconnected, will allow the development of physical spiking neural networks.

It is worth noting that AFM neurons belong to a broader class of technology known as spintronics, which have been implemented widely in data storage, among other novel applications. 
There is much research exploring spintronic devices for novel computing architectures\cite{Fukami_Ohno_2018, Grollier_Querlioz_Camsari_EverschorSitte_Fukami_Stiles_2020, krzysteczko2012memristive, Kurenkov_DuttaGupta_Zhang_Fukami_Horio_Ohno_2019, Pan_Naeemi_2018, zhang2020antiferromagnet,cai2019voltage, Sengupta_Roy_2015, Zhang_Tserkovnyak_2020, Louis_Lisenkov_Nikitov_Tyberkevych_Slavin_2016, kurenkov2020neuromorphic}.

This paper is organized as follows. 
In section \ref{arweqwef}, the physics of the AFM neuron will be discussed in simple terms, and the differential equations governing their behavior will be introduced. 
Then, section \ref{wrthe} demonstrates the operation of a single AFM neuron. 
After this, section \ref{rhthwse} will demonstrate the interconnection of neurons and explains how these neurons could be employed in a spiking neural network (SNN). 
Finally, section \ref{gwrrt} demonstrates a few simple neuromorphic circuits that employ unique physical properties of AFM neurons. 
This will be followed by conclusion remarks.

\section{Physics of AFM neuron operation}\label{arweqwef}

This section sets out to explain how AFM neurons function, and the physical principles underlying their behavior. 
It begins by presenting a physical model of an AFM neuron, including equations describing its behavior. 
Then, realistic parameters for AFM neurons are presented, which will allow for simulation of AFM neurons and AFM neural networks.
After this is a brief section with estimates about the speed of performance, size, power consumption, and the feasibility of manufacture.

\subsection{Antiferromagnetic neuron physics}

Magnetic materials can be classified into different categories, including ferromagnetic (FM) materials and antiferromagnetic (AFM) materials. \cite{Baryakhtar, Coey_2010}. 
FM materials are familiar in everyday life as permanent magnets and have a single magnetic lattice with a magnetization $\bm{M}$ pointing in a certain direction\cite{Baryakhtar, Coey_2010}.

\begin{figure}
\includegraphics{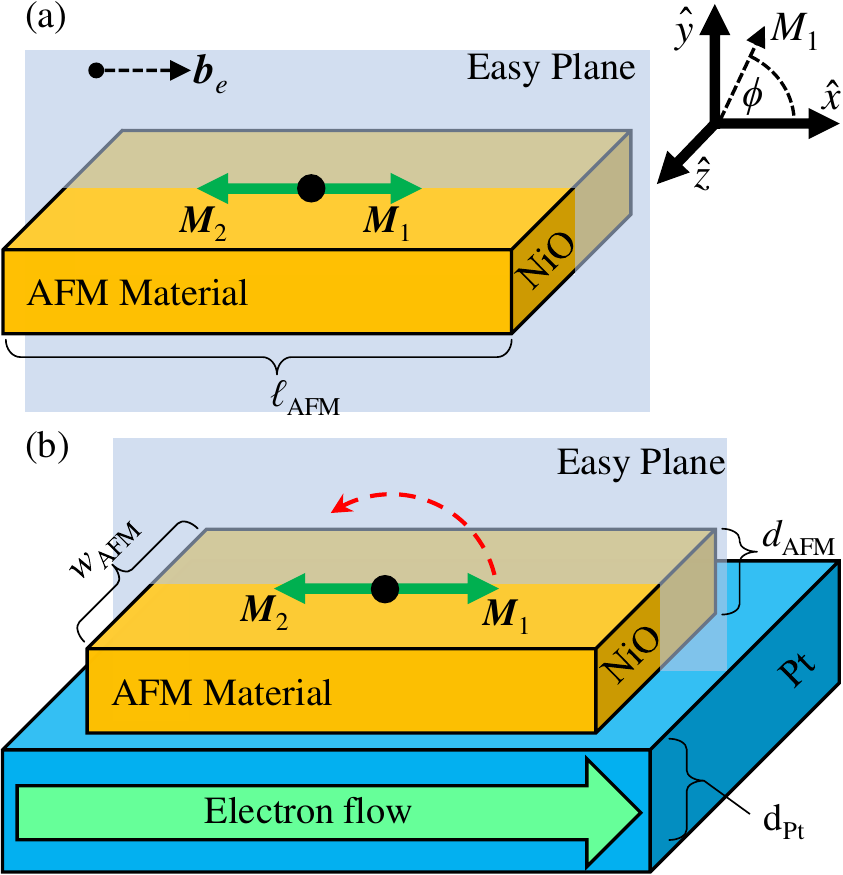}
\caption{\label{afm} AFM neuron configuration. (a) An antiferromagnetic material, NiO, with two anti-parallel magnetizations $\bm{M}_1$ and $\bm{M}_2$. At equilibrium, these two magnetizations are oriented along the easy axis $\bm{b}_e$. Also shown, by a semi-transparent blue color, is the easy plane. (b) The AFM material together with a Pt film. The electric current is flowing through the Pt. When the NiO easy plane is oriented as shown in this figure, the electric current creates a torque on the AFM that causes $\bm{M}_1$ and $\bm{M}_2$ to rotate. Rotation in the easy plane is shown by a dashed red arc.}
\end{figure}

Antiferromagnetic materials are similar to FM materials, except they can have two or more intrinsic magnetic sublattices with different magnetizations.
A visualization of a simple AFM material is shown in Fig. \ref{afm}(a), with two intrinsic magnetizations $\bm{M}_1$ and $\bm{M}_2$.
Due to a very strong exchange interaction, these two magnetizations are inclined to be anti-parallel.
Thus, if the direction of $\bm{M}_1$ is changed, $\bm{M}_2$ will also be reoriented, and vice-versa\cite{Baryakhtar, Coey_2010}.
Due to antiparallel orientation of $\bm{M}_1$ and $\bm{M}_2$, the net magnetic moment of an AFM material vanishes, and thus AFM materials do not create any stray magnetic fields. 
This allows for close packing of AFM elements in integrated circuits, thus greatly increasing areal density compared to FM-based circuits.

AFM materials are much more common in nature than FM materials. 
The great variety of AFM materials allows one to choose, for each particular application, a material with optimal parameters. 
In this article, we consider AFM neurons based on nickel oxide (NiO) --- a high-quality AFM dielectric (insulator), which is currently one of the most extensively studied materials in AFM spintronics\cite{moriyama2018spin,chen2018antidamping,moriyama2019intrinsic}.

AFM materials can be isotropic or anisotropic.
In isotropic AFM materials, the magnetizations $\bm{M}_1$ and $\bm{M}_2$ are free to rotate in any direction, as long as they stay antiparallel to each other. 
In anisotropic AFM materials, on the other hand, there are more and less preferable directions of magnetizations.
Nickel oxide is a bi-axial anisotropic AFM. 
The anisotropy in NiO causes $\bm{M}_1$ and $\bm{M}_2$ to remain in a plane called the ``easy plane'', as shown in Fig. \ref{afm}(a)\cite{Baryakhtar, Coey_2010}. 
Deviations from the easy plane cost a lot of energy and typically can be ignored. 
Within the easy plane, additional in-plane anisotropy in NiO aligns $\bm{M}_1$ and $\bm{M}_2$ with $\bm{b}_e$, the easy axis, which is also shown in Fig. \ref{afm}(a). 
The in-plane anisotropy in NiO is relatively weak, so magnetizations can rotate in the easy plane if disturbed by an external signal. 
This in-plane rotation of $\bm{M}_1$ and $\bm{M}_2$ is the basis of AFM neuron operation.

An AFM neuron can be formed if a metal film with a strong spin-orbit interaction, such as platinum (Pt), is grown onto the AFM material. 
Platinum is an electrical conductor. 
When an electric current flows through the Pt in the geometry depicted in Fig. \ref{afm}(b), the AFM magnetizations can be made to rotate in the easy plane due to the spin Hall effect and spin transfer torque (STT)\cite{Khymyn_Lisenkov_Tiberkevich_Ivanov_Slavin_2017, Khymyn_Tiberkevich_Slavin_2017}. 
As $\bm{M}_1$ rotates with respect to the easy plane, it has an azimuthal angle $\phi$ with the easy axis. 
The azimuthal angle $\phi$ is shown in the coordinate axis at the top right corner of Fig. \ref{afm}.
The dynamic change in $\phi$ can be modeled with the differential equation\cite{Khymyn_Lisenkov_Tiberkevich_Slavin_Ivanov_2016, Khymyn_Lisenkov_Tiberkevich_Ivanov_Slavin_2017, Khymyn_Lisenkov_Voorheis_Sulymenko_Prokopenko_Tiberkevich_Akerman_Slavin_2018}:
\begin{equation} \frac{1}{\omega_{\rm ex}}\ddot{ \phi} + \alpha \dot{\phi} + \frac{\omega_{\rm e}}{2}\sin 2\phi = \sigma I .\label{vteq}\end{equation} 
In this equation, $\omega_{\rm ex} = 2\pi f_{\rm ex}$ is the exchange frequency, $\alpha $ is the dimensionless effective damping parameter, $ \omega_{\rm e} = 2\pi f_{\rm e} $ is the easy axis anisotropy frequency, $\sigma$ is the spin-torque efficiency, and $I$ is the input electric current for this system.
Further details about the physics of AFM neurons and the derivation of equation (\ref{vteq}) can be found in \cite{Khymyn_Lisenkov_Tiberkevich_Slavin_Ivanov_2016, Khymyn_Lisenkov_Tiberkevich_Ivanov_Slavin_2017}. 
In this paper, we refer to (\ref{vteq}) as the artificial neuron equation.

When $\bm{M}_1$ and $\bm{M}_2$ have a non-zero angular velocity $\dot{\phi}$, an impulse of electric voltage is generated via the inverse spin Hall effect\cite{Khymyn_Lisenkov_Tiberkevich_Slavin_Ivanov_2016, Hoffmann_2013, Sinova_Valenzuela_Wunderlich_Back_Jungwirth_2015}. 
This is the output signal of the AFM neuron. 
It is given by a simple equation\cite{Khymyn_Lisenkov_Tiberkevich_Ivanov_Slavin_2017, Khymyn_Lisenkov_Voorheis_Sulymenko_Prokopenko_Tiberkevich_Akerman_Slavin_2018}:
\begin{equation} v(t) = \beta \dot{\phi}(t) ,\label{volts}\end{equation}
where $v(t)$ is a time dependent voltage, and $\beta$ is the spin pumping efficiency.

It is notable that equation (\ref{vteq}) is analogous to a model of the superconducting phase of a Josephson Junction that is capacitively shunted under a bias current \cite{Meckbach_2013, anderson1964lectures} and an artificial neuron based on a Josephson Junction\cite{Schneider_Donnelly_Russek_2018, Crotty_Schult_Segall_2010}.

\subsection{Mechanical analog: simple pendulum}

\begin{figure}
\includegraphics{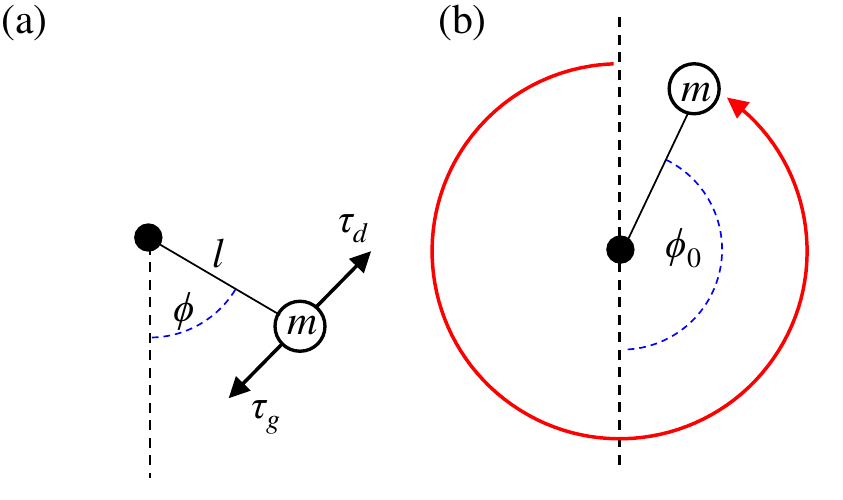}
\caption{\label{wash} Simple pendulum as a mechanical analog of an AFM neuron. 
(a) The mass $m$ is affixed at a distance $l$ from an axis, and is displaced from vertical by an angle $\phi$. It is subject to a torque from gravity $\tau_g$, an external driving torque $\tau_d$, and a frictional torque $\tau_f$. 
(b) The driving torque has displaced the mass by $\phi_0 = 155^\circ$. With a small push, it will pass the vertical line and revolve once about its axis, which is shown by a red arc.}
\end{figure}

To develop an intuitive understanding of (\ref{vteq}), this section investigates a mechanical system that is also described by the artificial neuron equation.
The system is a simple pendulum with an applied torque\cite{Khymyn_Lisenkov_Voorheis_Sulymenko_Prokopenko_Tiberkevich_Akerman_Slavin_2018}.
A schematic of the pendulum is shown in Fig. \ref{wash}(a).
In this schematic, an object with a mass $m$ is situated at a distance $l$ from its axis of rotation. 
It is attached to the axis of rotation by a thin rod, which is assumed to be both rigid and weightless. 
The mass is subject to a gravitational force with an acceleration $g$.
The pendulum is displaced from the vertical axis by the angle $\phi$, as depicted in Fig. \ref{wash}(a). 
It is assumed that the mass is free to rotate continually about its axis, although the rotation is subject to viscous friction.

In this model, the mass is affected by three torques. 
Firstly, the mass is subject to a gravitational force; so the torque due to gravity is given by $\tau_g = -mgl\sin\phi$.
Secondly, it is assumed that there is an external driving torque applied to the mass, which we define as $\tau_d$. 
Thirdly, there is a frictional torque, which is given by $\tau_f=-b \dot{\phi}$, where $b$ is a damping constant, and $\dot{\phi}$ is the angular velocity of the mass.
Thus, the damping torque $\tau_f$ is proportional to the angular velocity and acts against the motion of the pendulum.

Before deriving the equation of motion, it is useful to develop an intuitive feel for the motion of the pendulum. 
Three cases will be considered here.
First, when there is no driving torque, $\tau_d = 0$, the mass will be in equilibrium with an equilibrium angle $\phi_0$ that is zero; $\phi_0 = 0$. 
If the position of the pendulum is momentarily displaced, the mass will oscillate back and forth until friction returns the mass to its equilibrium position.

Consider a second situation, where there is a very large driving torque.
In this case, the mass will revolve repeatedly around the axis of rotation, with a velocity $\dot{\phi}$ that varies depending on instantaneous angle $\phi$ of the mass. 
Specifically, when $\phi>180^\circ$, the angular velocity will increase under the influence of gravity, and when $\phi<180^\circ$, gravity will act to slow the angular velocity of the pendulum. 
These changes in velocity will be approximately sinusoidal in form. 

The two previous situations are relevant for understanding the mechanics of this system.
The third situation is analogous to the functioning of the AFM neuron, as will be explained with an example that is depicted in Fig. \ref{wash}(b).
In this example, a moderate torque $\tau_d$, which is not sufficient to cause permanent rotations, is applied. 
Instead, the torque will shift the equilibrium angle $\phi_0$ of the mass to a point where $\tau_d$ is compensated by the gravity torque, $\tau_g(\phi_0) = \tau_d$; (in Fig. 3(b) $\phi_0 = 155^\circ$).
If the mass is subject to a small perturbation, the pendulum will oscillate about $\phi_0$ until damping brings the mass back to its equilibrium position.
However, if the perturbation is large enough that the mass passes the threshold angle $\phi_{\rm th} = 180^\circ$ (vertically above the axis of rotation), the mass will follow the path depicted by a solid red arc in Fig. \ref{wash}(b). 
When $\phi > \phi_{\rm th}$, $\tau_d$ and $\tau_g$ are in the same direction, and the mass will accelerate rapidly. 
This rapid acceleration continues until the mass passes $\phi = 0^\circ$.
After this, when $\phi>0^\circ$, $\tau_g$ works in opposition to $\tau_d$.
For a moderate perturbation, the gravitational torque and friction will be sufficient to slow the mass, and the mass will return to $\phi_0 = 155^\circ$. 
Of course, if the momentary perturbation is large enough, the pendulum may rotate two or more times.

It is clear from this mechanical analog that, the closer the equilibrium angle $\phi_0$ is to the threshold angle $\phi_{th}$, a smaller perturbation will cause a single pendulum rotation. 
Thus, by applying torque $\tau_d$ close to its threshold value, one can substantially increase the pendulum ``sensitivity''. 
At the same time, the angular speed of the pendulum during the rotation is determined, mostly, by the gravity and bias torque and is almost independent of the amplitude of the initial ``kick''. 
In this regime the pendulum works as a threshold element, producing the same rotation every time input perturbations exceed a certain critical value and, in this sense, is very similar to a biological neuron.

The three torques can be used to derive an equation of motion, which can be found from Newton's second law for rotation, $I_m \ddot{\phi} = \sum{\tau}$.
Here $I_m$ is the moment of inertia of the system, $\ddot{\phi}$ is the angular acceleration of the mass, and $\sum{\tau}$ is the sum of the torques applied to the system.
Thus, we can explicitly write the equations of motion by substituting the torques in this equation:
\begin{equation} I_m \ddot{\phi} + b \dot{\phi}+mgl \sin\phi = \tau_d .\label{pen2}\end{equation}
By comparing this equation with (\ref{vteq}), it is evident that this system is analogous to that of an AFM neuron.
It is useful to describe each term in this equation individually, in order to gain a better understanding of this system.

The first term on the left-hand side of (\ref{pen2}), $I_m\ddot{\phi}$, contains the moment of inertia. 
The moment of inertia describes how difficult it is to change the angular velocity of an object rotating about an axis. 
When comparing the mechanical analog in (\ref{pen2}) with the artificial neuron equation (\ref{vteq}), it is evident that the first term on the left-hand side of (\ref{vteq}) models the inertia of the system.
This implies that $(1 / \omega_{ex})$ is analogous to the moment of inertia.
Thus, the exchange interaction in the AFM material leads to the AFM neuron having an effective inertia. 
This effective inertia is one of the factors that makes AFM neurons unique, giving AFM neurons properties like a response latency and a finite refraction time.

The second term on the left-hand side of (\ref{pen2}), $b\dot{\phi}$, represents the damping due to friction experienced by the pendulum. 
Friction acts to slow any movement by the pendulum. 
The magnitude of $b$ determines how much friction is present. 
In the pendulum model, we consider $b$ to be related to air resistance. 
To decrease $b$, the air can be removed, so that the pendulum moves with no friction in a perfect vacuum. 
Likewise, $b$ can be increased by immersing the pendulum in a viscous fluid like water or honey. 
By comparison with the artificial neuron equation, the second term on the left-hand side of (\ref{vteq}), $\alpha\dot{\phi}$, represents damping. 
When $\bm{M}_1$ has a non-zero angular velocity $\dot{\phi}$, damping will slow this rotation. 
The magnitude of $\alpha$ determines how much damping is present in this system. 
The damping acts like a frictional force to slow $\bm{M}_1$ and $\bm{M}_2$, similar to the damping in a pendulum. 
For the AFM neuron, $\alpha$ to a certain extent can be controlled in fabrication. 
Bulk NiO, for example, has an intrinsic effective damping of about $ 10^{-4}$; in a thin film $\alpha$ will be larger, depending on fabrication characteristics\cite{moriyama2019intrinsic}. 
The effective damping will increase when NiO is fabricated as a thin film.
It will also increase when covered by a conducting thin film, due to spin pumping\cite{Verba_Tiberkevich_Slavin_2018}.
Thus, how the AFM neuron is fabricated will impact the effective damping.

The third term on the left-hand side of (\ref{pen2}) is $mgl\sin\phi$.
This is the gravitational potential for the pendulum. 
This potential defines a preferable orientation (i.e., anisotropy) of the pendulum ($\phi = 0^\circ$) and governs how the velocity of the mass will change as it revolves around the axis.
Likewise, the third term on the left-hand side of (\ref{vteq}), $ (\omega_{\rm e}/2)\sin 2\phi$, models the anisotropy of the AFM material. 
This anisotropy governs how the velocity of $\bm{M}_1$ and $\bm{M}_2$ will change as it rotates. 
It is notable that $\omega_e$, the easy axis anisotropy frequency, is present in this term.
When $\bm{M}_1$ and $\bm{M}_2$ are rotated under the action of the external spin current, the anisotropy of the AFM acts to realign $\bm{M}_1$ and $\bm{M}_2$ with the easy axis. 

Finally, the term on the right-hand side of (\ref{pen2}), $\tau_d$, is the externally applied driving torque to the pendulum. 
When at equilibrium, this torque displaces the pendulum to a position of $\phi_0$. 
When in motion, this torque acts to increase the angular velocity.
Likewise, in the model of the AFM neuron, the term on the right-hand side of (\ref{vteq}) represents the spin transfer torque applied by the spin current to $\bm{M}_1$ and $\bm{M}_2$ in the AFM neuron. 
STT is analogous to the driving torque in the pendulum. 
The STT depends on the current $I=I_{\rm dc} + i_{\rm p}(t)$ that flows through the Pt substrate. 
Here, $I_{\rm dc}$ is a bias current that brings the angle of the magnetization to an initial angular displacement of $\phi_0$, and $i_{\rm p}(t)$ represents a momentary perturbation, that incites the magnetization to rotate. 

The magnitude of the bias current determines the regime in which the AFM neuron is operating.
That is, if the bias current exceeds some threshold current $I_{\rm th}$, the system will behave as an auto-oscillator, with magnetization $\bm{M}_1$ and $\bm{M}_2$ continually rotating in the easy plane. 
In a biaxial AFM material like NiO, the transition to the auto-oscillatory regime occurs when the magnetizations cross the threshold angle $\phi_{\rm th} = 45^\circ$, making the threshold current\cite{Khymyn_Lisenkov_Voorheis_Sulymenko_Prokopenko_Tiberkevich_Akerman_Slavin_2018}
\begin{equation}I_{\rm th} = \frac{\omega_e}{2 \sigma} \label{trrfgv} . \end{equation}
Thus, for this system to be in the neuronal regime and thus exhibit behavior like that of a typical biological neuron, it is required that the bias current be sub-threshold, $I_{\rm dc}<I_{\rm th}$. 
When $I_{\rm dc}<I_{\rm th}$, the initial angular displacement of the magnetization will be $\phi_0 = \arcsin(I/I_{\rm th})/2$. 

\subsection{Physical implementation and simulation parameters} 

This subsection will discuss the physical implementation of an AFM neuron.
It begins by presenting physical parameters that allow for the realistic simulation of AFM neuron behavior.
After this, the feasibility of implementing an AFM neuron is discussed.

A summary of material constants related to equations (\ref{vteq}) and (\ref{volts}) is given in Table \ref{tehfsa}.
Many of the values in this table are the same as those presented in an earlier work\cite{Khymyn_Lisenkov_Tiberkevich_Ivanov_Slavin_2017}, and are specific for an AFM neuron composed of a NiO/Pt bilayer.
The spin-torque efficiency $\sigma$, and the voltage phase proportionality constant $\beta$, are given by\cite{Khymyn_Lisenkov_Tiberkevich_Ivanov_Slavin_2017}
\begin{equation}\sigma = \eta\frac{|\gamma|}{M_s d_{\rm AFM} w_{\rm AFM}d_{\rm Pt} } , \label{sigs}\end{equation}
\begin{equation}\beta = \eta\frac{\ell_{\rm Pt}}{d_{\rm Pt}} , \label{bets}\end{equation}
where $\gamma$ is the gyromagnetic ratio, $M_s$ is the saturation magnetization of one NiO sublattice, $d_{\rm AFM}$ is the thickness of the NiO, $w_{\rm AFM}$ is the width of the NiO/Pt interface, $d_{\rm Pt}$ is the thickness of the Pt, and $\ell_{\rm Pt}$ is the lenght of the NiO/Pt interface. 
These physical dimensions were shown in Fig. \ref{afm}.
The constant $\eta$ is given by\cite{Khymyn_Lisenkov_Tiberkevich_Ivanov_Slavin_2017}
\begin{equation} \eta= \Big[ \theta_{\rm SH} \frac{g_r e \lambda \rho }{2 \pi } \Big] \tanh\frac{d_{Pt} }{2\lambda} , \label{thewfwew}\end{equation}
where $\theta_{\rm SH}$ is the spin Hall angle, $g_r$ is the spin mixing conductance, $e$ is the magnitude of the fundamental electric charge, $\lambda$ is the spin-diffusion length in Pt, and $\rho$ is the resistivity of thin film Pt.

\begin{table}[tbp]
\begin{tabular}{lll}
\multicolumn{1}{l|}{Parameter~~~} & \multicolumn{1}{l|}{Description} & Simulation value \\ \hline
\multicolumn{1}{l|}{$f_{\rm ex}$} & \multicolumn{1}{l|}{Exchange frequency} & 27.5 ~THz \\
\multicolumn{1}{l|}{$\alpha$} & \multicolumn{1}{l|}{Effective damping} & $0.001$ -- 0.1 \\
\multicolumn{1}{l|}{$f_{\rm e}$} & \multicolumn{1}{l|}{Easy axis anisotropy frequency} & 1.75 ~GHz \\
\multicolumn{1}{l|}{~} & \multicolumn{1}{l|}{~} & ~ \\
\multicolumn{1}{l|}{$|\gamma|/2\pi$} & \multicolumn{1}{l|}{Gyromagnetic ratio} & 28 ~GHz/T \\
\multicolumn{1}{l|}{$M_s$} & \multicolumn{1}{l|}{Saturation magnetization} & 351 ~kA/m \\
\multicolumn{1}{l|}{$\theta_{SH}$} & \multicolumn{1}{l|}{Spin Hall angle} & 0.1 \\
\multicolumn{1}{l|}{$g_r$} & \multicolumn{1}{l|}{Spin mixing conductance} & $6.9 \times 10^{18}$ m$^{-2}$ \\
\multicolumn{1}{l|}{$e$} & \multicolumn{1}{l|}{Elementary charge} & $1.6 \times 10^{-19}$ C \\
\multicolumn{1}{l|}{$\lambda$} & \multicolumn{1}{l|}{Pt spin diffusion length} & 7.3~nm \\
\multicolumn{1}{l|}{$\rho$} & \multicolumn{1}{l|}{Resistivity of Pt} & $4.8\times10^{-7}~\Omega \cdot$m \\
\multicolumn{1}{l|}{~} & \multicolumn{1}{l|}{~} & ~ \\
\multicolumn{1}{l|}{$d_{\rm AFM}$} & \multicolumn{1}{l|}{NiO thickness} & 5 nm \\
\multicolumn{1}{l|}{$w_{\rm AFM}$} & \multicolumn{1}{l|}{NiO/Pt interface width} & 10 nm \\
\multicolumn{1}{l|}{$\ell_{\rm AFM}$} & \multicolumn{1}{l|}{NiO/Pt interface length} & 40 nm \\
\multicolumn{1}{l|}{$d_{\rm Pt}$} & \multicolumn{1}{l|}{Pt thickness} & 20 nm \\
\multicolumn{1}{l|}{~} & \multicolumn{1}{l|}{~} & ~ \\
\multicolumn{1}{l|}{$\eta$} & \multicolumn{1}{l|}{~ } & $5.4\times 10^{-17}$ V$\cdot$s \\
\multicolumn{1}{l|}{$\sigma$} & \multicolumn{1}{l|}{Spin-torque efficiency} & $27.1\times 10^{12}$ rad/A$\cdot$s \\
\multicolumn{1}{l|}{$\beta$} & \multicolumn{1}{l|}{Spin pumping efficiency} & $0.11\times 10^{-15}$ ~V$\cdot$s \\
\multicolumn{1}{l|}{$I_{\rm th}$} & \multicolumn{1}{l|}{Threshold current } & 0.203 mA \\
\end{tabular}
\caption{AFM material parameters, fundamental constants, and physical dimensions used in simulation.}
\label{tehfsa}
\end{table}

It should be noted in equations (\ref{vteq}) and (\ref{volts}), that by increasing the values of $\eta$, $\sigma$, and $\beta$, the performance of the AFM neuron will be improved. 
Therefore, the performance characteristics of the AFM neuron depend on the physical dimensions of the AFM neuron; specifically, $d_{\rm AFM}$, $w_{\rm AFM}$, and $d_{\rm Pt}$.
For example, in (\ref{sigs}), the value of $\sigma$ increases for small values of $d_{\rm AFM}$ and $w_{\rm AFM}$. 
Therefore, we have chosen $d_{\rm AFM} = 5$~nm and $w_{\rm AFM} = 10$~nm as simulation parameters. 
These values are small, yet reasonable for nano-fabrication.

Likewise, $d_{\rm Pt}$ appears in (\ref{sigs}), (\ref{bets}), and (\ref{thewfwew}). 
To maximize $\eta$, a thickness with $d_{Pt}> 2\lambda$ should be chosen.
However, a thin $d_{\rm Pt}$ is beneficial to maximize $\sigma$ and $\beta$.
Taking this into consideration, $d_{Pt} \sim 20 $ nm was chosen as a simulation parameter. 
This thickness is reasonable for nano-fabrication. 

The size of the entire structure must be considered for the choice of $\ell_{\rm AFM}$, the length of the NiO/Pt interface. 
In general, it is preferred for modern micro-electronic structures to be as small as possible.
Thus, the NiO material should be small enough to be a single crystal with a single domain, yet large enough to have a magnetization that is not disturbed by thermal noise. 
That is, the volume of the AFM material should be large enough that the magnetization will not spontaneously rotate in the easy plane due to thermal energy. 
The energy required for thermal noise to rotate the magnetization is equal to $V_{\rm AFM} B_e M_s$, where $V_{\rm AFM}$ is the volume of the AFM material and $B_e = \omega_e/|\gamma|$ is the easy axis anisotropy field. 
The thermal energy is equal to $k_B T$, where $k_B$ is the Boltzmann constant and the temperature is $T =300$~K. 
Thus, a reasonable volume would be $V_{\rm AFM}>10 (k_B T/B_e M_s)$, which corresponds to an AFM material with a volume of $V_{\rm AFM}\sim 2000$ nm$^3$.
For our chosen $d_{\rm AFM}$, this corresponds to a NiO/Pt interface with a surface area of $400$~nm$^2$. 
This surface area is on par with the size of CMOS transistors. 
With this surface area, and $w_{\rm AFM}$ defined above, it is appropriate to use a simulation parameter for the length of the NiO/Pt interface as $\ell_{\rm AFM}=40$ nm. 

With these size parameters, the spin torque efficiency is approximately $\sigma = 27.1\times 10^{12}$~rad/A$\cdot$s, and the spin pumping efficiency is $\beta \sim 0.11\times 10^{-15}$ V$\cdot$s/rad.
Considering that a typical value for $\dot{\phi}$ is about 1~rad/ps, we can assume that altogether, voltage spikes produced by an AFM neuron will have a magnitude on the order of $\beta\dot{\phi}\approx100~\mu$V, with a duration as short as 1 ps.
This was demonstrated in Figure 1. Also, with this $\sigma$, the threshold current is $I_{\rm th} \sim 0.2$~mA.

The effective damping $\alpha$ for NiO is shown in Table \ref{tehfsa} to vary between 0.001 and 0.1. 
Damping in NiO depends on the technology of film preparation, and therefore there is some flexibility in how $\alpha$ can be varied in simulation. 
Smaller damping parameters lead to a shorter spike duration. 

It is important to note that AFM neurons require a DC electric current $I_{\rm dc}$ to be constantly flowing through the Pt substrate. 
The purpose of this current, as stated earlier, is to bias the magnetization so that $\phi_0$ is close to $\phi_{\rm th}$, so it can generate a spike with the receipt of a small current impulse $i_{\rm p}(t)$. 
The energy consumption of an AFM neuron can be estimated by considering $I_{\rm dc}$.
Previously, it was estimated that a current near the threshold bias current of $I_{\rm th} = 0.2$~mA would be required to drive the rotation of $\bm{M}_1$ and $\bm{M}_2$. 
If the resistance of the platinum beneath the NiO is given by $R_{\rm pt}=\rho \ell_{\rm AFM}/ d_{\rm Pt} w_{\rm AFM}$, then the power consumption of this structure can be estimated as $I_{\rm th}^2R_{\rm Pt} \sim 4$~\textmu W. 
If the time per synaptic operation is conservatively estimated as $100$~ps, the energy consumption of a single AFM neuron is $\sim 10^{-3}$~pJ per synaptic operation. 
This power consumption can be compared with other spiking neuromorphic hardware \cite{Frenkel_Lefebvre_Legat_Bol_2018}. 
For example, the energy per synaptic operation for the Intel Loihi Neuromorphic Chip was reported to be 20~pJ \cite{Esser_Merolla_Arthur_Cassidy_Appuswamy_Andreopoulos_Berg_McKinstry_Melano_Barch2016}. 
The efficiency of an AFM neuron can be evaluated in units of SOPS/W, where SOPS stands for synaptic operations per second \cite{kendall2020building}. 
Using this measure, a single AFM neuron will have a performance of 10 TSOPS and an efficiency of about 2500 TSOPS/W, which far more efficient than comparable systems reported in\cite{kendall2020building}.
The high efficiency of AFM neurons is a direct result of its high speed of operation, which is a result of the high exchange frequency of NiO and other antiferromagnetic materials. 

It is worth noting that platinum was chosen for simulations because it is relatively straightforward to fabricate and has an acceptable spin Hall angle $\theta_{SH} \sim 0.1 $. 
However, it may be possible to employ a topological insulator with a much higher spin Hall angle, for example $\theta_{SH} \sim 50$ for BiSb\cite{khang2018conductive}. 
The use of a material with a higher spin Hall angle will greatly improve the power efficiency of this system. 
For example, use of material with $\theta_{SH} \sim 10$ will decrease power consumption to $\sim 10^{-7}$ pJ per synaptic operation and, at the same time, increase output spike amplitude to $\sim$ 10 mV.

At present, AFM neurons have not yet been fabricated.
The reason relates to the difficulty in fabricating an insulating structure with nanometer dimensions. 
To function properly, the AFM structure should be fabricated with several important characteristics.
As previously mentioned, the structure should be large enough so that the magnetization will not rotate spontaneously due to thermal noise, but small enough that it is monocrystalline and of a single domain.
Additionally, the easy plane anisotropy in the AFM film must be oriented correctly with respect to the Pt film plane, and a clean interface between the substrate and the AFM material is required for an efficient interfacial effect. 
These technical challenges will need to be overcome prior to the experimental demonstration of an AFM neuron. 
However, we believe that the performance of this system justifies expanded research in this area. 
In addition, it has been suggested that a neuron with properties similar to an AFM neuron can be constructed from a magnetic tunnel junction using available fabrication technology\cite{louis2022Magnonics, louis2022, liu2020synthetic}. 

Active research related to AFM materials is not limited to AFM neurons; at present much research is ongoing for both fundamental science and applied technology related to AFM materials. 
There have been recent proposals to employ AFM materials in THz signal generators, detectors, spectrum analysis, among other ideas\cite{ Sulymenko_Prokopenko_Tiberkevich_Slavin_Ivanov_Khymyn_2017, Sulymenko_Prokopenko_Tiberkevich_Slavin_2018, Artemchuk_Sulymenko_Louis_Li_Khymyn_Bankowski_Meitzler_Tyberkevych_Slavin_Prokopenko_2020, parthasarathy2021precessional, gomonay2018narrow}.
There also has been much research related to AFM based memory applications, which is relevant to the development of AFM neurons\cite{wadley2016electrical,vaidya2020subterahertz, chen2018antidamping, boventer2021room, puliafito2019micromagnetic, shi2020electrical, ni2021temperature, baldrati2019mechanism, chen2019electric, gray2019spin}.

\section{Single AFM neuron behavior}\label{wrthe}

This section discusses the dynamic behavior of an AFM neuron that is modeled by equation (\ref{vteq}).
It will begin by demonstrating the spiking behavior of an AFM neuron by numerical simulation.
After this, the refractory properties of AFM neurons are discussed, as will the response of AFM neurons to polarity changes. 
This section ends by demonstrating the utility of AFM neurons for Boolean logic. 

\subsection{AFM neuron spiking demonstration}\label{wefasdfaef}

Together, equations (\ref{vteq}) and (\ref{volts}) describe the input-output behavior of an AFM neuron; the input is the current $I$ and the output is the voltage $v(t)$. 
In most cases, there is no general analytical solution; thus (\ref{vteq}) can only be solved numerically. 

\begin{figure}
\includegraphics{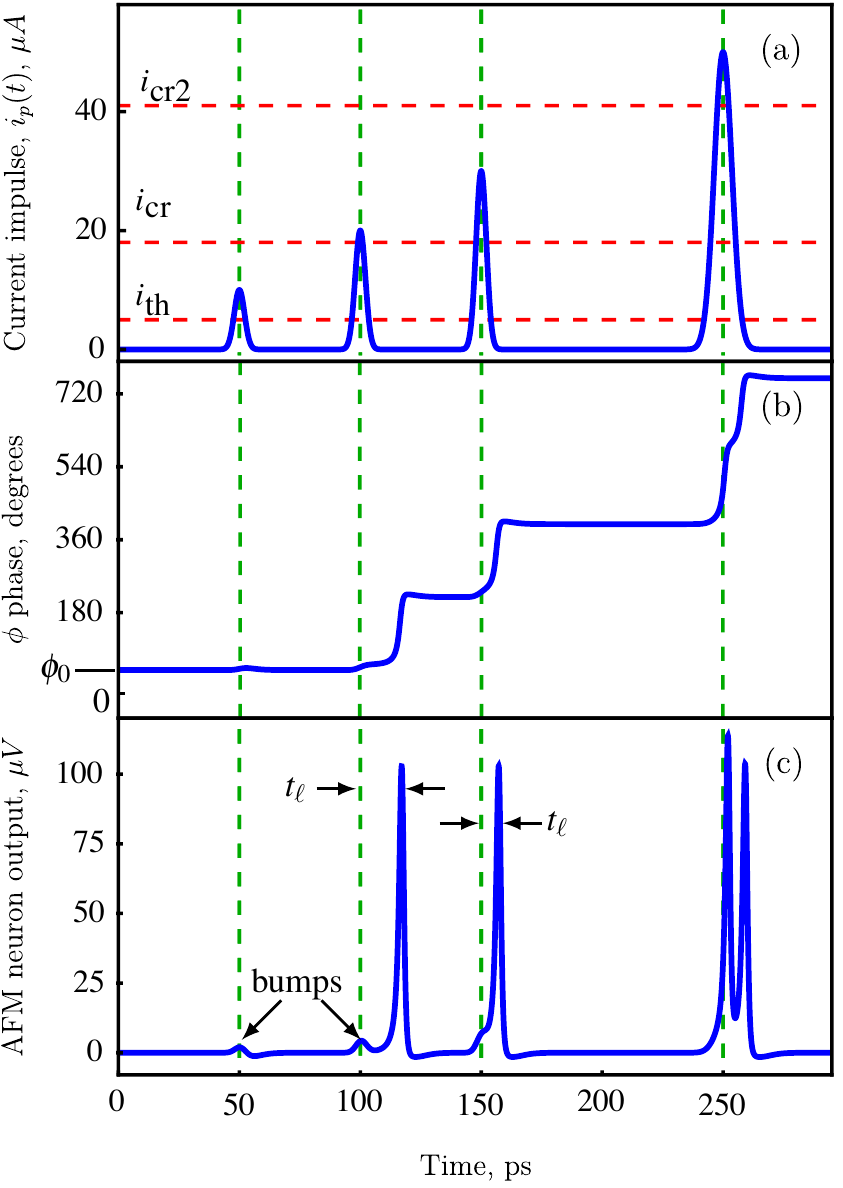}
\caption{\label{spikes}
Input current and simulated AFM neuron output. 
(a) The simulated input current. The blue curve represents the input current $i_p(t)$, the lower dashed pink line represents the threshold current $i_{\rm th}$, the center red dashed line represents critical current, $i_{\rm cr}$, and the top red dashed line represents the critical current for bursting behavior, $i_{\rm cr2}$. 
(b) The simulated azimuthal angle $\phi$ of $M_1$ in the easy plane. Green vertical dashed lines show the time of input perturbations. 
(c) Voltage output of the AFM neuron. The latency $t_\ell$ for two spikes is labeled. In this simulation, $\alpha = 0.009$.}
\end{figure}

Numerical simulations of (\ref{vteq}) and (\ref{volts}) were performed to demonstrate an AFM neuron generating spikes. 
Results of these simulations are shown in Fig. \ref{spikes}.
The simulations were performed with a bias current of $I_{\rm dc} = 198$~\textmu A, which is 5~\textmu A below the threshold current $I_{\rm th}= 203$~\textmu A. 
As mentioned earlier, $I_{\rm th}$ is the threshold current that determines when the AFM neuron is in the neuronal regime ($I_{\rm dc}<I_{\rm th}$), and when it behaves as an auto-oscillator ($I_{\rm dc}>I_{\rm th}$). 

The input to the AFM neuron, current impulse $i_p(t)$, is shown in Fig. \ref{spikes}(a).
In this simulation, $i_p(t)$ has four peaks. 
The threshold current $i_{\rm th}=I_{\rm th}-I_{\rm dc}$ which is 5~\textmu A above the bias current, is represented in the figure by a red dashed line. 
Note, that the \emph{threshold} current $i_{\rm th}$ is the input signal amplitude that leads to magnetization rotation in DC regime, i.e., for a \emph{very long} duration of the input signal. 
To induce rotation with a \emph{short} input pulse, its amplitude should exceed a certain \emph{critical} value $i_{\rm cr}$, which depends on the input duration, and for the parameters of Fig. \ref{spikes}, equals $i_{\rm cr} = 18$~\textmu A. 
For further increase of the input current amplitude ($i_p > i_{\rm cr2} = 41$~\textmu A), one input pulse will generate burst spikes, as explained below. 
The four input pulses in Fig. 4(a) have different amplitudes: one is below the critical current ($i_{\rm th} < i_p < i_{\rm cr}$), two pulses of different amplitudes in the range $i_{\rm cr} < i_p < i_{\rm cr2}$, and the last spike in the burst range $i_p > i_{\rm cr2}$.

Fig. \ref{spikes}(b) and (c) show the response of the AFM neuron to the current impulse.
Fig. \ref{spikes}(b) shows the azimuthal angle $\phi$ of $\bm{M}_1$, and Fig. \ref{spikes}(c) shows the output voltage according to equation (\ref{volts}).
It is evident from this figure that for the first 50~ps, in the absence of a current impulse, that $\phi$ remains constant, and thus $\dot{\phi}=0$.
The output voltage, which is proportional to $\dot{\phi}$, remains at zero during this interval.

At $t=50$~ps, there is a current impulse with an amplitude of 10~\textmu A. 
This impulse is larger than $i_{\rm th}$ but is is less than $i_{\rm cr}$.
Because the impulse is less than the critical current, the magnetization does not rotate, and there is no output spike. 
This current impulse, however does cause a small movement in $\bm{M}_1$, as can be seen in the small bump at $t=50$~ps in Fig. \ref{spikes}(c). 


Next, at $t=100$~ps, there is a current impulse of 20 \textmu A which exceeds $i_{\rm cr}$.
The response to this input current impulse consists of, first, a small bump at $t = 100$~ps (see Fig. \ref{spikes}(c)).
Then, after a delay of $t_\ell \sim 10$~ps, the neuron fires. 
This can be seen at time $t=110$~ps in the 180$^\circ$ rotation of $\phi$ in Fig. \ref{spikes}(b), and in the voltage spike in Fig. \ref{spikes}(c).

The behavior of the AFM neuron at $t=100$~ps demonstrates two properties which closely resemble the behavior of biological neurons. 
Firstly, AFM neurons follow the ``all-or-nothing'' law\cite{squire2012fundamental}.
That is, these neurons only fire when they receive sufficient stimulus.
Biological neurons only elicit action potentials when the membrane potential rises above a threshold, which then depolarizes the neural membrane.
In the AFM neuron, this is equivalent to the input current being momentarily larger than the critical current.

Secondly, concerning the delay $t_\ell$. 
In biological neurons, this delay is called the \emph{neuronal response latency}, which is the time between the input stimulus and the activation potential \cite{levakova2015review}. 
In AFM neurons, the delay is due to the time it takes for the magnetization to rotate past the threshold angle. 
Visualized as a simple pendulum, the perturbation provides just enough energy for $\phi$ to pass $\phi_{\rm th}$.
However, the velocity will be very slow as $\phi$ approaches the critical angle. 
Then, once $\phi$ is past $\phi_{\rm th}$, the mass will accelerate.


The duration of the response latency depends on the magnitude of the input impulse. 
This is demonstrated by the response of the AFM neuron at time $t=150$~ps.
At this time, the current impulse is 30~\textmu A, which is larger than the previous spike.
Once again, the AFM neuron magnetization rotates by 180$^\circ$, as shown in Fig. \ref{spikes}(b), and the rotation brings about a voltage spike, as shown in Fig. \ref{spikes}(c). 
Please note that this voltage spike has the same amplitude and duration as the previous voltage spike.
However, the response latency has a shorter duration because the AFM neuron received a larger input current impulse. 

Lastly, consider the behavior of the AFM neuron at $t=250$~ps.
At this time, the momentary increase in the instantaneous electric current is larger than $i_{\rm cr2}$. 
When the electric current is larger than $i_{\rm cr2}$, the AFM neuron will exhibit bursting behavior. 
This can be seen in Fig. \ref{spikes}(c), where the AFM neuron shows a double peak, and in Fig. \ref{spikes}(b), where $\vb{M}_1$ rotates by 360$^\circ$, twice as much as for the previous two impulses. 
In biological neurons, this behavior also known as bursting\cite{overton1997burst, krahe2004burst, izhikevich2004model, gerstner2014neuronal, aoyagi2001bursting}. 
This is analogous to the pendulum in Fig. \ref{wash} rotating twice about its axis. 

\begin{figure}
\includegraphics{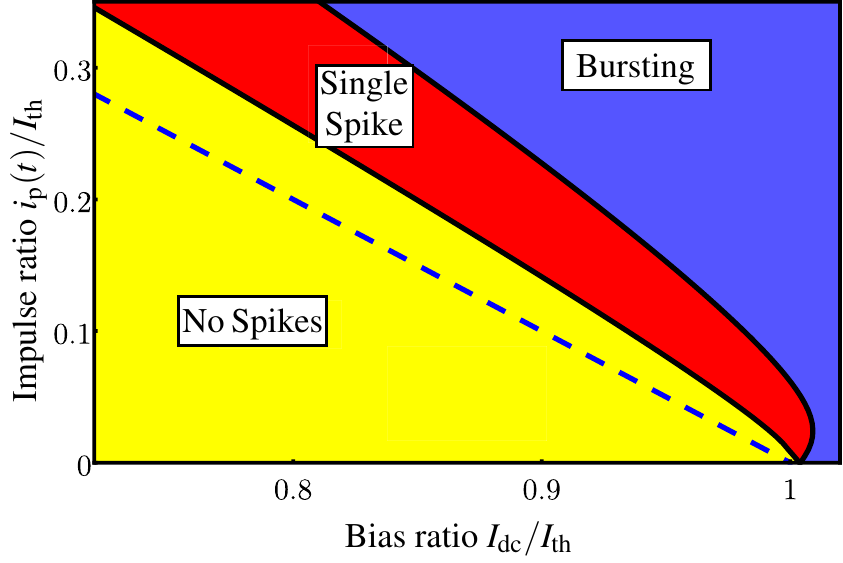}
\caption{\label{phase} AFM neuron spike characteristics. For the yellow region, the AFM neuron produces no spikes. In the red region, the AFM neuron generates a single spike. In the blue region, the AFM neuron exhibits bursting behavior. The behavior depends on the relative values of $i_p(t)$ and $I_{\rm dc}$. In this simulation, $\alpha = 0.01$. Figure reproduced from \cite{Khymyn_Lisenkov_Voorheis_Sulymenko_Prokopenko_Tiberkevich_Akerman_Slavin_2018}.}
\end{figure}

For AFM neurons, $i_{\rm cr} \ne i_{\rm th}$; thus, it is expected that a change in the relative magnitudes of $I_{\rm dc}$ and $i_{\rm p}(t)$ will lead to different behaviors. 
This is demonstrated in Fig. \ref{phase}, which was made via simulation with a sinusoidal input for $i_{\rm p}(t)$ and a bias current $I_{\rm dc}$\cite{Khymyn_Lisenkov_Voorheis_Sulymenko_Prokopenko_Tiberkevich_Akerman_Slavin_2018}.
The axes are scaled as a ratio of $I_{\rm dc}/I_{\rm th}$ and $i_{\rm p}(t)/I_{\rm th}$.
In this figure, there are three regions: single spike (red), bursting (blue), and no spikes (yellow). 
The boundary between the red and yellow regions represents the critical current $i_{\rm cr}$, which is the minimum current impulse required to generate a spike. 
The boundary between the generation of single spiking signals and bursting signals, $i_{\rm cr2}$, corresponds to a line between the red and blue regions. 
Additionally, there is a dashed line in the yellow region, which represents the threshold current $i_p = i_{\rm th} = I_{\rm th} - I_{\rm dc}$ for very long pulses.
One noteworthy characteristic in this plot is that as $I_{\rm dc}$ decreases, the separation between $i_{\rm cr}$ and $i_{\rm th}$ increases, and there is an increased range where single spiking signals can occur.

It can also be seen from Fig. \ref{phase} that when $I_{\rm dc}>I_{\rm th}$, the AFM magnetizations will rotate more than once. 
In this regime, AFM neurons are capable of generating a continuous train of spikes that has a frequency that depends on the amplitude of $I_{\rm dc}$.
Interestingly, biological neurons subject to sustained suprathreshold stimuli will also elicit a train of action potentials whose frequency depends on the strength of the stimulus\cite{gerstner2014neuronal, silverthorn2015human}. 
Biological neurons can exhibit adaptation in spike trains; they feature a change in the duration of interspike intervals.
They can also exhibit stuttering, which are spike trains with inconsistent rhythms\cite{gerstner2014neuronal}.
AFM neurons can also exhibit adaptation by tuning the bias current, and can exhibit stuttering with the introduction of noise to the bias current. 
For both types of neurons, suprathreshold spike generation characteristics are dependent on the size and variation of the stimulus.

\subsection{AFM neuron refraction}\label{evasrvsr}

Due to the effective inertia modeled by the first term in (\ref{vteq}), AFM neurons intrinsically exhibit refraction, a property that will be explored in this subsection. 
This property, that bears close resemblance to the behavior of biological neurons, is a positive attribute for AFM neurons.
Refraction is not intrinsically present in CMOS spiking neuron, although it can be emulated\cite{Frenkel_Lefebvre_Legat_Bol_2018}.

\begin{figure}
\includegraphics{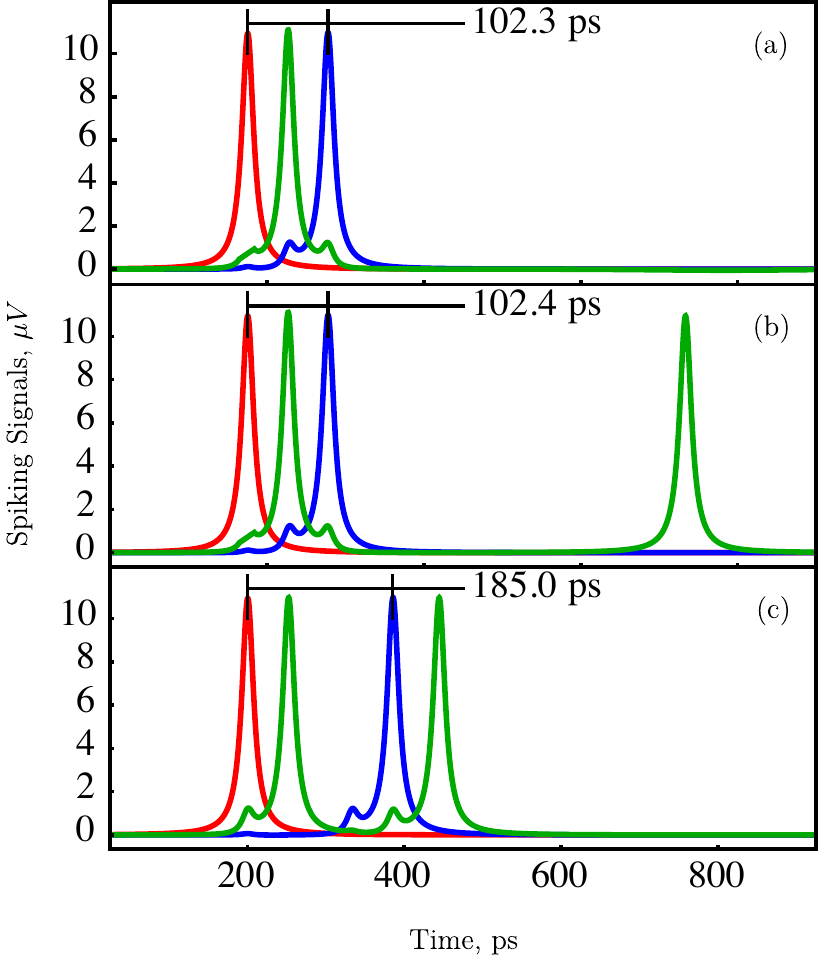}
\caption{\label{refpol} Simulation demonstrating the refractory period of the AFM neuron. For these simulations, inputs to the neuron are shown with red and blue curves, while the output of the neuron is shown with a green curve. 
(a) The red input spike leads to a green output spike with $t_\ell =50$~ps. The blue input spike occurs after an interval of 102.3 ps, during the absolute refractory period. The neuron does not spike a second time.
(b) The blue input spike occurs after an interval of 102.4~ps, during the relative refractory period. The second output spike occurs after a long delay. 
(c) The blue input spike occurs after an interval of 185~ps. The second output spike has the same latency as the first spike, because the AFM neuron is not in its refractory period. 
In these simulations, $\alpha = 0.1$.}
\end{figure}

In AFM neurons, there is a period of time when the AFM magnetizations are in the act of rotation, and the AFM neuron is unable to fire even if it receives an additional impulse $i_{\rm p}(t)$. 
In biological systems, this is called the \emph{absolute refractory period}, which is defined as the time interval after the neuron fires where it cannot fire again\cite{squire2012fundamental}. 
There are also times where the AFM magnetizations are in the act of recovery, and the AFM neuron responds with modified behavior.
In biological systems, this is generally called the \emph{relative refractory period}\cite{squire2012fundamental}. These two refractory properties of an AFM neuron are demonstrated in Fig. \ref{refpol}, and will be discussed below. 

The absolute refractory period will be considered first in Fig. \ref{refpol}(a). 
In this simulation, the first spike arrives at an AFM neuron at $t=200$~ps, as shown by a red curve. 
After a latency of about 50~ps, the neuron fires, as shown by the green curve.
Then, as shown by a blue curve, a second input spike arrives at $t=302.3$~ps, 102.3 ps after the first spike. 
In this case, the second spike does not cause the AFM neuron to spike. 
This is because the AFM neuron is still within its absolute refractory period, and it is unable to fire a second time. 

The simulation results in Fig. \ref{refpol}(b) demonstrates the relative refractory period of AFM neurons.
In this case, a second input arrives at $t=302.4$~ps, which is 102.4 ps after the first spike and 0.1~ps later than above.
In this case, the AFM neuron generates a second spike after a latency of about 400 ps.
Thus, during the relative refractory period the AFM neuron can generate a spike, but with an extended latency

A simulation showing the response of an AFM neuron without refraction is considered in Fig. \ref{refpol}(c). 
Once again, the first spike arrives at $t=200$~ps and after a latency of about 50~ps, the neuron responds with a spike. 
The blue input then arrives at the AFM neuron at $t=385$~ps, which is 185~ps after the initial input. 
Then, once again, after a latency of about 50~ps, the neuron fires, as is shown by a green curve. 
Please note that the latency responses for both inputs are the same.
For the parameters simulated, 185~ps is the minimum time required for the AFM neuron to fire with the same latency. 

The length of the refractory period of an AFM neuron can be dynamically controlled by changing the bias current $I_{\rm dc}$. 
The refractory period also depends on the effective damping, the anisotropy, and the exchange frequency of the AFM material.
Please note that for an AFM neuron with a lower damping parameter, the refractory time will be substantially shorter.

\begin{figure}
\includegraphics{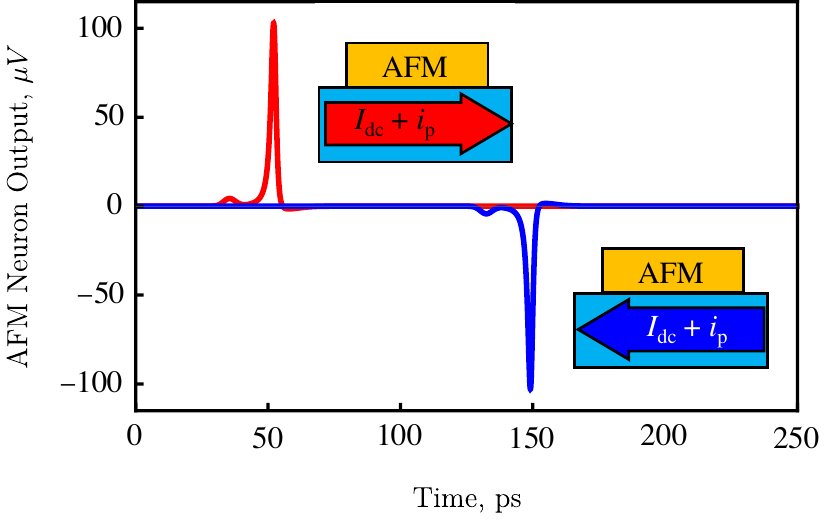}
\caption{\label{pols}
Reversible polarity of a single AFM neuron. The red curve shows the response of the neuron to a positive current that surpasses $i_{\rm cr}$, while the blue curve shows the response of the same AFM neuron to a negative current that surpasses $-i_{\rm cr}$. In these simulations, $\alpha = 0.009$.}
\end{figure}

\subsection{AFM neuron polarity}\label{asdawwfe}

Interestingly, AFM neurons can also reverse their polarity; specifically, they can spike with both positive and negative voltages.
This can be achieved by biasing an AFM neuron with DC currents of different polarity, which induces magnetization rotation in the opposite direction.

AFM neuron behavior for different bias polarities is illustrated in Fig. \ref{pols}. 
Here red curves show responses of AFM neurons with positive current, and blue curves show the response of AFM neurons to negative current.
It is evident from the figure that the polarity of AFM neuron spikes can be reversed in response to a change in current direction.

%

Unlike AFM neurons, biological neurons are unable to produce negative action potentials in response to negative stimuli.
However, biological neurons can be inhibited by hyper-polarization stimuli\cite{Purves_Williams_2001}. 
A spike employing negative polarity to inhibit a neighboring neuron will be demonstrated in section \ref{gwrrt}. 

The bi-polar character of AFM neuron dynamics may find useful applications in developing new types of SNNs consisting of two competing subnetworks (positively- and negatively-biased). 
The neurons in one subnetwork will stimulate neurons in the same network, but inhibit neurons of the opposite network.

\subsection{AFM neurons as gates for Boolean logic}\label{trhwet}

Previous subsections demonstrated AFM neuron spiking, refraction, and polarity.
This subsection will describe how AFM neurons can be used as logic gates to perform Boolean logic.

\begin{figure}
\includegraphics{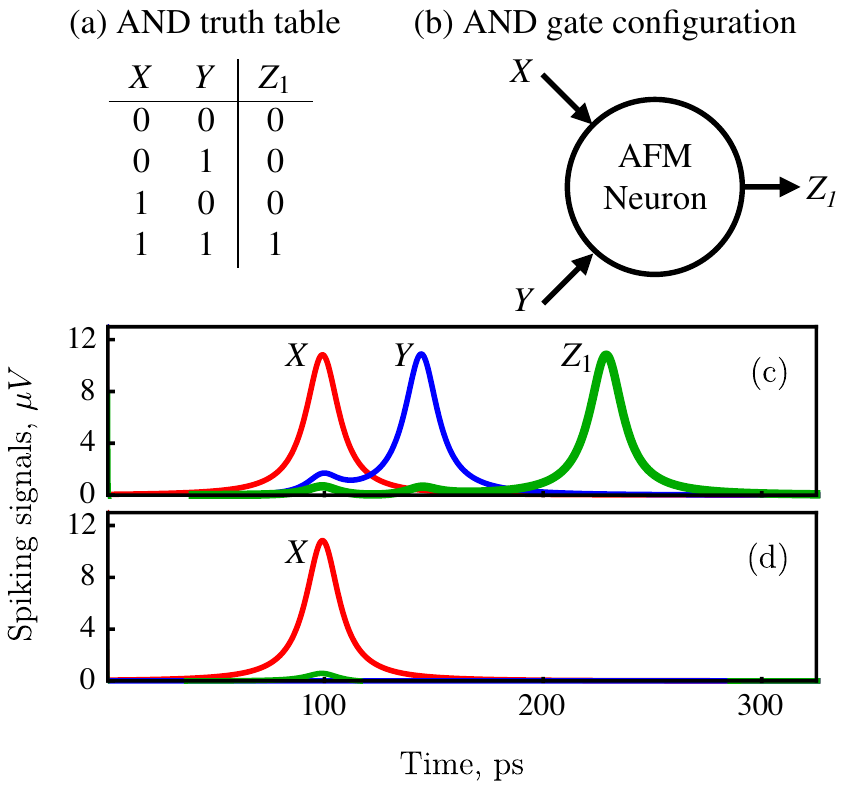}
\caption{\label{and} AND gate implemented with an AFM neuron. (a) Truth table for an AND gate. (b) Schematic of the AFM neuron for an AND gate. Arrows signify inputs and outputs. (c) Simulation with a logical inputs $X=1$ and $Y=1$. $X$ is shown with a red curve, and $Y$ is shown with a blue curve. The output $Z_1=1$ is shown with a green curve. (d) Simulation with a logical inputs $X=1$ and $Y=0$. $X$ is shown with a red curve, and the $Y=0$ curve is not visible. The $Z_1 = 0$ curve does not show a spike, but instead a small bump. In these simulations, $\alpha = 0.1$.}
\end{figure}

The most quintessential Boolean logic gate might be the AND gate\cite{Tocci_Widmer_Moss_2007}. 
The truth table for an AND gate is shown in Fig. \ref{and}(a).
This table shows $X$ and $Y$ as inputs, and $Z_1$= $X \cup Y$ as an output. 
On this table, 1 represents true, and 0 represents false. 
Fig. \ref{and}(b) shows the neural network for an AND gate. 
As before, $X$ and $Y$ are inputs, and $Z_1$ is an output. 
To configure this AFM neuron as an AND gate, the magnitude of each input is individually smaller than the critical current $i_{\rm cr}$. 
However, when both $X$ and $Y$ spike at about the same time, the magnetizations in the AFM material rotate, and the neuron generates a spike.

Numerical simulations of an AFM neuron configured as an AND gate were performed, with results shown in Fig. \ref{and}(c) and (d).
The AND gate in Fig. \ref{and}(c) has inputs $X=1$ and $Y=1$. 
The $X$ input is shown as a function of time with a red curve, and it has a spike at $t=100$~ps. 
This spiking signal represents $X=1$.
Likewise, the $Y$ input is shown as a function of time with a blue curve, and has a spiking signal at $t=150$~ps. 
This spike represents a $Y=1$. 
Together, these two inputs provide enough energy that the neuron is able to generate an output spike at $t=220$~ps, which is represented by a green curve. 
This output spike represents $Z_1=1$, thus demonstrating that this portion of the AND gate truth table can be realized by an AFM neuron.

Fig. \ref{and}(d) shows the results of numerical simulation for an AND gate when the inputs are $X=1$ and $Y=0$.
The $X$ input is shown as a function of time with a red curve, and it has a spike at $t=100$~ps. 
This red spiking signal represents an $X= 1$ input.
In this plot, $Y$ is represented by a blue curve, which does not spike at all, and hence it properly represents $Y=0$.
In Fig. \ref{and}(d) there is no green curve spike, only a small bump at $t = 100$~ps, and thus for this case the output is $Z_1=0$. 
In Fig. \ref{and}(d) there is no green curve spike, only a small bump at $t = 100$~ps, and thus for this case the output is $Z_1=0$. 
This is as expected for these inputs.
Additional simulations were performed to confirm that the same AFM neuron will fulfill lines 1 and 2 of the AND truth table, thus showing that a single AFM neuron can act as an AND gate.

An AFM neuron with a configuration similar to Fig. \ref{neuron} can act as an OR gate.
The only change required is to increase the amplitudes of the inputs signals so that the neuron will fire from one or more inputs.

\begin{figure}
\includegraphics{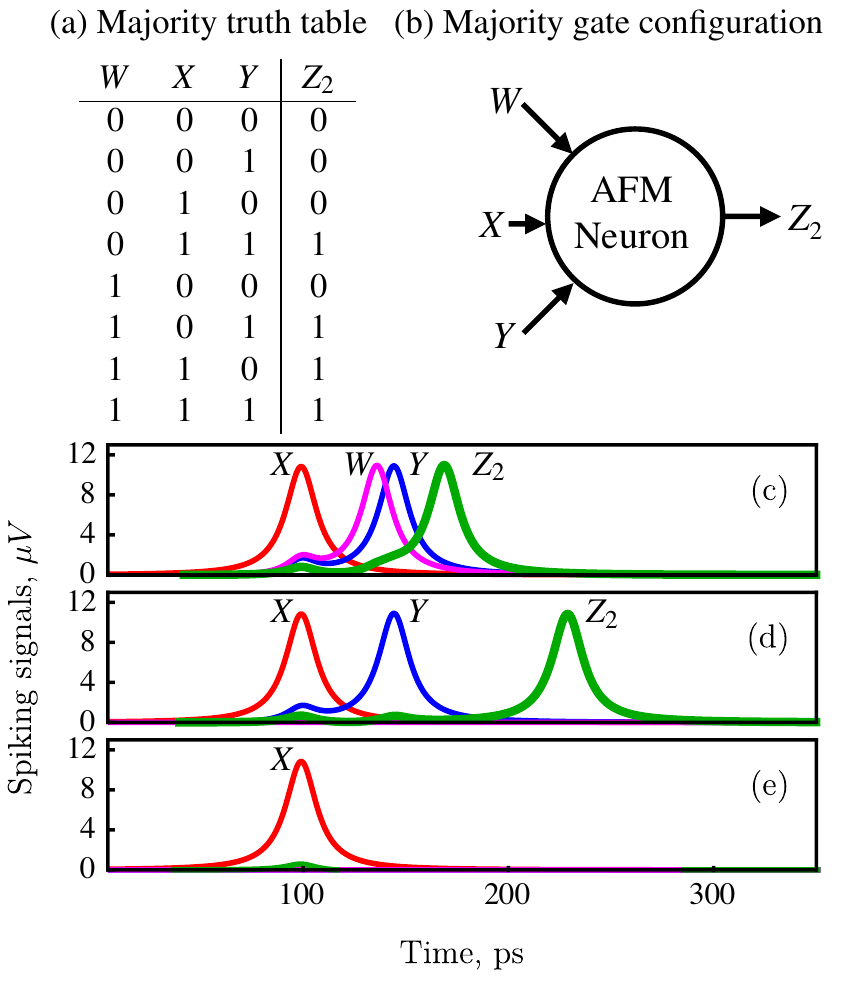}
\caption{\label{maj} Majority gate implemented with a single neuron. (a) Truth table for a majority gate. (b) Schematic of the AFM neuron for the majority gate, with inputs $W$, $X$ and $Y$, and output $Z_2$. (c) Simulation of a majority gate with $W=1$ in magenta, $X=1$ in red, and $Y=1$ in blue, as labeled. Because there are $\geq 2$ true inputs, $Z_2=1$, as shown by the green spike. (d) Simulation of the majority gate for $W=0$, $X=1$, and $Y=1$. Because $\geq 2$ inputs are true, the output $Z_2$ is also true, as evidenced by the green curve. (e) Simulation of the majority gate for $W=0$, $X=1$, and $Y=0$. Because $<2$ inputs are true, $Z_2=0$, and hence there is no spike on the green curve. In these simulations, $\alpha = 0.1$.}
\end{figure}

A single AFM neuron can also be used to represent a majority gate. 
A majority gate is a Boolean circuit that outputs a true when more than half of its inputs are true.
The truth table for a 3-input majority gate is shown in Fig. \ref{maj}(a), and the configuration for an AFM neuron implementation of this majority gate is shown in Fig. \ref{maj}(b). 
For comparison, a 3-input majority gate would typically require 4 NAND gates\cite{Tocci_Widmer_Moss_2007}.
To configure the AFM neuron as a majority gate, a single input would be smaller than $i_{\rm cr}$, while two inputs together would cause the AFM neuron to generate a single spike. 
Likewise, three inputs together also will create a single spike, and have a threshold current that does not induce bursting behavior. 

Numerical simulations of the majority gate are shown in Fig. \ref{maj}(c), (d), and (e). 
Simulations for the last line of the truth table, when $W=1$, $X=1$, and $Y=1$, are shown in Fig. \ref{maj}(c). 
These inputs are labeled on the graph.
The output $Z_2=1$ is on the same graph as labeled.

Fig. \ref{maj}(d) shows simulations for $W=0$, $X=1$, and $Y=1$, and Fig. \ref{maj}(e) shows simulations for $W=0$, $X=1$, and $Y=0$. 
In both cases, $Z_2$ responds as expected. 
It is evident that a single spike input is insufficient to generate an output spike, while two or more inputs spikes can generate an output spike, thus confirming that a single AFM neuron can act as a majority gate.

A previous work provides an in-depth analysis of performing Boolean logic with AFM neurons, and includes a description of a Q-gate and a full-adder that consists of just three AFM neurons\cite{Sulymenko_Prokopenko_Lisenkov_kerman_Tyberkevych_Slavin_Khymyn_2018}.

\section{Interconnecting AFM neurons}\label{rhthwse}

For AFM neurons to be useful, they should be interconnected into a neural network.
The previous sections described the behavior of individual AFM neurons.
This section will discuss how multiple AFM neurons can be interconnected to form a physical neural network. 

\begin{figure}
\includegraphics{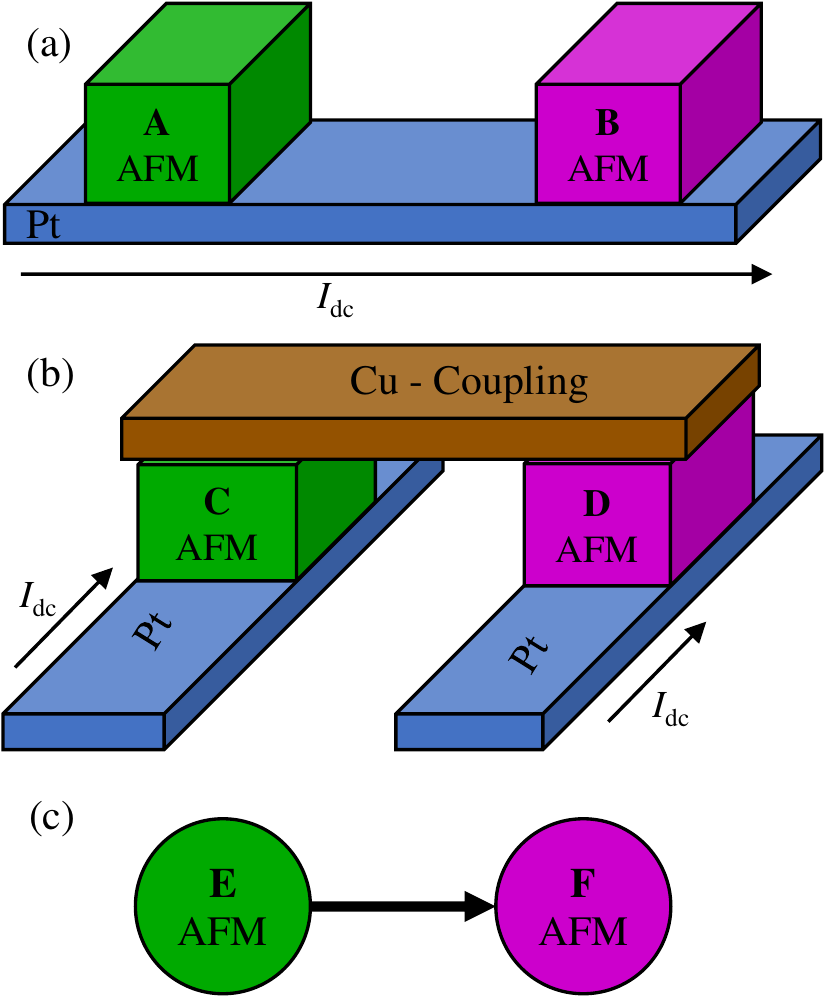}
\caption{\label{connections} AFM neuron interconnection. 
(a) Schematic of electrical interconnection. Here, neuron `A' is green and neuron `B' is magenta, and the two neurons are connected via the blue platinum substrate. 
(b) Schematic of spin current interconnection. Here neuron `C' and `D' are biased independently by two Pt substrates, and they are interconnected via a Copper coupling, shown in brown. 
(c) Schematic of a generic interconnection. Neuron `E' is represented by a green circle, and neuron `F' is represented by a magenta circle. These two neurons are interconnected by a generic synapse, which is represented by an arrow.}
\end{figure}

A summary of interconnection schemes is shown in Fig. \ref{connections}. 
In Fig. \ref{connections}(a), neurons `A' and `B' are connected by a single piece of platinum. 
With this connection, both neurons are subject to the same bias current. 
Likewise, spikes generated by neuron `A', will travel downstream to induce a spike in neuron `B'. 

A second simple method of interconnecting neurons is shown in Fig. \ref{connections}(b). 
In this case, both AFM neurons connected to separate Pt stips. 
This allows a bias current to be provided to each AFM neuron independently. 
In addition, a copper waveguide interconnects the neurons to transmit spin current between neurons `C' and `D'.
By changing the width, thickness, and length of the copper waveguide, it is possible to change the coupling strength between the neurons.

Unfortunately, the strengths of the inter-neuron connections in Fig. \ref{connections}(a) and (b) are fixed, and cannot be adjusted. 
To use AFM neurons in neural networks for machine learning and AI, it is required that the connection between neurons be mediated by an adjustable synapse. 
This is depicted schematically in Fig. \ref{connections}(c). 
In this figure, both neurons are assumed to be properly biased with a DC current. 
When a momentary impulse of sufficient amplitude is applied to neuron `E', it will generate a spike. 
This spike will travel through a synapse, which is represented in this figure by an arrow, to synapse `F'. 
If the signal that arrives at neuron F is of sufficient amplitude, neuron F will also generate a spike.

All synapses simulated in this paper are assumed to be ``ideal''; meaning that the synaptic weights can be adjusted instantaneously, and be of any value.
This paper is focused primarily on presenting AFM neurons as realizable hardware; thus, we do not delve into the details of how synapses will be realized. 
It is worth mentioning that electrical synapses can be physically realized in different ways\cite{zhu2020comprehensive}, for example, using memristors\cite{jo2010nanoscale}, multiferroic heterostructures\cite{lu2019artificial}, Josephson junctions\cite{Schneider_Donnelly_Russek_Baek_Pufall_Hopkins_Dresselhaus_Benz_Rippard_2018}, or even implemented in CMOS\cite{bartolozzi2007synaptic}.

Artificial synapses can also be constructed using spintronic devices; i.e. devices that use magnetic elements and their dynamics. 
For example, spin torque nano oscillators \cite{leroux2021hardware, leroux2021radio}, domain walls\cite{akinola2019three, narasimman2016low, bhowmik2019chip}, and skyrmions \cite{song2020skyrmion, huang2017magnetic, das2022bilayer} have all been shown to act like synapses to interconnect artificial neurons. 
Spintronic devices have many distinct advantages over their electrical counterparts, with an ability to closely resemble biological synapses and a low power consumption. 
Spintronic synapses have been shown to exhibit the potentiation and depression behaviors of biological synapses \cite{song2020skyrmion, huang2017magnetic}, and also shown to exhibit spike time dependent plasticity, the biological process that adjusts the weights between neurons\cite{akinola2019three, narasimman2016low}.

\subsection{Mathematical model for AFM neural network}

A mathematical model for a single AFM neuron was given in equation (\ref{vteq}). 
In contrast, when multiple AFM neurons are interconnected into a neural network, the neurons will interact with each other via spiking output voltages. 
The interaction can be modeled by a system of differential equations:
\begin{equation} \frac{1}{\omega_{ex}}\ddot{ \phi_i} + \alpha \dot{\phi_i} + \frac{\omega_{\rm e}}{2}\sin 2\phi_i = \sigma I_i + \sum_{i\ne k} \kappa_{ik} \dot{\phi}_k .\label{systemeq}\end{equation} 
In this equation, $i$ and $k$ are indices that represent the $i$-th and $k$-th neuron, and $\kappa_{ik}$ represents a matrix of coupling coefficients. 
Like all neural networks, we envision that the interaction between neurons will be mediated by synaptic weights. 
In this equation, synapses are represented by the coupling coefficients $\kappa_{ik}$.
In a few words, by simulating (\ref{systemeq}), one can simulate an entire spiking neural network of AFM neurons. 
It is hoped that, eventually, instead of simulations, fabricated AFM neurons will be available to perform computations at a picosecond timescale.

\subsection{Simple neuron chain}

\begin{figure}
\includegraphics{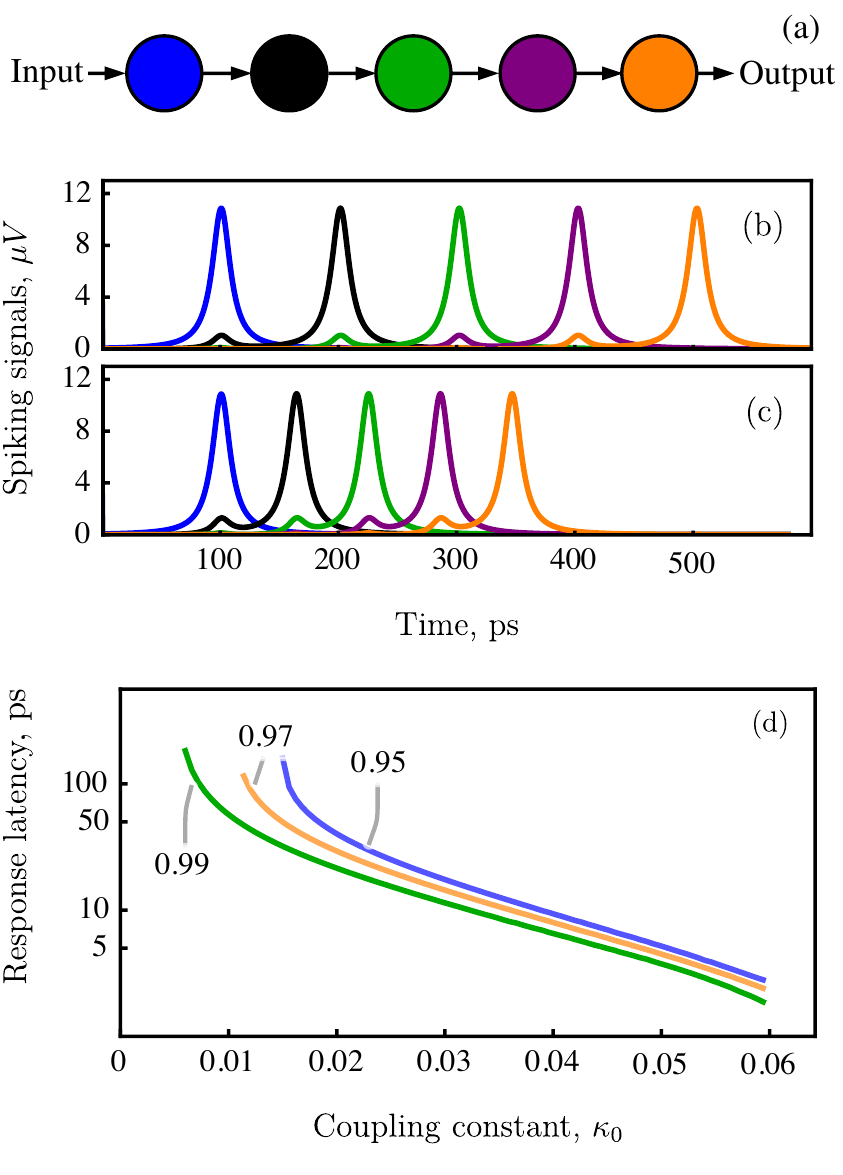}\\
\caption{\label{chain} Simulation results for five AFM neurons connected in a sequential chain. (a) The neural network for a five-neuron chain. Circles represent AFM neurons, and arrows represent synapses. (b) Voltage output with $\kappa_0 = 0.011$. The color of each curve represents the voltage output of each neuron. (c) Voltage output with $\kappa_0 = 0.015$. (d) The relationship between $\kappa_0$ and response latency. Different colored lines show $I_{\rm dc}/I_{\rm th}$ values of 0.99 (green), 0.97 (orange), and 0.95 (blue). In these simulations, $\alpha = 0.1$.}
\end{figure}

Simulations were performed on a chain of five interconnected neurons, as represented in Fig. \ref{chain}(a).
In this figure, five AFM neurons are represented by circles, and electrical connections with synapses are represented by arrows. 

The neuron chain is intended to function as follows. 
First, there is a current impulse that originates on the left that will cause the magnetizations of the blue neuron to rotate and thus generate a voltage spike. 
The spike from the blue neuron then travels through a synapse to the black neuron, where it will have sufficient amplitude to cause the magnetization of black neuron to rotate, and thus generate a voltage spike. 
After this, the spike from the black neuron will flow to the next neuron, thus continuing down the chain.

Numerical simulations were run for this system according to equation (\ref{systemeq}). 
In this simulation, every neuron is considered to have a damping constant of $\alpha = 0.1$ and a bias current of $I_{\rm dc} = 198~\mu$A, which is $\sim 0.98 I_{\rm th}$. 
Weights in the synaptic connections are assigned to the matrix $\kappa_{ik}$, which in this case is a $5 \times 5$ coupling matrix given by:
\begin{equation} \kappa_{ik} = \kappa_0\mqty[ 0 & 0 & 0 & 0 & 0 \\ 1 & 0 & 0 & 0 & 0 \\ 0 & 1 & 0 & 0 & 0 \\ 0 & 0 & 1 & 0 & 0 \\ 0 & 0 & 0 & 1 & 0 ] ,\label{esgge} \end{equation} where $\kappa_0$ is the coupling constant that determines the overall strength of neuronal interconnects. 

Consider first the simulation results with $\kappa_0 = 0.011$, shown in Fig. \ref{chain}(b). 
For this simulation, a momentary stimulus is provided to the leftmost (blue) neuron, leading to a spike at time $t=100$~ps, as can be seen from the blue curve in Fig. \ref{chain}(b). 
The voltage produced by the blue neuron, which is about 10 $\mu$V, then causes the black AFM neuron to spike at $t=190$~ps, as shown by the black curve. 
The spike from the black neuron, which has the same amplitude, continues towards the green neuron. 
This leads the green neuron to generate a spike, which then leads the purple and orange neurons to generate spikes in succession.
Thus, a spiking signal can propagate through a chain of interconnected neurons. 

Note that in Fig. \ref{chain}(b), there is a uniform time delay of about $90$~ps between two neighboring neuron spikes. 
This delay is the neuron response latency $t_\ell$ which was discussed previously. 
As before, the duration of the latency can be adjusted by increasing the amplitude of the perturbation $i_p(t)$ that initiates the magnetization revolution in the AFM material. 
In a neural network, the amplitude of a spike incident on an AFM neuron can be changed by adjusting the value of the synaptic weight. 

Thus, the duration of the response latency can be adjusted by changing the weights in the coupling matrix. 
Simulation results for a chain of neurons, with a larger synaptic weight, is demonstrated in Fig. \ref{chain}(c). 
Here, the weights were changed to $\kappa_0 = 0.015$, while the bias current, the effective damping, and the coupling matrix (\ref{esgge}) remained unchanged. 
With the increased synaptic weights, the neurons fire as before, with a shortened latency of about 50~ps. 
The time between neuron spikes can thus be controlled by adjusting the synaptic weight. 
The relationship between synaptic weights and response latency may be useful in the implementation of machine learning algorithms for SNNs like SpikeProp, where spike timing plays a critical role\cite{Kasabov_2018}. 


Of course, the bias current also plays a role in determining the latency in a sequential chain of neurons. 
This is examined in Fig. \ref{chain}(d).
The figure shows how the latency and the coupling coefficient $\kappa_0$ are related, for three values of bias current $I_{\rm dc}/I_{\rm th}$. 
Two trends can be ascertained from this graph.
First, for all three bias currents, as $\kappa_0$ increases, the latency decreases.
Second, as the bias current increases, a smaller coupling constant is able to induce a shorter latency. 

Thus, this section has established that AFM neurons can be interconnected, and that spikes can be transmitted between neurons through these interconnections. 
It was also established that the weights can be used to change the response latency of neuron spikes.

\subsection{Symmetric coupling}


Interestingly, it is possible to transmit spin current directly between neurons with a non-magnetic metal like copper. 
Pure spin current can be defined as the net flow of spin angular momentum without the net flow of charge carriers\cite{Maekawa_Valenzuela_Saitoh_2017}. \color{black}
This idea was illustrated schematically in Fig. \ref{connections}(b). 
In that figure, when neuron `C' spikes, it generates spin current that can flow into the copper connector and travel to neuron `D'. 
If the spin current is of sufficient magnitude, it can induce neuron D to generate a spike. 
The spike generated by neuron D can then re-enter the copper connecter, and flow back to neuron C.
That is, the coupling between these two neurons is bi-directional; hence this type of coupling can be called \emph{symmetric coupling}.

The concept of symmetric coupling in a system of bio-inspired neurons is novel; therefore, it will be considered in detail in this subsection.
Unfortunately, machine algorithms have not been developed that can use bi-directional signal propagation. 
Therefore, it will be beneficial if AFM neurons could be interconnected via symmetric coupling yet transmit in just a single direction.
This can be done by exploiting the refraction properties of AFM neurons to induce one-way spike transmission.

\begin{figure}
\includegraphics{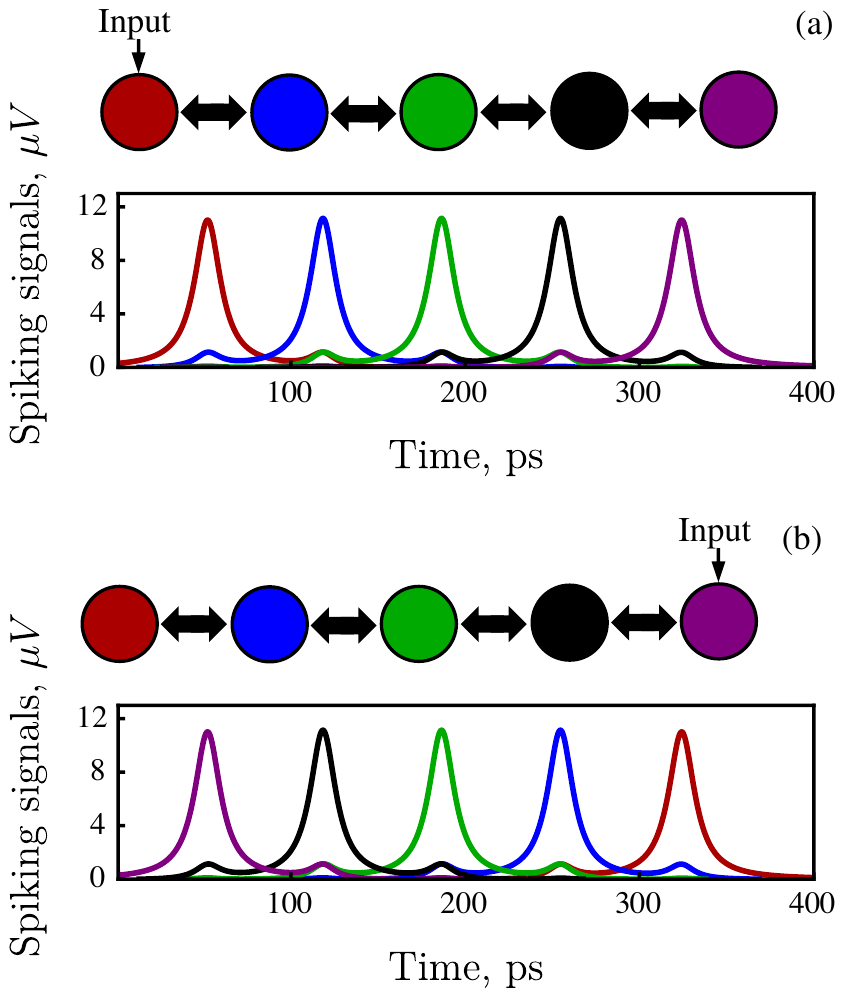}\\
\caption{\label{symChain} Symmetric coupling and unidirectional signal propagation. (a) Simulation result for a symmetrically coupled neural network, with an initial signal on the left, and (b) with an initial signal on the right. In these simulations, $\alpha = 0.1$, $\kappa_0 = 0.011$.}
\end{figure}

Figure \ref{symChain} will demonstrate how the refraction properties of AFM neurons can be used to send a signal unidirectionally through a chain of symmetrically coupled neurons. 
In Fig. \ref{symChain}(a) the signal will originate in the red neuron on the left, and in Fig. \ref{symChain}(b) the signal will originate in the purple neuron on the right. 
In this schematic, all four synapses are fully bidirectional; for example, a signal generated by the blue neuron will flow to both the red neuron and the green neuron. 
For this neuron chain, the coupling matrix is a symmetric matrix,
\begin{equation} \kappa_{ik} = \kappa_0\mqty[ 0 & 1 & 0 & 0 & 0 \\ 1 & 0 & 1 & 0 & 0 \\ 0 & 1 & 0 & 1 & 0 \\ 0 & 0 & 1 & 0 & 1 \\ 0 & 0 & 0 & 1 & 0 ]\label{esggetg}. \end{equation} 
In spite of the fact that all connections are bi-directional, simulations show that the signal will flow in only one direction.

First, consider a simulation performed where the first spike was initiated in the red neuron on the left. 
The simulation results are shown in Figure \ref{symChain}(a). 
In this simulation, the red neuron generates a spike a $t=50$~ps. 
This spike propagates to the blue neuron, which generates a spike at $t=120$~ps.
The spike from the blue neuron propagates symmetrically in two directions; and arrives at both the red neuron and the green neuron at $t=120$~ps.
At $t=120$~ps the red neuron is in its absolute refractory period and will not generate a spike.
In contrast, the green neuron will generate a spike as a result of the incoming spin current at $t= 190$~ps. 
After this, the green, black, and purple neurons spike in sequence.
In each case, the signal will propagate symmetrically after a neuron spikes, but the refractory properties of AFM neurons prevent them from spiking, thus ensuring unidirectional signal propagation.

Consider now the network shown schematically in Fig. \ref{symChain}(b). 
This network is identical to that in Fig. \ref{symChain}(a), except that in this case the first spike is initiated in the purple neuron on the right. 
Results of simulating this neural network are also shown in Figure \ref{symChain}(b), with the signal propagating from the purple neuron to the red neuron.

It is important to note that the neural networks in Fig. \ref{symChain}(a) and (b) are identical and unchanged; the only difference is where the initial spike was delivered. 
It should also be emphasized that spikes only propagate in one direction due to the refraction properties of the AFM neuron.

\begin{figure}
\includegraphics{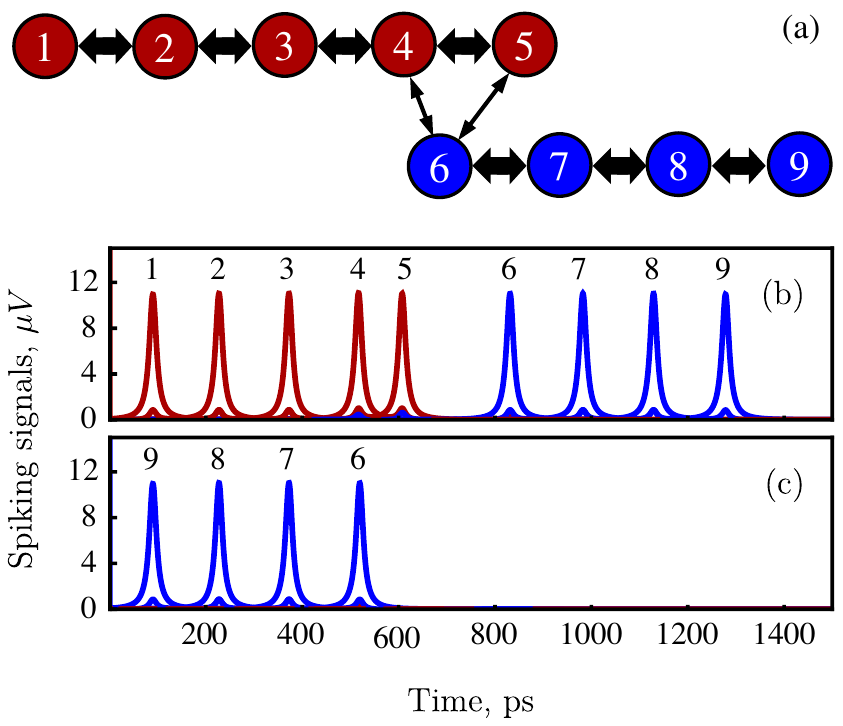}\\
\caption{\label{isolate} Isolator composed of symmetrically coupled neurons. (a) Schematic of the isolator circuit, with two sets of neurons, red and blue. Within these two sets, couplings are identical and fully symmetric. There is also a weak symmetric coupling between neuron 6 and neurons 4 and 5. (b) Simulation for a signal flowing from neuron 1 to neuron 9. The synaptic weights between 4-5-6 are such that the combined spikes of 4 and 5 provide sufficient energy for neuron 6 to generate a spike. (c) Simulation for a signal flowing from neuron 9 to neuron 6. The signal is unable to continue flowing through the red chain due to the weak connections between neurons 6 and 4/5. In these simulations, $\alpha = 0.1$.}
\end{figure}

It is possible to form an isolator with a symmetrically connected neural network. 
An isolator serves a function similar to that of a diode or a check valve, and restricts the signal flow to a single direction.
A circuit that contains an isolator is shown in Fig. \ref{isolate}(a). 
The circuit consists of two sets of symmetrically coupled neurons, shown in red for neurons 1 through 5, and shown in blue for neurons 6 through 9. 
The portion of this circuit that functions as an isolator are neurons 4, 5, and 6, and includes the two weak synaptic couplings that converge on neuron 6. 
In this configuration, signals can flow in just one direction, from neuron 1 to neuron 9, but cannot flow in the reverse direction. 
This will be demonstrated via simulation.

Results of the first simulations performed on this circuit are shown in Fig. \ref{isolate}(b).
In this simulation, neuron 1 spikes at $t\sim 100 $~ps, and this signal propagates from neuron 1 towards neuron 5. 
At about $t=500 $~ps, neurons 4 and 5 generate spikes in rapid succession. 
Together, the two spikes generated by these two neurons have sufficient strength to cause neuron 6 to spike, despite the weak synaptic coupling between the two chains. 
After neuron 6 generates a spike, the signal continues down the chain towards neuron 9.

Simulations were performed on the same neural network with a signal originating at neuron 9.
Results of this simulation are shown in Fig. \ref{isolate}(c). 
In this simulation, the signal at neuron 9 travels to neuron 6, which spikes at $t=500$~ps. 
Due to the the weak synaptic coupling, the spike generated by neuron 6 has insufficient strength to cause either neuron 4 or 5 to generate a spike. 
Thus, the signal is unable to journey from neuron 9 to neuron 1; spikes are prevented from propagating in the ``backward'' direction.

\begin{figure}
\includegraphics{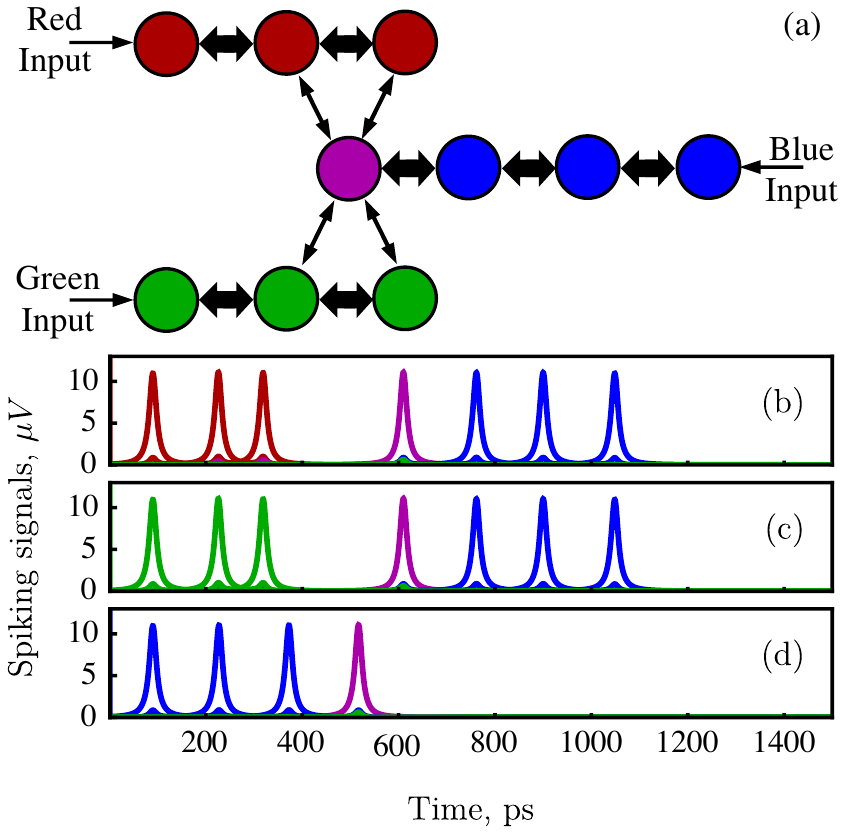}\\
\caption{\label{comby} Combiner composed of symmetrically coupled neurons. (a) Schematic of the combiner circuit, with 3 sets of neurons. Input neurons are red and green, while output neurons are blue. Arrow widths represent coupling strength. (b) Simulation result for a signal that begins in the red neurons. (c) Simulation results for a signal that begins in the green neurons. (d) Simulation result for a signal that begins in the blue neurons. In these simulations, $\alpha = 0.1$.}
\end{figure}

The design of the isolator can be extended into a combiner circuit, whose schematic is shown in Fig. \ref{comby}(a). 
This circuit has 3 sets of neurons, shown in red, green, and blue. 
The red and green neurons represent input branches, and the blue neurons represent an output branch. 
With this circuit, signals can flow from the red to blue neurons, or from the green to blue neurons. 
However, this configuration ensures that there is no signal leakage between the red and green branches.

Correct operation of this circuit is confirmed by simulation, with results shown in Fig. \ref{comby}(b), (c) and (d).
In Fig. \ref{comby}(b), a signal propagates from the red input through the red neurons to the magenta neuron, then to the blue output neurons. 
There is no leakage to the green neurons with this architecture.
Likewise, in Fig. \ref{comby}(c), a signal that propagates from the green input to green neurons and travels to the blue neurons, without leakage to the red neurons.
Lastly, in Fig. \ref{comby}(d), a signal that begins with the blue input will propagate through the blue neurons, but will not propagate to the green or red neurons. 
Thus, this circuit can allow signals from the red and green branch to combine, without crosstalk between branches, and without signals flowing backwards.

\section{Memory loops and inhibition in AFM neural networks}\label{gwrrt} 

This section will demonstrate neural networks that employ loops as memory, and provides control with inhibition.
This section begins with a simple loop, then continues with a demonstration of negative inhibition and positive inhibition.
After this, the two concepts are combined to a controllable memory circuit.

\begin{figure}
\includegraphics{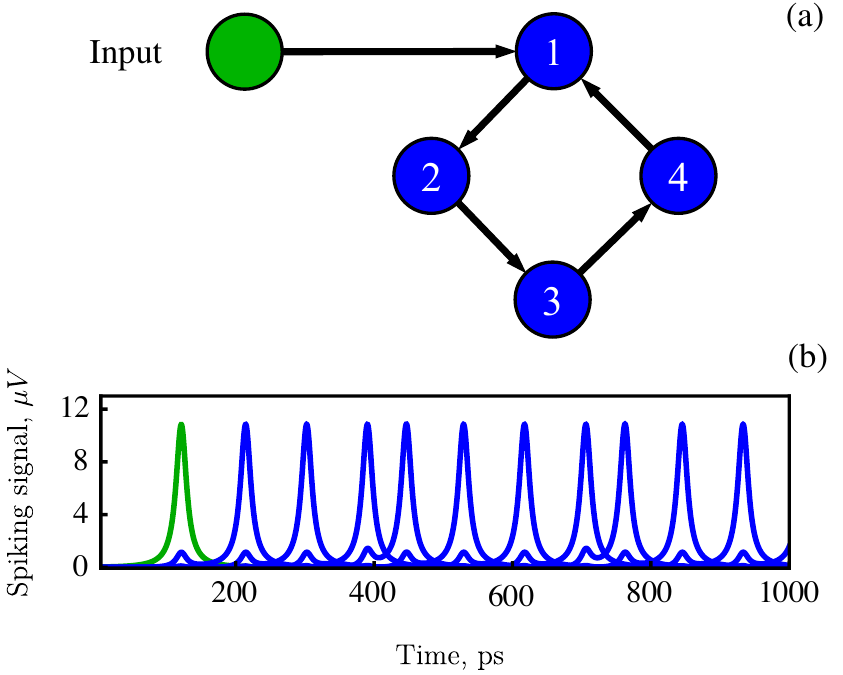}
\caption{\label{loop} Simple memory loop. (a) Schematic of the neural network for the loop, with a green input neuron and 4 labeled blue loop neurons. (b) Simulation of the neural network, where the green curve represents a spike from the green input neuron, and the blue curves represent spikes from the four loop neurons. In these simulations, $\alpha = 0.1$.}
\end{figure}

%
Consider first the neural network shown in Fig. \ref{loop}(a).
Simulation results for this neural network are shown in Fig. \ref{loop}(b).
The simulation begins by first initiating a spike in the green neuron. 
This leads to a spike in neuron 1, which leads to a spike in neuron 2, which leads to a spike in neuron 3, which then continues around the loop back to neuron 1. 
As can be seen in the simulation results, the spike will continue around this loop indefinitely. 

\begin{figure}
\includegraphics{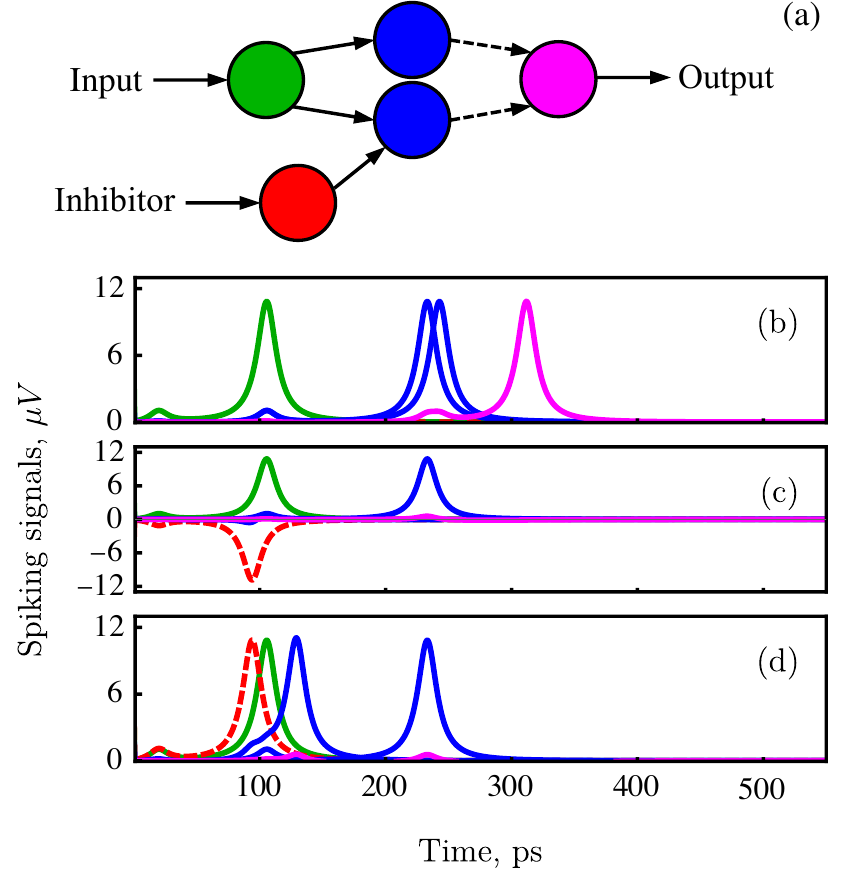}
\caption{\label{inhib}Negative inhibition and positive inhibition. 
(a) The neural network for demonstrating inhibition, with the input neuron (green), intermediate neurons (blue), output neuron (magenta), and the inhibition neuron (red). The synapses are weighted such that solid arrows represent a full connection, while dashed arrows represent a half connection. (b) Simulation of uninhibited network. The input neuron, with a green curve, fires. This leads the two intermediate neurons, with blue curves, to fire. Together, the spikes from the intermediate neurons provide enough current for the magenta output neuron to fire. (c) Simulation demonstrating negative inhibition. Inhibitor neuron generates a spike with negative polarity, which prevents the lower intermediate neuron from spiking. (d) Simulation demonstrating positive inhibition.  The red and green neurons fire, leading the lower intermediate neuron to quickly fire. After a delay, the upper intermediate neuron fires. Because the intermediate neurons fire at different times, the output neuron does not fire. In these simulations, $\alpha = 0.1$.}
\end{figure}

Inhibition functions are demonstrated in Fig. \ref{inhib}. 
As shown in Fig. \ref{inhib}(a), this neural network consists of 5 neurons, including input (green) and output (magenta) neurons. 
A signal can be carried from the input neuron to the output neuron through two intermediate neurons (blue). 
Please note that the synapses are adjusted so that the signal from the input neuron is strong enough to initiate a spike in both intermediate neurons. 
However, for the magenta neuron to fire, both blue intermediate neurons must spike at nearly the same time. 

A simulation of the signal being carried from the input neuron to the output neuron is shown in Fig. \ref{inhib}(b). 
First, the green input neuron generates a spike. 
Then, the two intermediate neurons spike, which together deliver enough energy to allow the output neuron to generate a spike. 

The diagram in Fig. \ref{inhib}(a) also shows a red ``inhibitor'' neuron. 
The utility of the inhibitor neuron is demonstrated by simulation in Fig. \ref{inhib}(c).
In this simulation, the red inhibitor neuron fires at $t=90$~ps, as shown by the red dashed line.
As can be seen in the figure, the spike generated by the inhibitor neuron has a negative polarity, as was described in section \ref{asdawwfe}. 
When the green input neuron fires at $t=100$~ps, only the upper blue intermediate neuron fires in response, which does not provide sufficient energy to cause the magenta neuron to fire. 
The lower blue neuron does not fire, as the inhibitory spike with negative polarity suppressed the rotation of its $\vb{M}_1$ magnetization. 
Essentially, when the negative inhibitory signal arrives at the lower blue intermediate neuron, it effectively cancels the positive signal and prevents the AFM magnetization from rotating in that neuron.
The inhibition demonstrated in Fig. \ref{inhib}(c) can be called \emph{negative inhibition}. 
This form of inhibition resembles hyper-polarizing stimuli in biological neurons\cite{Purves_Williams_2001}. 

A different form of inhibition, which can be called \emph{positive inhibition} is demonstrated by simulation in Fig. \ref{inhib}(d). 
Once again, in this simulation the red inhibitor neuron fires at $t=90$~ps, as shown by the red dashed line. 
Here, the spike generated by the inhibitor neuron has a positive polarity. 
After the inhibitor neuron fires, the green neuron generates a spike at $t=100$~ps. 
The combines spikes from the red and green neurons causes the lower intermediate neuron to fire at a different time than in the previous simulation.
Because the lower intermediate neuron fires at a different time than the upper intermediate neuron, there is insufficient energy arriving at the magenta output neuron, which does not fire. 
Thus, a positive signal is able to inhibit an output. 
It is notable that positive inhibition plays an important role in with biological neurons, and that this feature of AFM neurons may be employed in future AI tasks.

\begin{figure}
\includegraphics{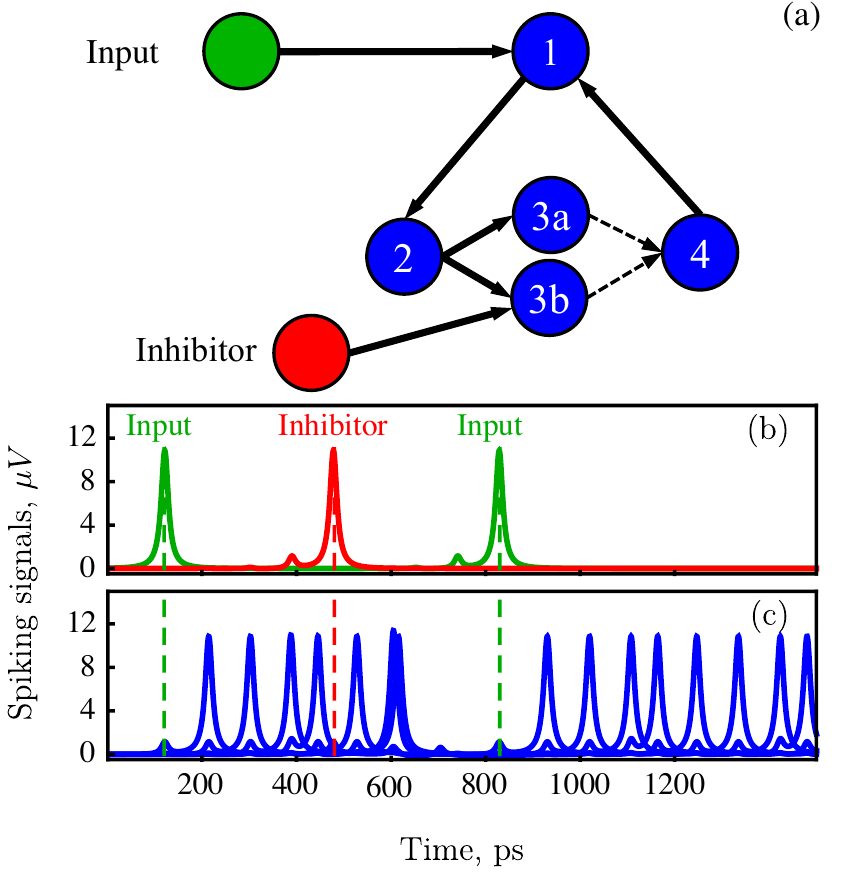}
\caption{\label{mem} AFM neuron memory cell. (a) The memory cell neural network. The input neuron is colored green, the inhibitor neuron is colored red, and the memory neurons are colored blue. (b) Simulated input and inhibition signals (c) Simulated response to the input. The input signal at $t=100$~ps initiates the signal to flow from 1 to 4 and cyclically repeat until $t=500$~ps when the inhibitor signal stops the cycle. The cycle is restarted by an input spike at $t=800$~ps. In these simulations, $\alpha = 0.1$.}
\end{figure}

This inhibitor functionality can have application in a memory cell, which is shown in Fig. \ref{mem}(a).
The memory cell consists of an input neuron, an inhibitor neuron, and 5 intermediate neurons, as shown in the figure. 
Operation of the memory cell will be demonstrated by simulation. 
Initially, the entire system is at rest. 
Then, at $t=100$~ps, the green input neuron generates a spike. 
This is shown by a green curve in Fig. \ref{mem}(b). 
This signal from the input neuron then enters the blue neuron loop at the neuron labeled ``1'', as shown in Fig. \ref{mem}(b). 
After this, the spiking signal will travel through the neuron loop, from neuron ``1'' to ``2'', then to neurons ``3a'' and ``3b'', then to neuron ``4''. 
These neurons, spiking in succession, are shown by multiple blue curves in Fig. \ref{mem}(c).
Please note that the synapses have been adjusted so that both neurons 3a and 3b must spike at nearly the same time in order for neuron 4 to generate a spike. 
 
The inhibitor neuron fires at $t=500$~ps, as shown by a red curve in Fig. \ref{mem}(b). 
The signal from the inhibitor neuron causes neuron 3b to fire at a time earlier than neuron 3a. 
Due to this variance in timing and the weak synaptic connection between neurons 3 and 4, there is insufficient energy for neuron 4 to generate a spike, which stops the signal from circulating. 
Thus, the properties of positive inhibition were used to control a spiking signal propagating in a memory loop.
After this, the input signal fires once more at $t=800$~ps, and the memory signal circulates once again, as shown in the figure.

\section{Comparison of AFM neurons and biological neurons}

In previous sections, characteristics that AFM neuron and biological neurons shared were presented. 
This section will provide a summary of these shared characteristics. 

First, AFM neurons generate spiking voltages that closely resemble the action potentials of biological neurons. Furthermore, AFM neurons follow the ``all-or-nothing'' law. 
That is, AFM neurons only generate spikes when their input surpasses a critical current, in much the same way that an action potential is elicited from biological neurons. 
These two properties were demonstrated in section \ref{wefasdfaef}.

Second, AFM neurons have a response latency, as was also demonstrated in section \ref{wefasdfaef}. 
Basically, in both biological neurons and AFM neurons, after sufficient stimulus has been delivered to the neuron, there is a time delay before the spike occurs. 
Latency of a stimulated biological neuron varies, depending on both the frequency and the intensity of its stimulus. 
With AFM neurons, the duration of the response latency can be controlled dynamically by adjusting the amplitude of the stimulus $i_p(t)$, and by adjusting the magnitude of the bias current $I_{\rm dc}$. 
The latency period also depends on the effective damping, the anisotropy, and the exchange frequency of the AFM material. 

Third, both AFM neurons and biological neurons have a refractory period. 
Specifically, both types of neurons have an absolute refractory period and a relative refractory period. 
This was discussed in section \ref{evasrvsr}. 

Fourth, both AFM neurons and biological neurons have variable spiking modes. 
That is, AFM neurons can generate single spikes, bursting spikes, or spike trains. 
They can also exhibit adaptation and stuttering.
This was discussed in section \ref{evasrvsr}. 

Fifth, the generation of action potentials in both AFM neurons and biological neurons can be suppressed through inhibition. 
This was demonstrated in section \ref{gwrrt}. 
In AFM neurons, both negative inhibition and positive inhibition can be used to suppress neuronal activity. 
The duration of inhibition for AFM neurons depends on the characteristics of their refractory period. 

Lastly, AFM neurons can be interconnected into a neural network where connections between neurons are mediated by synapses. 
Two methods of interconnecting neurons, via electrical current and via spin current, along with examples of spintronic synapses, were discussed in section \ref{rhthwse}.

\section{Conclusion}

AFM neurons are nanometer scale electronic elements that exhibit behavior that resembles that of a biological neuron. 
Specifically, they have several properties that resemble the characteristics of biological neurons, including a finite refraction time, response latency, and bursting behavior.
In addition, AFM neurons operate much faster than the state of the art, with a ps spike time, and with an ultra-low energy consumption of about $10^{-3}$ pJ per synaptic operation.
In this paper, it was demonstrated by simulation that AFM neurons can be interconnected into a neural network, with response latency, refraction, and inhibition as tools to potentially be used in machine learning algorithms. 
It was also demonstrated that AFM neurons can perform Boolean operations, have a polarity that can be reversed, can exhibit inhibition, and can perform memory functions. 
In addition, a one-way signal transmission can be implemented with symmetric coupling, along with an isolator and a combiner.

\nocite{*}
\bibliography{AFMNeuronReviewBib}

\begin{thebibliography}{94}%
\makeatletter
\providecommand \@ifxundefined [1]{%
 \@ifx{#1\undefined}
}%
\providecommand \@ifnum [1]{%
 \ifnum #1\expandafter \@firstoftwo
 \else \expandafter \@secondoftwo
 \fi
}%
\providecommand \@ifx [1]{%
 \ifx #1\expandafter \@firstoftwo
 \else \expandafter \@secondoftwo
 \fi
}%
\providecommand \natexlab [1]{#1}%
\providecommand \enquote  [1]{``#1''}%
\providecommand \bibnamefont  [1]{#1}%
\providecommand \bibfnamefont [1]{#1}%
\providecommand \citenamefont [1]{#1}%
\providecommand \href@noop [0]{\@secondoftwo}%
\providecommand \href [0]{\begingroup \@sanitize@url \@href}%
\providecommand \@href[1]{\@@startlink{#1}\@@href}%
\providecommand \@@href[1]{\endgroup#1\@@endlink}%
\providecommand \@sanitize@url [0]{\catcode `\\12\catcode `\$12\catcode
  `\&12\catcode `\#12\catcode `\^12\catcode `\_12\catcode `\%12\relax}%
\providecommand \@@startlink[1]{}%
\providecommand \@@endlink[0]{}%
\providecommand \url  [0]{\begingroup\@sanitize@url \@url }%
\providecommand \@url [1]{\endgroup\@href {#1}{\urlprefix }}%
\providecommand \urlprefix  [0]{URL }%
\providecommand \Eprint [0]{\href }%
\providecommand \doibase [0]{https://doi.org/}%
\providecommand \selectlanguage [0]{\@gobble}%
\providecommand \bibinfo  [0]{\@secondoftwo}%
\providecommand \bibfield  [0]{\@secondoftwo}%
\providecommand \translation [1]{[#1]}%
\providecommand \BibitemOpen [0]{}%
\providecommand \bibitemStop [0]{}%
\providecommand \bibitemNoStop [0]{.\EOS\space}%
\providecommand \EOS [0]{\spacefactor3000\relax}%
\providecommand \BibitemShut  [1]{\csname bibitem#1\endcsname}%
\let\auto@bib@innerbib\@empty
\bibitem [{\citenamefont {Baldominos}, \citenamefont {Saez},\ and\
  \citenamefont {Isasi}(2019)}]{baldominos2019survey}%
  \BibitemOpen
  \bibfield  {author} {\bibinfo {author} {\bibfnamefont {A.}~\bibnamefont
  {Baldominos}}, \bibinfo {author} {\bibfnamefont {Y.}~\bibnamefont {Saez}},\
  and\ \bibinfo {author} {\bibfnamefont {P.}~\bibnamefont {Isasi}},\ }\bibfield
   {title} {\enquote {\bibinfo {title} {A survey of handwritten character
  recognition with {MNIST} and {EMNIST}},}\ }\href@noop {} {\bibfield
  {journal} {\bibinfo  {journal} {Applied Sciences}\ }\textbf {\bibinfo
  {volume} {9}},\ \bibinfo {pages} {3169} (\bibinfo {year} {2019})}\BibitemShut
  {NoStop}%
\bibitem [{\citenamefont {LeCun}, \citenamefont {Bengio},\ and\ \citenamefont
  {Hinton}(2015)}]{lecun2015deep}%
  \BibitemOpen
  \bibfield  {author} {\bibinfo {author} {\bibfnamefont {Y.}~\bibnamefont
  {LeCun}}, \bibinfo {author} {\bibfnamefont {Y.}~\bibnamefont {Bengio}},\ and\
  \bibinfo {author} {\bibfnamefont {G.}~\bibnamefont {Hinton}},\ }\bibfield
  {title} {\enquote {\bibinfo {title} {Deep learning},}\ }\href@noop {}
  {\bibfield  {journal} {\bibinfo  {journal} {Nature}\ }\textbf {\bibinfo
  {volume} {521}},\ \bibinfo {pages} {436--444} (\bibinfo {year}
  {2015})}\BibitemShut {NoStop}%
\bibitem [{\citenamefont {Silver}\ \emph {et~al.}(2016)\citenamefont {Silver},
  \citenamefont {Huang}, \citenamefont {Maddison}, \citenamefont {Guez},
  \citenamefont {Sifre}, \citenamefont {Van Den~Driessche}, \citenamefont
  {Schrittwieser}, \citenamefont {Antonoglou}, \citenamefont {Panneershelvam},
  \citenamefont {Lanctot} \emph {et~al.}}]{silver2016mastering}%
  \BibitemOpen
  \bibfield  {author} {\bibinfo {author} {\bibfnamefont {D.}~\bibnamefont
  {Silver}}, \bibinfo {author} {\bibfnamefont {A.}~\bibnamefont {Huang}},
  \bibinfo {author} {\bibfnamefont {C.~J.}\ \bibnamefont {Maddison}}, \bibinfo
  {author} {\bibfnamefont {A.}~\bibnamefont {Guez}}, \bibinfo {author}
  {\bibfnamefont {L.}~\bibnamefont {Sifre}}, \bibinfo {author} {\bibfnamefont
  {G.}~\bibnamefont {Van Den~Driessche}}, \bibinfo {author} {\bibfnamefont
  {J.}~\bibnamefont {Schrittwieser}}, \bibinfo {author} {\bibfnamefont
  {I.}~\bibnamefont {Antonoglou}}, \bibinfo {author} {\bibfnamefont
  {V.}~\bibnamefont {Panneershelvam}}, \bibinfo {author} {\bibfnamefont
  {M.}~\bibnamefont {Lanctot}}, \emph {et~al.},\ }\bibfield  {title} {\enquote
  {\bibinfo {title} {Mastering the game of {Go} with deep neural networks and
  tree search},}\ }\href@noop {} {\bibfield  {journal} {\bibinfo  {journal}
  {Nature}\ }\textbf {\bibinfo {volume} {529}},\ \bibinfo {pages} {484--489}
  (\bibinfo {year} {2016})}\BibitemShut {NoStop}%
\bibitem [{\citenamefont {Egri-Nagy}\ and\ \citenamefont
  {T{\"o}rm{\"a}nen}(2020)}]{egri2020game}%
  \BibitemOpen
  \bibfield  {author} {\bibinfo {author} {\bibfnamefont {A.}~\bibnamefont
  {Egri-Nagy}}\ and\ \bibinfo {author} {\bibfnamefont {A.}~\bibnamefont
  {T{\"o}rm{\"a}nen}},\ }\bibfield  {title} {\enquote {\bibinfo {title} {The
  game is not over yet—{G}o in the post-alphago era},}\ }\href@noop {}
  {\bibfield  {journal} {\bibinfo  {journal} {Philosophies}\ }\textbf {\bibinfo
  {volume} {5}},\ \bibinfo {pages} {37} (\bibinfo {year} {2020})}\BibitemShut
  {NoStop}%
\bibitem [{\citenamefont {Sharir}, \citenamefont {Peleg},\ and\ \citenamefont
  {Shoham}(2020)}]{sharir2020cost}%
  \BibitemOpen
  \bibfield  {author} {\bibinfo {author} {\bibfnamefont {O.}~\bibnamefont
  {Sharir}}, \bibinfo {author} {\bibfnamefont {B.}~\bibnamefont {Peleg}},\ and\
  \bibinfo {author} {\bibfnamefont {Y.}~\bibnamefont {Shoham}},\ }\bibfield
  {title} {\enquote {\bibinfo {title} {The cost of training {NLP} models: A
  concise overview},}\ }\href@noop {} {\bibfield  {journal} {\bibinfo
  {journal} {arXiv preprint arXiv:2004.08900}\ } (\bibinfo {year}
  {2020})}\BibitemShut {NoStop}%
\bibitem [{\citenamefont {Strubell}, \citenamefont {Ganesh},\ and\
  \citenamefont {McCallum}(2019)}]{strubell2019energy}%
  \BibitemOpen
  \bibfield  {author} {\bibinfo {author} {\bibfnamefont {E.}~\bibnamefont
  {Strubell}}, \bibinfo {author} {\bibfnamefont {A.}~\bibnamefont {Ganesh}},\
  and\ \bibinfo {author} {\bibfnamefont {A.}~\bibnamefont {McCallum}},\
  }\bibfield  {title} {\enquote {\bibinfo {title} {Energy and policy
  considerations for deep learning in {NLP}},}\ }\href@noop {} {\bibfield
  {journal} {\bibinfo  {journal} {arXiv preprint arXiv:1906.02243}\ } (\bibinfo
  {year} {2019})}\BibitemShut {NoStop}%
\bibitem [{\citenamefont {Levy}\ and\ \citenamefont
  {Calvert}(2021)}]{levy2021communication}%
  \BibitemOpen
  \bibfield  {author} {\bibinfo {author} {\bibfnamefont {W.~B.}\ \bibnamefont
  {Levy}}\ and\ \bibinfo {author} {\bibfnamefont {V.~G.}\ \bibnamefont
  {Calvert}},\ }\bibfield  {title} {\enquote {\bibinfo {title} {Communication
  consumes 35 times more energy than computation in the human cortex, but both
  costs are needed to predict synapse number},}\ }\href@noop {} {\bibfield
  {journal} {\bibinfo  {journal} {Proceedings of the National Academy of
  Sciences}\ }\textbf {\bibinfo {volume} {118}} (\bibinfo {year}
  {2021})}\BibitemShut {NoStop}%
\bibitem [{\citenamefont {Grollier}\ \emph {et~al.}(2020)\citenamefont
  {Grollier}, \citenamefont {Querlioz}, \citenamefont {Camsari}, \citenamefont
  {Everschor-Sitte}, \citenamefont {Fukami},\ and\ \citenamefont
  {Stiles}}]{Grollier_Querlioz_Camsari_EverschorSitte_Fukami_Stiles_2020}%
  \BibitemOpen
  \bibfield  {author} {\bibinfo {author} {\bibfnamefont {J.}~\bibnamefont
  {Grollier}}, \bibinfo {author} {\bibfnamefont {D.}~\bibnamefont {Querlioz}},
  \bibinfo {author} {\bibfnamefont {K.~Y.}\ \bibnamefont {Camsari}}, \bibinfo
  {author} {\bibfnamefont {K.}~\bibnamefont {Everschor-Sitte}}, \bibinfo
  {author} {\bibfnamefont {S.}~\bibnamefont {Fukami}},\ and\ \bibinfo {author}
  {\bibfnamefont {M.~D.}\ \bibnamefont {Stiles}},\ }\bibfield  {title}
  {\enquote {\bibinfo {title} {Neuromorphic spintronics},}\ }\href
  {https://doi.org/10.1038/s41928-019-0360-9} {\bibfield  {journal} {\bibinfo
  {journal} {Nature Electronics}\ }\textbf {\bibinfo {volume} {3}},\ \bibinfo
  {pages} {360--370} (\bibinfo {year} {2020})}\BibitemShut {NoStop}%
\bibitem [{\citenamefont {Capra}\ \emph {et~al.}(2020)\citenamefont {Capra},
  \citenamefont {Bussolino}, \citenamefont {Marchisio}, \citenamefont
  {Shafique}, \citenamefont {Masera},\ and\ \citenamefont
  {Martina}}]{Capra_Bussolino_Marchisio_Shafique_Masera_Martina_2020}%
  \BibitemOpen
  \bibfield  {author} {\bibinfo {author} {\bibfnamefont {M.}~\bibnamefont
  {Capra}}, \bibinfo {author} {\bibfnamefont {B.}~\bibnamefont {Bussolino}},
  \bibinfo {author} {\bibfnamefont {A.}~\bibnamefont {Marchisio}}, \bibinfo
  {author} {\bibfnamefont {M.}~\bibnamefont {Shafique}}, \bibinfo {author}
  {\bibfnamefont {G.}~\bibnamefont {Masera}},\ and\ \bibinfo {author}
  {\bibfnamefont {M.}~\bibnamefont {Martina}},\ }\bibfield  {title} {\enquote
  {\bibinfo {title} {An updated survey of efficient hardware architectures for
  accelerating deep convolutional neural networks},}\ }\href
  {https://doi.org/10.3390/fi12070113} {\bibfield  {journal} {\bibinfo
  {journal} {Future Internet}\ }\textbf {\bibinfo {volume} {12}},\ \bibinfo
  {pages} {113} (\bibinfo {year} {2020})}\BibitemShut {NoStop}%
\bibitem [{\citenamefont {Belkhir}\ and\ \citenamefont
  {Elmeligi}(2018)}]{Belkhir_Elmeligi_2018}%
  \BibitemOpen
  \bibfield  {author} {\bibinfo {author} {\bibfnamefont {L.}~\bibnamefont
  {Belkhir}}\ and\ \bibinfo {author} {\bibfnamefont {A.}~\bibnamefont
  {Elmeligi}},\ }\bibfield  {title} {\enquote {\bibinfo {title} {Assessing
  {ICT} global emissions footprint: {T}rends to 2040 \& recommendations},}\
  }\href {https://doi.org/10.1016/j.jclepro.2017.12.239} {\bibfield  {journal}
  {\bibinfo  {journal} {Journal of Cleaner Production}\ }\textbf {\bibinfo
  {volume} {177}},\ \bibinfo {pages} {448--463} (\bibinfo {year}
  {2018})}\BibitemShut {NoStop}%
\bibitem [{\citenamefont {Masanet}\ \emph {et~al.}(2020)\citenamefont
  {Masanet}, \citenamefont {Shehabi}, \citenamefont {Lei}, \citenamefont
  {Smith},\ and\ \citenamefont
  {Koomey}}]{Masanet_Shehabi_Lei_Smith_Koomey_2020}%
  \BibitemOpen
  \bibfield  {author} {\bibinfo {author} {\bibfnamefont {E.}~\bibnamefont
  {Masanet}}, \bibinfo {author} {\bibfnamefont {A.}~\bibnamefont {Shehabi}},
  \bibinfo {author} {\bibfnamefont {N.}~\bibnamefont {Lei}}, \bibinfo {author}
  {\bibfnamefont {S.}~\bibnamefont {Smith}},\ and\ \bibinfo {author}
  {\bibfnamefont {J.}~\bibnamefont {Koomey}},\ }\bibfield  {title} {\enquote
  {\bibinfo {title} {Recalibrating global data center energy-use estimates},}\
  }\href {https://doi.org/10.1126/science.aba3758} {\bibfield  {journal}
  {\bibinfo  {journal} {Science}\ }\textbf {\bibinfo {volume} {367}},\ \bibinfo
  {pages} {984--986} (\bibinfo {year} {2020})}\BibitemShut {NoStop}%
\bibitem [{\citenamefont {Kendall}\ and\ \citenamefont
  {Kumar}(2020)}]{kendall2020building}%
  \BibitemOpen
  \bibfield  {author} {\bibinfo {author} {\bibfnamefont {J.~D.}\ \bibnamefont
  {Kendall}}\ and\ \bibinfo {author} {\bibfnamefont {S.}~\bibnamefont
  {Kumar}},\ }\bibfield  {title} {\enquote {\bibinfo {title} {The building
  blocks of a brain-inspired computer},}\ }\href@noop {} {\bibfield  {journal}
  {\bibinfo  {journal} {Applied Physics Reviews}\ }\textbf {\bibinfo {volume}
  {7}},\ \bibinfo {pages} {011305} (\bibinfo {year} {2020})}\BibitemShut
  {NoStop}%
\bibitem [{\citenamefont {Davies}\ \emph {et~al.}(2018)\citenamefont {Davies},
  \citenamefont {Srinivasa}, \citenamefont {Lin}, \citenamefont {Chinya},
  \citenamefont {Cao}, \citenamefont {Choday}, \citenamefont {Dimou},
  \citenamefont {Joshi}, \citenamefont {Imam}, \citenamefont {Jain},\ and\
  \citenamefont
  {et~al.}}]{Davies_Srinivasa_Lin_Chinya_Cao_Choday_Dimou_Joshi_Imam_Jain_et2018}%
  \BibitemOpen
  \bibfield  {author} {\bibinfo {author} {\bibfnamefont {M.}~\bibnamefont
  {Davies}}, \bibinfo {author} {\bibfnamefont {N.}~\bibnamefont {Srinivasa}},
  \bibinfo {author} {\bibfnamefont {T.-H.}\ \bibnamefont {Lin}}, \bibinfo
  {author} {\bibfnamefont {G.}~\bibnamefont {Chinya}}, \bibinfo {author}
  {\bibfnamefont {Y.}~\bibnamefont {Cao}}, \bibinfo {author} {\bibfnamefont
  {S.~H.}\ \bibnamefont {Choday}}, \bibinfo {author} {\bibfnamefont
  {G.}~\bibnamefont {Dimou}}, \bibinfo {author} {\bibfnamefont
  {P.}~\bibnamefont {Joshi}}, \bibinfo {author} {\bibfnamefont
  {N.}~\bibnamefont {Imam}}, \bibinfo {author} {\bibfnamefont {S.}~\bibnamefont
  {Jain}},\ and\ \bibinfo {author} {\bibnamefont {et~al.}},\ }\bibfield
  {title} {\enquote {\bibinfo {title} {Loihi: {A} neuromorphic manycore
  processor with on-chip learning},}\ }\href
  {https://doi.org/10.1109/MM.2018.112130359} {\bibfield  {journal} {\bibinfo
  {journal} {IEEE Micro}\ }\textbf {\bibinfo {volume} {38}},\ \bibinfo {pages}
  {82--99} (\bibinfo {year} {2018})}\BibitemShut {NoStop}%
\bibitem [{\citenamefont {Davies}\ \emph {et~al.}(2021)\citenamefont {Davies},
  \citenamefont {Wild}, \citenamefont {Orchard}, \citenamefont {Sandamirskaya},
  \citenamefont {Guerra}, \citenamefont {Joshi}, \citenamefont {Plank},\ and\
  \citenamefont {Risbud}}]{davies2021advancing}%
  \BibitemOpen
  \bibfield  {author} {\bibinfo {author} {\bibfnamefont {M.}~\bibnamefont
  {Davies}}, \bibinfo {author} {\bibfnamefont {A.}~\bibnamefont {Wild}},
  \bibinfo {author} {\bibfnamefont {G.}~\bibnamefont {Orchard}}, \bibinfo
  {author} {\bibfnamefont {Y.}~\bibnamefont {Sandamirskaya}}, \bibinfo {author}
  {\bibfnamefont {G.~A.~F.}\ \bibnamefont {Guerra}}, \bibinfo {author}
  {\bibfnamefont {P.}~\bibnamefont {Joshi}}, \bibinfo {author} {\bibfnamefont
  {P.}~\bibnamefont {Plank}},\ and\ \bibinfo {author} {\bibfnamefont {S.~R.}\
  \bibnamefont {Risbud}},\ }\bibfield  {title} {\enquote {\bibinfo {title}
  {Advancing neuromorphic computing with {L}oihi: A survey of results and
  outlook},}\ }\href@noop {} {\bibfield  {journal} {\bibinfo  {journal}
  {Proceedings of the IEEE}\ }\textbf {\bibinfo {volume} {109}},\ \bibinfo
  {pages} {911--934} (\bibinfo {year} {2021})}\BibitemShut {NoStop}%
\bibitem [{\citenamefont {Akopyan}\ \emph {et~al.}(2015)\citenamefont
  {Akopyan}, \citenamefont {Sawada}, \citenamefont {Cassidy}, \citenamefont
  {Alvarez-Icaza}, \citenamefont {Arthur}, \citenamefont {Merolla},
  \citenamefont {Imam}, \citenamefont {Nakamura}, \citenamefont {Datta},
  \citenamefont {Nam},\ and\ \citenamefont
  {et~al.}}]{Akopyan_Sawada_Cassidy_AlvarezIcaza_Arthur_Merolla_Imam_Nakamura_Datta_Nam2015}%
  \BibitemOpen
  \bibfield  {author} {\bibinfo {author} {\bibfnamefont {F.}~\bibnamefont
  {Akopyan}}, \bibinfo {author} {\bibfnamefont {J.}~\bibnamefont {Sawada}},
  \bibinfo {author} {\bibfnamefont {A.}~\bibnamefont {Cassidy}}, \bibinfo
  {author} {\bibfnamefont {R.}~\bibnamefont {Alvarez-Icaza}}, \bibinfo {author}
  {\bibfnamefont {J.}~\bibnamefont {Arthur}}, \bibinfo {author} {\bibfnamefont
  {P.}~\bibnamefont {Merolla}}, \bibinfo {author} {\bibfnamefont
  {N.}~\bibnamefont {Imam}}, \bibinfo {author} {\bibfnamefont {Y.}~\bibnamefont
  {Nakamura}}, \bibinfo {author} {\bibfnamefont {P.}~\bibnamefont {Datta}},
  \bibinfo {author} {\bibfnamefont {G.-J.}\ \bibnamefont {Nam}},\ and\ \bibinfo
  {author} {\bibnamefont {et~al.}},\ }\bibfield  {title} {\enquote {\bibinfo
  {title} {Truenorth: Design and tool flow of a 65 m{W} 1 million neuron
  programmable neurosynaptic chip},}\ }\href
  {https://doi.org/10.1109/TCAD.2015.2474396} {\bibfield  {journal} {\bibinfo
  {journal} {IEEE Transactions on Computer-Aided Design of Integrated Circuits
  and Systems}\ }\textbf {\bibinfo {volume} {34}},\ \bibinfo {pages}
  {1537--1557} (\bibinfo {year} {2015})}\BibitemShut {NoStop}%
\bibitem [{\citenamefont {Merolla}\ \emph {et~al.}(2014)\citenamefont
  {Merolla}, \citenamefont {Arthur}, \citenamefont {Alvarez-Icaza},
  \citenamefont {Cassidy}, \citenamefont {Sawada}, \citenamefont {Akopyan},
  \citenamefont {Jackson}, \citenamefont {Imam}, \citenamefont {Guo},
  \citenamefont {Nakamura},\ and\ \citenamefont
  {et~al.}}]{Merolla_Arthur_AlvarezIcaza_Cassidy_Sawada_Akopyan_Jackson_Imam_Guo_Nakamura2014}%
  \BibitemOpen
  \bibfield  {author} {\bibinfo {author} {\bibfnamefont {P.~A.}\ \bibnamefont
  {Merolla}}, \bibinfo {author} {\bibfnamefont {J.~V.}\ \bibnamefont {Arthur}},
  \bibinfo {author} {\bibfnamefont {R.}~\bibnamefont {Alvarez-Icaza}}, \bibinfo
  {author} {\bibfnamefont {A.~S.}\ \bibnamefont {Cassidy}}, \bibinfo {author}
  {\bibfnamefont {J.}~\bibnamefont {Sawada}}, \bibinfo {author} {\bibfnamefont
  {F.}~\bibnamefont {Akopyan}}, \bibinfo {author} {\bibfnamefont {B.~L.}\
  \bibnamefont {Jackson}}, \bibinfo {author} {\bibfnamefont {N.}~\bibnamefont
  {Imam}}, \bibinfo {author} {\bibfnamefont {C.}~\bibnamefont {Guo}}, \bibinfo
  {author} {\bibfnamefont {Y.}~\bibnamefont {Nakamura}},\ and\ \bibinfo
  {author} {\bibnamefont {et~al.}},\ }\bibfield  {title} {\enquote {\bibinfo
  {title} {Artificial brains. {A} million spiking-neuron integrated circuit
  with a scalable communication network and interface},}\ }\href
  {https://doi.org/10.1126/science.1254642} {\bibfield  {journal} {\bibinfo
  {journal} {Science}\ }\textbf {\bibinfo {volume} {345}},\ \bibinfo {pages}
  {668--673} (\bibinfo {year} {2014})}\BibitemShut {NoStop}%
\bibitem [{\citenamefont {Esser}\ \emph {et~al.}(2016)\citenamefont {Esser},
  \citenamefont {Merolla}, \citenamefont {Arthur}, \citenamefont {Cassidy},
  \citenamefont {Appuswamy}, \citenamefont {Andreopoulos}, \citenamefont
  {Berg}, \citenamefont {McKinstry}, \citenamefont {Melano}, \citenamefont
  {Barch},\ and\ \citenamefont
  {et~al.}}]{Esser_Merolla_Arthur_Cassidy_Appuswamy_Andreopoulos_Berg_McKinstry_Melano_Barch2016}%
  \BibitemOpen
  \bibfield  {author} {\bibinfo {author} {\bibfnamefont {S.~K.}\ \bibnamefont
  {Esser}}, \bibinfo {author} {\bibfnamefont {P.~A.}\ \bibnamefont {Merolla}},
  \bibinfo {author} {\bibfnamefont {J.~V.}\ \bibnamefont {Arthur}}, \bibinfo
  {author} {\bibfnamefont {A.~S.}\ \bibnamefont {Cassidy}}, \bibinfo {author}
  {\bibfnamefont {R.}~\bibnamefont {Appuswamy}}, \bibinfo {author}
  {\bibfnamefont {A.}~\bibnamefont {Andreopoulos}}, \bibinfo {author}
  {\bibfnamefont {D.~J.}\ \bibnamefont {Berg}}, \bibinfo {author}
  {\bibfnamefont {J.~L.}\ \bibnamefont {McKinstry}}, \bibinfo {author}
  {\bibfnamefont {T.}~\bibnamefont {Melano}}, \bibinfo {author} {\bibfnamefont
  {D.~R.}\ \bibnamefont {Barch}},\ and\ \bibinfo {author} {\bibnamefont
  {et~al.}},\ }\bibfield  {title} {\enquote {\bibinfo {title} {Convolutional
  networks for fast, energy-efficient neuromorphic computing},}\ }\href
  {https://doi.org/10.1073/pnas.1604850113} {\bibfield  {journal} {\bibinfo
  {journal} {Proceedings of the National Academy of Sciences}\ }\textbf
  {\bibinfo {volume} {113}},\ \bibinfo {pages} {11441--11446} (\bibinfo {year}
  {2016})}\BibitemShut {NoStop}%
\bibitem [{\citenamefont {Furber}\ \emph {et~al.}(2014)\citenamefont {Furber},
  \citenamefont {Galluppi}, \citenamefont {Temple},\ and\ \citenamefont
  {Plana}}]{Furber_Galluppi_Temple_Plana_2014}%
  \BibitemOpen
  \bibfield  {author} {\bibinfo {author} {\bibfnamefont {S.~B.}\ \bibnamefont
  {Furber}}, \bibinfo {author} {\bibfnamefont {F.}~\bibnamefont {Galluppi}},
  \bibinfo {author} {\bibfnamefont {S.}~\bibnamefont {Temple}},\ and\ \bibinfo
  {author} {\bibfnamefont {L.~A.}\ \bibnamefont {Plana}},\ }\bibfield  {title}
  {\enquote {\bibinfo {title} {The {S}pi{NN}aker project},}\ }\href
  {https://doi.org/10.1109/JPROC.2014.2304638} {\bibfield  {journal} {\bibinfo
  {journal} {Proceedings of the IEEE}\ }\textbf {\bibinfo {volume} {102}},\
  \bibinfo {pages} {652--665} (\bibinfo {year} {2014})}\BibitemShut {NoStop}%
\bibitem [{\citenamefont {Neckar}\ \emph {et~al.}(2018)\citenamefont {Neckar},
  \citenamefont {Fok}, \citenamefont {Benjamin}, \citenamefont {Stewart},
  \citenamefont {Oza}, \citenamefont {Voelker}, \citenamefont {Eliasmith},
  \citenamefont {Manohar},\ and\ \citenamefont {Boahen}}]{neckar2018braindrop}%
  \BibitemOpen
  \bibfield  {author} {\bibinfo {author} {\bibfnamefont {A.}~\bibnamefont
  {Neckar}}, \bibinfo {author} {\bibfnamefont {S.}~\bibnamefont {Fok}},
  \bibinfo {author} {\bibfnamefont {B.~V.}\ \bibnamefont {Benjamin}}, \bibinfo
  {author} {\bibfnamefont {T.~C.}\ \bibnamefont {Stewart}}, \bibinfo {author}
  {\bibfnamefont {N.~N.}\ \bibnamefont {Oza}}, \bibinfo {author} {\bibfnamefont
  {A.~R.}\ \bibnamefont {Voelker}}, \bibinfo {author} {\bibfnamefont
  {C.}~\bibnamefont {Eliasmith}}, \bibinfo {author} {\bibfnamefont
  {R.}~\bibnamefont {Manohar}},\ and\ \bibinfo {author} {\bibfnamefont
  {K.}~\bibnamefont {Boahen}},\ }\bibfield  {title} {\enquote {\bibinfo {title}
  {Braindrop: A mixed-signal neuromorphic architecture with a dynamical
  systems-based programming model},}\ }\href@noop {} {\bibfield  {journal}
  {\bibinfo  {journal} {Proceedings of the IEEE}\ }\textbf {\bibinfo {volume}
  {107}},\ \bibinfo {pages} {144--164} (\bibinfo {year} {2018})}\BibitemShut
  {NoStop}%
\bibitem [{\citenamefont {Hemsoth}(2017)}]{Hemsoth_2017}%
  \BibitemOpen
  \bibfield  {author} {\bibinfo {author} {\bibfnamefont {N.}~\bibnamefont
  {Hemsoth}},\ }\bibfield  {title} {\enquote {\bibinfo {title} {Stanford
  {B}rainstorm {C}hip to hints at neuromorphic computing future},}\ }\href
  {https://www.nextplatform.com/2017/03/27/stanford-brainstorm-chip-hints-neuromorphic-computing-future/}
  {\bibfield  {journal} {\bibinfo  {journal} {The Next Platform}\ } (\bibinfo
  {year} {2017})}\BibitemShut {NoStop}%
\bibitem [{\citenamefont {Smith}(2021)}]{Smith_2021}%
  \BibitemOpen
  \bibfield  {author} {\bibinfo {author} {\bibfnamefont {M.}~\bibnamefont
  {Smith}},\ }\bibfield  {title} {\enquote {\bibinfo {title} {Brainchip
  releases latest episode in `{T}his is our mission' series},}\ }\href
  {https://www.businesswire.com/news/home/20210330006063/en/BrainChip-Releases-Latest-Episode-in-%E2%80%98This-is-our-Mission%E2%80%99-Series}
  {\bibfield  {journal} {\bibinfo  {journal} {Business Wire, JPR
  Communications}\ } (\bibinfo {year} {2021})}\BibitemShut {NoStop}%
\bibitem [{\citenamefont {Zhu}\ \emph {et~al.}(2020)\citenamefont {Zhu},
  \citenamefont {Zhang}, \citenamefont {Yang},\ and\ \citenamefont
  {Huang}}]{zhu2020comprehensive}%
  \BibitemOpen
  \bibfield  {author} {\bibinfo {author} {\bibfnamefont {J.}~\bibnamefont
  {Zhu}}, \bibinfo {author} {\bibfnamefont {T.}~\bibnamefont {Zhang}}, \bibinfo
  {author} {\bibfnamefont {Y.}~\bibnamefont {Yang}},\ and\ \bibinfo {author}
  {\bibfnamefont {R.}~\bibnamefont {Huang}},\ }\bibfield  {title} {\enquote
  {\bibinfo {title} {A comprehensive review on emerging artificial neuromorphic
  devices},}\ }\href@noop {} {\bibfield  {journal} {\bibinfo  {journal}
  {Applied Physics Reviews}\ }\textbf {\bibinfo {volume} {7}},\ \bibinfo
  {pages} {011312} (\bibinfo {year} {2020})}\BibitemShut {NoStop}%
\bibitem [{\citenamefont {Khymyn}\ \emph {et~al.}(2018)\citenamefont {Khymyn},
  \citenamefont {Lisenkov}, \citenamefont {Voorheis}, \citenamefont
  {Sulymenko}, \citenamefont {Prokopenko}, \citenamefont {Tiberkevich},
  \citenamefont {Akerman},\ and\ \citenamefont
  {Slavin}}]{Khymyn_Lisenkov_Voorheis_Sulymenko_Prokopenko_Tiberkevich_Akerman_Slavin_2018}%
  \BibitemOpen
  \bibfield  {author} {\bibinfo {author} {\bibfnamefont {R.}~\bibnamefont
  {Khymyn}}, \bibinfo {author} {\bibfnamefont {I.}~\bibnamefont {Lisenkov}},
  \bibinfo {author} {\bibfnamefont {J.}~\bibnamefont {Voorheis}}, \bibinfo
  {author} {\bibfnamefont {O.}~\bibnamefont {Sulymenko}}, \bibinfo {author}
  {\bibfnamefont {O.}~\bibnamefont {Prokopenko}}, \bibinfo {author}
  {\bibfnamefont {V.}~\bibnamefont {Tiberkevich}}, \bibinfo {author}
  {\bibfnamefont {J.}~\bibnamefont {Akerman}},\ and\ \bibinfo {author}
  {\bibfnamefont {A.}~\bibnamefont {Slavin}},\ }\bibfield  {title} {\enquote
  {\bibinfo {title} {Ultra-fast artificial neuron: generation of
  picosecond-duration spikes in a current-driven antiferromagnetic
  auto-oscillator},}\ }\href@noop {} {\bibfield  {journal} {\bibinfo  {journal}
  {Scientific Reports}\ }\textbf {\bibinfo {volume} {8}},\ \bibinfo {pages}
  {1--9} (\bibinfo {year} {2018})}\BibitemShut {NoStop}%
\bibitem [{\citenamefont {Sulymenko}\ \emph
  {et~al.}(2018{\natexlab{a}})\citenamefont {Sulymenko}, \citenamefont
  {Prokopenko}, \citenamefont {Lisenkov}, \citenamefont {{\AA}kerman},
  \citenamefont {Tyberkevych}, \citenamefont {Slavin},\ and\ \citenamefont
  {Khymyn}}]{Sulymenko_Prokopenko_Lisenkov_kerman_Tyberkevych_Slavin_Khymyn_2018}%
  \BibitemOpen
  \bibfield  {author} {\bibinfo {author} {\bibfnamefont {O.}~\bibnamefont
  {Sulymenko}}, \bibinfo {author} {\bibfnamefont {O.}~\bibnamefont
  {Prokopenko}}, \bibinfo {author} {\bibfnamefont {I.}~\bibnamefont
  {Lisenkov}}, \bibinfo {author} {\bibfnamefont {J.}~\bibnamefont
  {{\AA}kerman}}, \bibinfo {author} {\bibfnamefont {V.}~\bibnamefont
  {Tyberkevych}}, \bibinfo {author} {\bibfnamefont {A.~N.}\ \bibnamefont
  {Slavin}},\ and\ \bibinfo {author} {\bibfnamefont {R.}~\bibnamefont
  {Khymyn}},\ }\bibfield  {title} {\enquote {\bibinfo {title} {Ultra-fast logic
  devices using artificial ``neurons'' based on antiferromagnetic pulse
  generators},}\ }\href {https://doi.org/10.1063/1.5042348} {\bibfield
  {journal} {\bibinfo  {journal} {Journal of Applied Physics}\ }\textbf
  {\bibinfo {volume} {124}},\ \bibinfo {pages} {152115} (\bibinfo {year}
  {2018}{\natexlab{a}})}\BibitemShut {NoStop}%
\bibitem [{\citenamefont {Sulymenko}\ and\ \citenamefont
  {Prokopenko}(2019)}]{Sulymenko_Prokopenko_2019}%
  \BibitemOpen
  \bibfield  {author} {\bibinfo {author} {\bibfnamefont {O.}~\bibnamefont
  {Sulymenko}}\ and\ \bibinfo {author} {\bibfnamefont {O.}~\bibnamefont
  {Prokopenko}},\ }\bibfield  {title} {\enquote {\bibinfo {title} {Logic
  circuits based on neuron-like antiferromagnetic spin {H}all oscillators},}\
  }in\ \href {https://doi.org/10.1109/ELNANO.2019.8783266} {\emph {\bibinfo
  {booktitle} {2019 IEEE 39th International Conference on Electronics and
  Nanotechnology (ELNANO)}}}\ (\bibinfo  {publisher} {IEEE},\ \bibinfo {year}
  {2019})\ pp.\ \bibinfo {pages} {132--137}\BibitemShut {NoStop}%
\bibitem [{\citenamefont {Zahedinejad}\ \emph {et~al.}(2018)\citenamefont
  {Zahedinejad}, \citenamefont {Mazraati}, \citenamefont {Fulara},
  \citenamefont {Yue}, \citenamefont {Jiang}, \citenamefont {Awad},\ and\
  \citenamefont {{\AA}kerman}}]{zahedinejad2018cmos}%
  \BibitemOpen
  \bibfield  {author} {\bibinfo {author} {\bibfnamefont {M.}~\bibnamefont
  {Zahedinejad}}, \bibinfo {author} {\bibfnamefont {H.}~\bibnamefont
  {Mazraati}}, \bibinfo {author} {\bibfnamefont {H.}~\bibnamefont {Fulara}},
  \bibinfo {author} {\bibfnamefont {J.}~\bibnamefont {Yue}}, \bibinfo {author}
  {\bibfnamefont {S.}~\bibnamefont {Jiang}}, \bibinfo {author} {\bibfnamefont
  {A.}~\bibnamefont {Awad}},\ and\ \bibinfo {author} {\bibfnamefont
  {J.}~\bibnamefont {{\AA}kerman}},\ }\bibfield  {title} {\enquote {\bibinfo
  {title} {{CMOS} compatible {W/CoFeB/MgO} spin {Hall} nano-oscillators with
  wide frequency tunability},}\ }\href@noop {} {\bibfield  {journal} {\bibinfo
  {journal} {Applied Physics Letters}\ }\textbf {\bibinfo {volume} {112}},\
  \bibinfo {pages} {132404} (\bibinfo {year} {2018})}\BibitemShut {NoStop}%
\bibitem [{\citenamefont {Fukami}\ and\ \citenamefont
  {Ohno}(2018)}]{Fukami_Ohno_2018}%
  \BibitemOpen
  \bibfield  {author} {\bibinfo {author} {\bibfnamefont {S.}~\bibnamefont
  {Fukami}}\ and\ \bibinfo {author} {\bibfnamefont {H.}~\bibnamefont {Ohno}},\
  }\bibfield  {title} {\enquote {\bibinfo {title} {Perspective: Spintronic
  synapse for artificial neural network},}\ }\href
  {https://doi.org/10.1063/1.5042317} {\bibfield  {journal} {\bibinfo
  {journal} {Journal of Applied Physics}\ }\textbf {\bibinfo {volume} {124}},\
  \bibinfo {pages} {151904} (\bibinfo {year} {2018})}\BibitemShut {NoStop}%
\bibitem [{\citenamefont {Krzysteczko}\ \emph {et~al.}(2012)\citenamefont
  {Krzysteczko}, \citenamefont {M{\"u}nchenberger}, \citenamefont
  {Sch{\"a}fers}, \citenamefont {Reiss},\ and\ \citenamefont
  {Thomas}}]{krzysteczko2012memristive}%
  \BibitemOpen
  \bibfield  {author} {\bibinfo {author} {\bibfnamefont {P.}~\bibnamefont
  {Krzysteczko}}, \bibinfo {author} {\bibfnamefont {J.}~\bibnamefont
  {M{\"u}nchenberger}}, \bibinfo {author} {\bibfnamefont {M.}~\bibnamefont
  {Sch{\"a}fers}}, \bibinfo {author} {\bibfnamefont {G.}~\bibnamefont
  {Reiss}},\ and\ \bibinfo {author} {\bibfnamefont {A.}~\bibnamefont
  {Thomas}},\ }\bibfield  {title} {\enquote {\bibinfo {title} {The memristive
  magnetic tunnel junction as a nanoscopic synapse-neuron system},}\
  }\href@noop {} {\bibfield  {journal} {\bibinfo  {journal} {Advanced
  Materials}\ }\textbf {\bibinfo {volume} {24}},\ \bibinfo {pages} {762--766}
  (\bibinfo {year} {2012})}\BibitemShut {NoStop}%
\bibitem [{\citenamefont {Kurenkov}\ \emph {et~al.}(2019)\citenamefont
  {Kurenkov}, \citenamefont {DuttaGupta}, \citenamefont {Zhang}, \citenamefont
  {Fukami}, \citenamefont {Horio},\ and\ \citenamefont
  {Ohno}}]{Kurenkov_DuttaGupta_Zhang_Fukami_Horio_Ohno_2019}%
  \BibitemOpen
  \bibfield  {author} {\bibinfo {author} {\bibfnamefont {A.}~\bibnamefont
  {Kurenkov}}, \bibinfo {author} {\bibfnamefont {S.}~\bibnamefont
  {DuttaGupta}}, \bibinfo {author} {\bibfnamefont {C.}~\bibnamefont {Zhang}},
  \bibinfo {author} {\bibfnamefont {S.}~\bibnamefont {Fukami}}, \bibinfo
  {author} {\bibfnamefont {Y.}~\bibnamefont {Horio}},\ and\ \bibinfo {author}
  {\bibfnamefont {H.}~\bibnamefont {Ohno}},\ }\bibfield  {title} {\enquote
  {\bibinfo {title} {Artificial neuron and synapse realized in an
  antiferromagnet/ferromagnet heterostructure using dynamics of spin--orbit
  torque switching},}\ }\href {https://doi.org/10.1002/adma.201900636}
  {\bibfield  {journal} {\bibinfo  {journal} {Advanced Materials}\ }\textbf
  {\bibinfo {volume} {31}},\ \bibinfo {pages} {1900636} (\bibinfo {year}
  {2019})}\BibitemShut {NoStop}%
\bibitem [{\citenamefont {Pan}\ and\ \citenamefont
  {Naeemi}(2018)}]{Pan_Naeemi_2018}%
  \BibitemOpen
  \bibfield  {author} {\bibinfo {author} {\bibfnamefont {C.}~\bibnamefont
  {Pan}}\ and\ \bibinfo {author} {\bibfnamefont {A.}~\bibnamefont {Naeemi}},\
  }\bibfield  {title} {\enquote {\bibinfo {title} {Complementary logic
  implementation for antiferromagnet field-effect transistors},}\ }\href
  {https://doi.org/10.1109/JXCDC.2018.2878635} {\bibfield  {journal} {\bibinfo
  {journal} {IEEE Journal on Exploratory Solid-State Computational Devices and
  Circuits}\ }\textbf {\bibinfo {volume} {4}},\ \bibinfo {pages} {69--75}
  (\bibinfo {year} {2018})}\BibitemShut {NoStop}%
\bibitem [{\citenamefont {Zhang}\ and\ \citenamefont
  {Tserkovnyak}(2020{\natexlab{a}})}]{zhang2020antiferromagnet}%
  \BibitemOpen
  \bibfield  {author} {\bibinfo {author} {\bibfnamefont {S.}~\bibnamefont
  {Zhang}}\ and\ \bibinfo {author} {\bibfnamefont {Y.}~\bibnamefont
  {Tserkovnyak}},\ }\bibfield  {title} {\enquote {\bibinfo {title}
  {Antiferromagnet-based neuromorphics using dynamics of topological
  charges},}\ }\href@noop {} {\bibfield  {journal} {\bibinfo  {journal}
  {Physical Review Letters}\ }\textbf {\bibinfo {volume} {125}},\ \bibinfo
  {pages} {207202} (\bibinfo {year} {2020}{\natexlab{a}})}\BibitemShut
  {NoStop}%
\bibitem [{\citenamefont {Cai}\ \emph {et~al.}(2019)\citenamefont {Cai},
  \citenamefont {Fang}, \citenamefont {Zhang}, \citenamefont {Lv},
  \citenamefont {Zhang}, \citenamefont {Zhou}, \citenamefont {Finocchio},\ and\
  \citenamefont {Zeng}}]{cai2019voltage}%
  \BibitemOpen
  \bibfield  {author} {\bibinfo {author} {\bibfnamefont {J.}~\bibnamefont
  {Cai}}, \bibinfo {author} {\bibfnamefont {B.}~\bibnamefont {Fang}}, \bibinfo
  {author} {\bibfnamefont {L.}~\bibnamefont {Zhang}}, \bibinfo {author}
  {\bibfnamefont {W.}~\bibnamefont {Lv}}, \bibinfo {author} {\bibfnamefont
  {B.}~\bibnamefont {Zhang}}, \bibinfo {author} {\bibfnamefont
  {T.}~\bibnamefont {Zhou}}, \bibinfo {author} {\bibfnamefont {G.}~\bibnamefont
  {Finocchio}},\ and\ \bibinfo {author} {\bibfnamefont {Z.}~\bibnamefont
  {Zeng}},\ }\bibfield  {title} {\enquote {\bibinfo {title} {Voltage-controlled
  spintronic stochastic neuron based on a magnetic tunnel junction},}\
  }\href@noop {} {\bibfield  {journal} {\bibinfo  {journal} {Physical Review
  Applied}\ }\textbf {\bibinfo {volume} {11}},\ \bibinfo {pages} {034015}
  (\bibinfo {year} {2019})}\BibitemShut {NoStop}%
\bibitem [{\citenamefont {Sengupta}\ and\ \citenamefont
  {Roy}(2015)}]{Sengupta_Roy_2015}%
  \BibitemOpen
  \bibfield  {author} {\bibinfo {author} {\bibfnamefont {A.}~\bibnamefont
  {Sengupta}}\ and\ \bibinfo {author} {\bibfnamefont {K.}~\bibnamefont {Roy}},\
  }\bibfield  {title} {\enquote {\bibinfo {title} {Spin-transfer torque
  magnetic neuron for low power neuromorphic computing},}\ }in\ \href
  {https://doi.org/10.1109/IJCNN.2015.7280306} {\emph {\bibinfo {booktitle}
  {2015 International Joint Conference on Neural Networks (IJCNN)}}}\ (\bibinfo
   {publisher} {IEEE},\ \bibinfo {year} {2015})\ pp.\ \bibinfo {pages}
  {1--7}\BibitemShut {NoStop}%
\bibitem [{\citenamefont {Zhang}\ and\ \citenamefont
  {Tserkovnyak}(2020{\natexlab{b}})}]{Zhang_Tserkovnyak_2020}%
  \BibitemOpen
  \bibfield  {author} {\bibinfo {author} {\bibfnamefont {S.}~\bibnamefont
  {Zhang}}\ and\ \bibinfo {author} {\bibfnamefont {Y.}~\bibnamefont
  {Tserkovnyak}},\ }\bibfield  {title} {\enquote {\bibinfo {title}
  {Antiferromagnet-based neuromorphics using dynamics of topological
  charges},}\ }\href {https://doi.org/10.1103/PhysRevLett.125.207202}
  {\bibfield  {journal} {\bibinfo  {journal} {Physical Review Letters}\
  }\textbf {\bibinfo {volume} {125}},\ \bibinfo {pages} {207202} (\bibinfo
  {year} {2020}{\natexlab{b}})}\BibitemShut {NoStop}%
\bibitem [{\citenamefont {Louis}\ \emph {et~al.}(2016)\citenamefont {Louis},
  \citenamefont {Lisenkov}, \citenamefont {Nikitov}, \citenamefont
  {Tyberkevych},\ and\ \citenamefont
  {Slavin}}]{Louis_Lisenkov_Nikitov_Tyberkevych_Slavin_2016}%
  \BibitemOpen
  \bibfield  {author} {\bibinfo {author} {\bibfnamefont {S.}~\bibnamefont
  {Louis}}, \bibinfo {author} {\bibfnamefont {I.}~\bibnamefont {Lisenkov}},
  \bibinfo {author} {\bibfnamefont {S.}~\bibnamefont {Nikitov}}, \bibinfo
  {author} {\bibfnamefont {V.}~\bibnamefont {Tyberkevych}},\ and\ \bibinfo
  {author} {\bibfnamefont {A.}~\bibnamefont {Slavin}},\ }\bibfield  {title}
  {\enquote {\bibinfo {title} {Bias-free spin-wave phase shifter for magnonic
  logic},}\ }\href {https://doi.org/10.1063/1.4953395} {\bibfield  {journal}
  {\bibinfo  {journal} {AIP Advances}\ }\textbf {\bibinfo {volume} {6}},\
  \bibinfo {pages} {065103} (\bibinfo {year} {2016})}\BibitemShut {NoStop}%
\bibitem [{\citenamefont {Kurenkov}, \citenamefont {Fukami},\ and\
  \citenamefont {Ohno}(2020)}]{kurenkov2020neuromorphic}%
  \BibitemOpen
  \bibfield  {author} {\bibinfo {author} {\bibfnamefont {A.}~\bibnamefont
  {Kurenkov}}, \bibinfo {author} {\bibfnamefont {S.}~\bibnamefont {Fukami}},\
  and\ \bibinfo {author} {\bibfnamefont {H.}~\bibnamefont {Ohno}},\ }\bibfield
  {title} {\enquote {\bibinfo {title} {Neuromorphic computing with
  antiferromagnetic spintronics},}\ }\href@noop {} {\bibfield  {journal}
  {\bibinfo  {journal} {Journal of Applied Physics}\ }\textbf {\bibinfo
  {volume} {128}},\ \bibinfo {pages} {010902} (\bibinfo {year}
  {2020})}\BibitemShut {NoStop}%
\bibitem [{\citenamefont {Baryakhtar}\ and\ \citenamefont
  {Ivanov}(1986)}]{Baryakhtar}%
  \BibitemOpen
  \bibfield  {author} {\bibinfo {author} {\bibfnamefont {V.}~\bibnamefont
  {Baryakhtar}}\ and\ \bibinfo {author} {\bibfnamefont {B.}~\bibnamefont
  {Ivanov}},\ }\href@noop {} {\emph {\bibinfo {title} {Modern Magnetism: A
  Primer}}}\ (\bibinfo  {publisher} {Nauka Publishers},\ \bibinfo {year}
  {1986})\BibitemShut {NoStop}%
\bibitem [{\citenamefont {Coey}(2010)}]{Coey_2010}%
  \BibitemOpen
  \bibfield  {author} {\bibinfo {author} {\bibfnamefont {J.}~\bibnamefont
  {Coey}},\ }\href@noop {} {\emph {\bibinfo {title} {Magnetism and Magnetic
  Materials}}}\ (\bibinfo  {publisher} {Cambridge University Press},\ \bibinfo
  {year} {2010})\BibitemShut {NoStop}%
\bibitem [{\citenamefont {Moriyama}\ \emph {et~al.}(2018)\citenamefont
  {Moriyama}, \citenamefont {Oda}, \citenamefont {Ohkochi}, \citenamefont
  {Kimata},\ and\ \citenamefont {Ono}}]{moriyama2018spin}%
  \BibitemOpen
  \bibfield  {author} {\bibinfo {author} {\bibfnamefont {T.}~\bibnamefont
  {Moriyama}}, \bibinfo {author} {\bibfnamefont {K.}~\bibnamefont {Oda}},
  \bibinfo {author} {\bibfnamefont {T.}~\bibnamefont {Ohkochi}}, \bibinfo
  {author} {\bibfnamefont {M.}~\bibnamefont {Kimata}},\ and\ \bibinfo {author}
  {\bibfnamefont {T.}~\bibnamefont {Ono}},\ }\bibfield  {title} {\enquote
  {\bibinfo {title} {Spin torque control of antiferromagnetic moments in
  {NiO}},}\ }\href@noop {} {\bibfield  {journal} {\bibinfo  {journal}
  {Scientific Reports}\ }\textbf {\bibinfo {volume} {8}},\ \bibinfo {pages}
  {1--6} (\bibinfo {year} {2018})}\BibitemShut {NoStop}%
\bibitem [{\citenamefont {Chen}\ \emph {et~al.}(2018)\citenamefont {Chen},
  \citenamefont {Zarzuela}, \citenamefont {Zhang}, \citenamefont {Song},
  \citenamefont {Zhou}, \citenamefont {Shi}, \citenamefont {Li}, \citenamefont
  {Zhou}, \citenamefont {Jiang}, \citenamefont {Pan} \emph
  {et~al.}}]{chen2018antidamping}%
  \BibitemOpen
  \bibfield  {author} {\bibinfo {author} {\bibfnamefont {X.}~\bibnamefont
  {Chen}}, \bibinfo {author} {\bibfnamefont {R.}~\bibnamefont {Zarzuela}},
  \bibinfo {author} {\bibfnamefont {J.}~\bibnamefont {Zhang}}, \bibinfo
  {author} {\bibfnamefont {C.}~\bibnamefont {Song}}, \bibinfo {author}
  {\bibfnamefont {X.}~\bibnamefont {Zhou}}, \bibinfo {author} {\bibfnamefont
  {G.}~\bibnamefont {Shi}}, \bibinfo {author} {\bibfnamefont {F.}~\bibnamefont
  {Li}}, \bibinfo {author} {\bibfnamefont {H.}~\bibnamefont {Zhou}}, \bibinfo
  {author} {\bibfnamefont {W.}~\bibnamefont {Jiang}}, \bibinfo {author}
  {\bibfnamefont {F.}~\bibnamefont {Pan}}, \emph {et~al.},\ }\bibfield  {title}
  {\enquote {\bibinfo {title} {Antidamping-torque-induced switching in biaxial
  antiferromagnetic insulators},}\ }\href@noop {} {\bibfield  {journal}
  {\bibinfo  {journal} {Physical Review Letters}\ }\textbf {\bibinfo {volume}
  {120}},\ \bibinfo {pages} {207204} (\bibinfo {year} {2018})}\BibitemShut
  {NoStop}%
\bibitem [{\citenamefont {Moriyama}\ \emph {et~al.}(2019)\citenamefont
  {Moriyama}, \citenamefont {Hayashi}, \citenamefont {Yamada}, \citenamefont
  {Shima}, \citenamefont {Ohya},\ and\ \citenamefont
  {Ono}}]{moriyama2019intrinsic}%
  \BibitemOpen
  \bibfield  {author} {\bibinfo {author} {\bibfnamefont {T.}~\bibnamefont
  {Moriyama}}, \bibinfo {author} {\bibfnamefont {K.}~\bibnamefont {Hayashi}},
  \bibinfo {author} {\bibfnamefont {K.}~\bibnamefont {Yamada}}, \bibinfo
  {author} {\bibfnamefont {M.}~\bibnamefont {Shima}}, \bibinfo {author}
  {\bibfnamefont {Y.}~\bibnamefont {Ohya}},\ and\ \bibinfo {author}
  {\bibfnamefont {T.}~\bibnamefont {Ono}},\ }\bibfield  {title} {\enquote
  {\bibinfo {title} {Intrinsic and extrinsic antiferromagnetic damping in
  {NiO}},}\ }\href@noop {} {\bibfield  {journal} {\bibinfo  {journal} {Physical
  Review Materials}\ }\textbf {\bibinfo {volume} {3}},\ \bibinfo {pages}
  {051402} (\bibinfo {year} {2019})}\BibitemShut {NoStop}%
\bibitem [{\citenamefont {Khymyn}\ \emph {et~al.}(2017)\citenamefont {Khymyn},
  \citenamefont {Lisenkov}, \citenamefont {Tiberkevich}, \citenamefont
  {Ivanov},\ and\ \citenamefont
  {Slavin}}]{Khymyn_Lisenkov_Tiberkevich_Ivanov_Slavin_2017}%
  \BibitemOpen
  \bibfield  {author} {\bibinfo {author} {\bibfnamefont {R.}~\bibnamefont
  {Khymyn}}, \bibinfo {author} {\bibfnamefont {I.}~\bibnamefont {Lisenkov}},
  \bibinfo {author} {\bibfnamefont {V.}~\bibnamefont {Tiberkevich}}, \bibinfo
  {author} {\bibfnamefont {B.~A.}\ \bibnamefont {Ivanov}},\ and\ \bibinfo
  {author} {\bibfnamefont {A.}~\bibnamefont {Slavin}},\ }\bibfield  {title}
  {\enquote {\bibinfo {title} {Antiferromagnetic {TH}z-frequency
  {J}osephson-like oscillator driven by spin current},}\ }\href
  {https://doi.org/10.1038/srep43705} {\bibfield  {journal} {\bibinfo
  {journal} {Scientific Reports}\ }\textbf {\bibinfo {volume} {7}},\ \bibinfo
  {pages} {43705} (\bibinfo {year} {2017})}\BibitemShut {NoStop}%
\bibitem [{\citenamefont {Khymyn}, \citenamefont {Tiberkevich},\ and\
  \citenamefont {Slavin}(2017)}]{Khymyn_Tiberkevich_Slavin_2017}%
  \BibitemOpen
  \bibfield  {author} {\bibinfo {author} {\bibfnamefont {R.}~\bibnamefont
  {Khymyn}}, \bibinfo {author} {\bibfnamefont {V.}~\bibnamefont
  {Tiberkevich}},\ and\ \bibinfo {author} {\bibfnamefont {A.}~\bibnamefont
  {Slavin}},\ }\bibfield  {title} {\enquote {\bibinfo {title}
  {Antiferromagnetic spin current rectifier},}\ }\href
  {https://doi.org/10.1063/1.4977974} {\bibfield  {journal} {\bibinfo
  {journal} {AIP Advances}\ }\textbf {\bibinfo {volume} {7}},\ \bibinfo {pages}
  {055931} (\bibinfo {year} {2017})}\BibitemShut {NoStop}%
\bibitem [{\citenamefont {Khymyn}\ \emph {et~al.}(2016)\citenamefont {Khymyn},
  \citenamefont {Lisenkov}, \citenamefont {Tiberkevich}, \citenamefont
  {Slavin},\ and\ \citenamefont
  {Ivanov}}]{Khymyn_Lisenkov_Tiberkevich_Slavin_Ivanov_2016}%
  \BibitemOpen
  \bibfield  {author} {\bibinfo {author} {\bibfnamefont {R.}~\bibnamefont
  {Khymyn}}, \bibinfo {author} {\bibfnamefont {I.}~\bibnamefont {Lisenkov}},
  \bibinfo {author} {\bibfnamefont {V.~S.}\ \bibnamefont {Tiberkevich}},
  \bibinfo {author} {\bibfnamefont {A.~N.}\ \bibnamefont {Slavin}},\ and\
  \bibinfo {author} {\bibfnamefont {B.~A.}\ \bibnamefont {Ivanov}},\ }\bibfield
   {title} {\enquote {\bibinfo {title} {Transformation of spin current by
  antiferromagnetic insulators},}\ }\href
  {https://doi.org/10.1103/PhysRevB.93.224421} {\bibfield  {journal} {\bibinfo
  {journal} {Physical Review B}\ }\textbf {\bibinfo {volume} {93}} (\bibinfo
  {year} {2016}),\ 10.1103/PhysRevB.93.224421}\BibitemShut {NoStop}%
\bibitem [{\citenamefont {Hoffmann}(2013)}]{Hoffmann_2013}%
  \BibitemOpen
  \bibfield  {author} {\bibinfo {author} {\bibfnamefont {A.}~\bibnamefont
  {Hoffmann}},\ }\bibfield  {title} {\enquote {\bibinfo {title} {Spin {H}all
  effects in metals},}\ }\href {https://doi.org/10.1109/TMAG.2013.2262947}
  {\bibfield  {journal} {\bibinfo  {journal} {IEEE Transactions on Magnetics}\
  }\textbf {\bibinfo {volume} {49}},\ \bibinfo {pages} {5172--5193} (\bibinfo
  {year} {2013})}\BibitemShut {NoStop}%
\bibitem [{\citenamefont {Sinova}\ \emph {et~al.}(2015)\citenamefont {Sinova},
  \citenamefont {Valenzuela}, \citenamefont {Wunderlich}, \citenamefont
  {Back},\ and\ \citenamefont
  {Jungwirth}}]{Sinova_Valenzuela_Wunderlich_Back_Jungwirth_2015}%
  \BibitemOpen
  \bibfield  {author} {\bibinfo {author} {\bibfnamefont {J.}~\bibnamefont
  {Sinova}}, \bibinfo {author} {\bibfnamefont {S.~O.}\ \bibnamefont
  {Valenzuela}}, \bibinfo {author} {\bibfnamefont {J.}~\bibnamefont
  {Wunderlich}}, \bibinfo {author} {\bibfnamefont {C.}~\bibnamefont {Back}},\
  and\ \bibinfo {author} {\bibfnamefont {T.}~\bibnamefont {Jungwirth}},\
  }\bibfield  {title} {\enquote {\bibinfo {title} {Spin {H}all effects},}\
  }\href {https://doi.org/10.1103/RevModPhys.87.1213} {\bibfield  {journal}
  {\bibinfo  {journal} {Reviews of Modern Physics}\ }\textbf {\bibinfo {volume}
  {87}},\ \bibinfo {pages} {1213--1260} (\bibinfo {year} {2015})}\BibitemShut
  {NoStop}%
\bibitem [{\citenamefont {Meckbach}(2013)}]{Meckbach_2013}%
  \BibitemOpen
  \bibfield  {author} {\bibinfo {author} {\bibfnamefont {J.~M.}\ \bibnamefont
  {Meckbach}},\ }\href@noop {} {\emph {\bibinfo {title} {Superconducting
  Multilayer Technology for {J}osephson Devices: Technology, Engineering,
  Physics, Applications}}},\ Karlsruher Schriftenreihe zur Supraleitung\
  (\bibinfo  {publisher} {KIT Scientific Publ},\ \bibinfo {year}
  {2013})\BibitemShut {NoStop}%
\bibitem [{\citenamefont {Anderson}(1964)}]{anderson1964lectures}%
  \BibitemOpen
  \bibfield  {author} {\bibinfo {author} {\bibfnamefont {P.}~\bibnamefont
  {Anderson}},\ }\href@noop {} {\emph {\bibinfo {title} {Lectures on the
  Many-body Problem}}}\ (\bibinfo  {publisher} {Academic Press},\ \bibinfo
  {year} {1964})\BibitemShut {NoStop}%
\bibitem [{\citenamefont {Schneider}, \citenamefont {Donnelly},\ and\
  \citenamefont {Russek}(2018)}]{Schneider_Donnelly_Russek_2018}%
  \BibitemOpen
  \bibfield  {author} {\bibinfo {author} {\bibfnamefont {M.~L.}\ \bibnamefont
  {Schneider}}, \bibinfo {author} {\bibfnamefont {C.~A.}\ \bibnamefont
  {Donnelly}},\ and\ \bibinfo {author} {\bibfnamefont {S.~E.}\ \bibnamefont
  {Russek}},\ }\bibfield  {title} {\enquote {\bibinfo {title} {Tutorial:
  High-speed low-power neuromorphic systems based on magnetic {J}osephson
  junctions},}\ }\href {https://doi.org/10.1063/1.5042425} {\bibfield
  {journal} {\bibinfo  {journal} {Journal of Applied Physics}\ }\textbf
  {\bibinfo {volume} {124}},\ \bibinfo {pages} {161102} (\bibinfo {year}
  {2018})}\BibitemShut {NoStop}%
\bibitem [{\citenamefont {Crotty}, \citenamefont {Schult},\ and\ \citenamefont
  {Segall}(2010)}]{Crotty_Schult_Segall_2010}%
  \BibitemOpen
  \bibfield  {author} {\bibinfo {author} {\bibfnamefont {P.}~\bibnamefont
  {Crotty}}, \bibinfo {author} {\bibfnamefont {D.}~\bibnamefont {Schult}},\
  and\ \bibinfo {author} {\bibfnamefont {K.}~\bibnamefont {Segall}},\
  }\bibfield  {title} {\enquote {\bibinfo {title} {Josephson junction
  simulation of neurons},}\ }\href {https://doi.org/10.1103/PhysRevE.82.011914}
  {\bibfield  {journal} {\bibinfo  {journal} {Physical Review E}\ }\textbf
  {\bibinfo {volume} {82}},\ \bibinfo {pages} {011914} (\bibinfo {year}
  {2010})}\BibitemShut {NoStop}%
\bibitem [{\citenamefont {Verba}, \citenamefont {Tiberkevich},\ and\
  \citenamefont {Slavin}(2018)}]{Verba_Tiberkevich_Slavin_2018}%
  \BibitemOpen
  \bibfield  {author} {\bibinfo {author} {\bibfnamefont {R.}~\bibnamefont
  {Verba}}, \bibinfo {author} {\bibfnamefont {V.}~\bibnamefont {Tiberkevich}},\
  and\ \bibinfo {author} {\bibfnamefont {A.}~\bibnamefont {Slavin}},\
  }\bibfield  {title} {\enquote {\bibinfo {title} {Damping of linear spin-wave
  modes in magnetic nanostructures: Local, nonlocal, and coordinate-dependent
  damping},}\ }\href {https://doi.org/10.1103/PhysRevB.98.104408} {\bibfield
  {journal} {\bibinfo  {journal} {Physical Review B}\ }\textbf {\bibinfo
  {volume} {98}},\ \bibinfo {pages} {104408} (\bibinfo {year}
  {2018})}\BibitemShut {NoStop}%
\bibitem [{\citenamefont {Frenkel}\ \emph {et~al.}(2018)\citenamefont
  {Frenkel}, \citenamefont {Lefebvre}, \citenamefont {Legat},\ and\
  \citenamefont {Bol}}]{Frenkel_Lefebvre_Legat_Bol_2018}%
  \BibitemOpen
  \bibfield  {author} {\bibinfo {author} {\bibfnamefont {C.}~\bibnamefont
  {Frenkel}}, \bibinfo {author} {\bibfnamefont {M.}~\bibnamefont {Lefebvre}},
  \bibinfo {author} {\bibfnamefont {J.-D.}\ \bibnamefont {Legat}},\ and\
  \bibinfo {author} {\bibfnamefont {D.}~\bibnamefont {Bol}},\ }\bibfield
  {title} {\enquote {\bibinfo {title} {A 0.086-mm $^2$ 12.7-{pJ/SOP}
  64k-synapse 256-neuron online-learning digital spiking neuromorphic processor
  in 28nm {CMOS}},}\ }\href {https://doi.org/10.1109/TBCAS.2018.2880425}
  {\bibfield  {journal} {\bibinfo  {journal} {IEEE Transactions on Biomedical
  Circuits and Systems}\ ,\ \bibinfo {pages} {1--1}} (\bibinfo {year}
  {2018})}\BibitemShut {NoStop}%
\bibitem [{\citenamefont {Khang}, \citenamefont {Ueda},\ and\ \citenamefont
  {Hai}(2018)}]{khang2018conductive}%
  \BibitemOpen
  \bibfield  {author} {\bibinfo {author} {\bibfnamefont {N.~H.~D.}\
  \bibnamefont {Khang}}, \bibinfo {author} {\bibfnamefont {Y.}~\bibnamefont
  {Ueda}},\ and\ \bibinfo {author} {\bibfnamefont {P.~N.}\ \bibnamefont
  {Hai}},\ }\bibfield  {title} {\enquote {\bibinfo {title} {A conductive
  topological insulator with large spin {H}all effect for ultralow power
  spin--orbit torque switching},}\ }\href@noop {} {\bibfield  {journal}
  {\bibinfo  {journal} {Nature Materials}\ }\textbf {\bibinfo {volume} {17}},\
  \bibinfo {pages} {808--813} (\bibinfo {year} {2018})}\BibitemShut {NoStop}%
\bibitem [{\citenamefont {Louis}\ \emph
  {et~al.}(2022{\natexlab{a}})\citenamefont {Louis}, \citenamefont {Bradley},
  \citenamefont {Slavin},\ and\ \citenamefont
  {Tyberkevych}}]{louis2022Magnonics}%
  \BibitemOpen
  \bibfield  {author} {\bibinfo {author} {\bibfnamefont {S.}~\bibnamefont
  {Louis}}, \bibinfo {author} {\bibfnamefont {H.}~\bibnamefont {Bradley}},
  \bibinfo {author} {\bibfnamefont {A.}~\bibnamefont {Slavin}},\ and\ \bibinfo
  {author} {\bibfnamefont {V.}~\bibnamefont {Tyberkevych}},\ }\bibfield
  {title} {\enquote {\bibinfo {title} {Artificial neuron based on a spin torque
  nano oscillator},}\ }\href@noop {} {\bibfield  {journal} {\bibinfo  {journal}
  {Book of Abstracts for 7th International Conference on Magnonics}\ ,\
  \bibinfo {pages} {C4--28}} (\bibinfo {year}
  {2022}{\natexlab{a}})}\BibitemShut {NoStop}%
\bibitem [{\citenamefont {Louis}\ \emph
  {et~al.}(2022{\natexlab{b}})\citenamefont {Louis}, \citenamefont {Bradley},
  \citenamefont {Slavin},\ and\ \citenamefont {Tyberkevych}}]{louis2022}%
  \BibitemOpen
  \bibfield  {author} {\bibinfo {author} {\bibfnamefont {S.}~\bibnamefont
  {Louis}}, \bibinfo {author} {\bibfnamefont {H.}~\bibnamefont {Bradley}},
  \bibinfo {author} {\bibfnamefont {A.}~\bibnamefont {Slavin}},\ and\ \bibinfo
  {author} {\bibfnamefont {V.}~\bibnamefont {Tyberkevych}},\ }\bibfield
  {title} {\enquote {\bibinfo {title} {Artificial neuron designed using a spin
  {Josephson} oscillator based on a synthetic antiferromagnet},}\ }\href@noop
  {} {\bibfield  {journal} {\bibinfo  {journal} {Abstracts of Joint
  MMM-Intermag 2022}\ ,\ \bibinfo {pages} {GPD--03}} (\bibinfo {year}
  {2022}{\natexlab{b}})}\BibitemShut {NoStop}%
\bibitem [{\citenamefont {Liu}\ \emph {et~al.}(2020)\citenamefont {Liu},
  \citenamefont {Barsukov}, \citenamefont {Barlas}, \citenamefont
  {Krivorotov},\ and\ \citenamefont {Lake}}]{liu2020synthetic}%
  \BibitemOpen
  \bibfield  {author} {\bibinfo {author} {\bibfnamefont {Y.}~\bibnamefont
  {Liu}}, \bibinfo {author} {\bibfnamefont {I.}~\bibnamefont {Barsukov}},
  \bibinfo {author} {\bibfnamefont {Y.}~\bibnamefont {Barlas}}, \bibinfo
  {author} {\bibfnamefont {I.~N.}\ \bibnamefont {Krivorotov}},\ and\ \bibinfo
  {author} {\bibfnamefont {R.~K.}\ \bibnamefont {Lake}},\ }\bibfield  {title}
  {\enquote {\bibinfo {title} {Synthetic antiferromagnet-based spin {Josephson}
  oscillator},}\ }\href@noop {} {\bibfield  {journal} {\bibinfo  {journal}
  {Applied Physics Letters}\ }\textbf {\bibinfo {volume} {116}},\ \bibinfo
  {pages} {132409} (\bibinfo {year} {2020})}\BibitemShut {NoStop}%
\bibitem [{\citenamefont {Sulymenko}\ \emph {et~al.}(2017)\citenamefont
  {Sulymenko}, \citenamefont {Prokopenko}, \citenamefont {Tiberkevich},
  \citenamefont {Slavin}, \citenamefont {Ivanov},\ and\ \citenamefont
  {Khymyn}}]{Sulymenko_Prokopenko_Tiberkevich_Slavin_Ivanov_Khymyn_2017}%
  \BibitemOpen
  \bibfield  {author} {\bibinfo {author} {\bibfnamefont {O.}~\bibnamefont
  {Sulymenko}}, \bibinfo {author} {\bibfnamefont {O.}~\bibnamefont
  {Prokopenko}}, \bibinfo {author} {\bibfnamefont {V.}~\bibnamefont
  {Tiberkevich}}, \bibinfo {author} {\bibfnamefont {A.}~\bibnamefont {Slavin}},
  \bibinfo {author} {\bibfnamefont {B.}~\bibnamefont {Ivanov}},\ and\ \bibinfo
  {author} {\bibfnamefont {R.}~\bibnamefont {Khymyn}},\ }\bibfield  {title}
  {\enquote {\bibinfo {title} {Terahertz-frequency spin {H}all auto-oscillator
  based on a canted antiferromagnet},}\ }\href
  {https://doi.org/10.1103/PhysRevApplied.8.064007} {\bibfield  {journal}
  {\bibinfo  {journal} {Physical Review Applied}\ }\textbf {\bibinfo {volume}
  {8}} (\bibinfo {year} {2017}),\ 10.1103/PhysRevApplied.8.064007}\BibitemShut
  {NoStop}%
\bibitem [{\citenamefont {Sulymenko}\ \emph
  {et~al.}(2018{\natexlab{b}})\citenamefont {Sulymenko}, \citenamefont
  {Prokopenko}, \citenamefont {Tiberkevich},\ and\ \citenamefont
  {Slavin}}]{Sulymenko_Prokopenko_Tiberkevich_Slavin_2018}%
  \BibitemOpen
  \bibfield  {author} {\bibinfo {author} {\bibfnamefont {O.}~\bibnamefont
  {Sulymenko}}, \bibinfo {author} {\bibfnamefont {O.~V.}\ \bibnamefont
  {Prokopenko}}, \bibinfo {author} {\bibfnamefont {V.}~\bibnamefont
  {Tiberkevich}},\ and\ \bibinfo {author} {\bibfnamefont {A.~N.}\ \bibnamefont
  {Slavin}},\ }\bibfield  {title} {\enquote {\bibinfo {title}
  {Terahertz-frequency signal source based on an antiferromagnetic tunnel
  junction},}\ }\href {https://doi.org/10.1109/LMAG.2018.2852291} {\bibfield
  {journal} {\bibinfo  {journal} {IEEE Magnetics Letters}\ ,\ \bibinfo {pages}
  {1--1}} (\bibinfo {year} {2018}{\natexlab{b}})}\BibitemShut {NoStop}%
\bibitem [{\citenamefont {Artemchuk}\ \emph {et~al.}(2020)\citenamefont
  {Artemchuk}, \citenamefont {Sulymenko}, \citenamefont {Louis}, \citenamefont
  {Li}, \citenamefont {Khymyn}, \citenamefont {Bankowski}, \citenamefont
  {Meitzler}, \citenamefont {Tyberkevych}, \citenamefont {Slavin},\ and\
  \citenamefont
  {Prokopenko}}]{Artemchuk_Sulymenko_Louis_Li_Khymyn_Bankowski_Meitzler_Tyberkevych_Slavin_Prokopenko_2020}%
  \BibitemOpen
  \bibfield  {author} {\bibinfo {author} {\bibfnamefont {P.~Y.}\ \bibnamefont
  {Artemchuk}}, \bibinfo {author} {\bibfnamefont {O.~R.}\ \bibnamefont
  {Sulymenko}}, \bibinfo {author} {\bibfnamefont {S.}~\bibnamefont {Louis}},
  \bibinfo {author} {\bibfnamefont {J.}~\bibnamefont {Li}}, \bibinfo {author}
  {\bibfnamefont {R.~S.}\ \bibnamefont {Khymyn}}, \bibinfo {author}
  {\bibfnamefont {E.}~\bibnamefont {Bankowski}}, \bibinfo {author}
  {\bibfnamefont {T.}~\bibnamefont {Meitzler}}, \bibinfo {author}
  {\bibfnamefont {V.~S.}\ \bibnamefont {Tyberkevych}}, \bibinfo {author}
  {\bibfnamefont {A.~N.}\ \bibnamefont {Slavin}},\ and\ \bibinfo {author}
  {\bibfnamefont {O.~V.}\ \bibnamefont {Prokopenko}},\ }\bibfield  {title}
  {\enquote {\bibinfo {title} {Terahertz frequency spectrum analysis with a
  nanoscale antiferromagnetic tunnel junction},}\ }\href
  {https://doi.org/10.1063/1.5140552} {\bibfield  {journal} {\bibinfo
  {journal} {Journal of Applied Physics}\ }\textbf {\bibinfo {volume} {127}},\
  \bibinfo {pages} {063905} (\bibinfo {year} {2020})}\BibitemShut {NoStop}%
\bibitem [{\citenamefont {Parthasarathy}\ \emph {et~al.}(2021)\citenamefont
  {Parthasarathy}, \citenamefont {Cogulu}, \citenamefont {Kent},\ and\
  \citenamefont {Rakheja}}]{parthasarathy2021precessional}%
  \BibitemOpen
  \bibfield  {author} {\bibinfo {author} {\bibfnamefont {A.}~\bibnamefont
  {Parthasarathy}}, \bibinfo {author} {\bibfnamefont {E.}~\bibnamefont
  {Cogulu}}, \bibinfo {author} {\bibfnamefont {A.~D.}\ \bibnamefont {Kent}},\
  and\ \bibinfo {author} {\bibfnamefont {S.}~\bibnamefont {Rakheja}},\
  }\bibfield  {title} {\enquote {\bibinfo {title} {Precessional spin-torque
  dynamics in biaxial antiferromagnets},}\ }\href@noop {} {\bibfield  {journal}
  {\bibinfo  {journal} {Physical Review B}\ }\textbf {\bibinfo {volume}
  {103}},\ \bibinfo {pages} {024450} (\bibinfo {year} {2021})}\BibitemShut
  {NoStop}%
\bibitem [{\citenamefont {Gomonay}, \citenamefont {Jungwirth},\ and\
  \citenamefont {Sinova}(2018)}]{gomonay2018narrow}%
  \BibitemOpen
  \bibfield  {author} {\bibinfo {author} {\bibfnamefont {O.}~\bibnamefont
  {Gomonay}}, \bibinfo {author} {\bibfnamefont {T.}~\bibnamefont {Jungwirth}},\
  and\ \bibinfo {author} {\bibfnamefont {J.}~\bibnamefont {Sinova}},\
  }\bibfield  {title} {\enquote {\bibinfo {title} {Narrow-band tunable
  terahertz detector in antiferromagnets via staggered-field and antidamping
  torques},}\ }\href@noop {} {\bibfield  {journal} {\bibinfo  {journal}
  {Physical Review B}\ }\textbf {\bibinfo {volume} {98}},\ \bibinfo {pages}
  {104430} (\bibinfo {year} {2018})}\BibitemShut {NoStop}%
\bibitem [{\citenamefont {Wadley}\ \emph {et~al.}(2016)\citenamefont {Wadley},
  \citenamefont {Howells}, \citenamefont {{\v{Z}}elezn{\`y}}, \citenamefont
  {Andrews}, \citenamefont {Hills}, \citenamefont {Campion}, \citenamefont
  {Nov{\'a}k}, \citenamefont {Olejn{\'\i}k}, \citenamefont {Maccherozzi},
  \citenamefont {Dhesi} \emph {et~al.}}]{wadley2016electrical}%
  \BibitemOpen
  \bibfield  {author} {\bibinfo {author} {\bibfnamefont {P.}~\bibnamefont
  {Wadley}}, \bibinfo {author} {\bibfnamefont {B.}~\bibnamefont {Howells}},
  \bibinfo {author} {\bibfnamefont {J.}~\bibnamefont {{\v{Z}}elezn{\`y}}},
  \bibinfo {author} {\bibfnamefont {C.}~\bibnamefont {Andrews}}, \bibinfo
  {author} {\bibfnamefont {V.}~\bibnamefont {Hills}}, \bibinfo {author}
  {\bibfnamefont {R.~P.}\ \bibnamefont {Campion}}, \bibinfo {author}
  {\bibfnamefont {V.}~\bibnamefont {Nov{\'a}k}}, \bibinfo {author}
  {\bibfnamefont {K.}~\bibnamefont {Olejn{\'\i}k}}, \bibinfo {author}
  {\bibfnamefont {F.}~\bibnamefont {Maccherozzi}}, \bibinfo {author}
  {\bibfnamefont {S.}~\bibnamefont {Dhesi}}, \emph {et~al.},\ }\bibfield
  {title} {\enquote {\bibinfo {title} {Electrical switching of an
  antiferromagnet},}\ }\href@noop {} {\bibfield  {journal} {\bibinfo  {journal}
  {Science}\ }\textbf {\bibinfo {volume} {351}},\ \bibinfo {pages} {587--590}
  (\bibinfo {year} {2016})}\BibitemShut {NoStop}%
\bibitem [{\citenamefont {Vaidya}\ \emph {et~al.}(2020)\citenamefont {Vaidya},
  \citenamefont {Morley}, \citenamefont {van Tol}, \citenamefont {Liu},
  \citenamefont {Cheng}, \citenamefont {Brataas}, \citenamefont {Lederman},\
  and\ \citenamefont {Del~Barco}}]{vaidya2020subterahertz}%
  \BibitemOpen
  \bibfield  {author} {\bibinfo {author} {\bibfnamefont {P.}~\bibnamefont
  {Vaidya}}, \bibinfo {author} {\bibfnamefont {S.~A.}\ \bibnamefont {Morley}},
  \bibinfo {author} {\bibfnamefont {J.}~\bibnamefont {van Tol}}, \bibinfo
  {author} {\bibfnamefont {Y.}~\bibnamefont {Liu}}, \bibinfo {author}
  {\bibfnamefont {R.}~\bibnamefont {Cheng}}, \bibinfo {author} {\bibfnamefont
  {A.}~\bibnamefont {Brataas}}, \bibinfo {author} {\bibfnamefont
  {D.}~\bibnamefont {Lederman}},\ and\ \bibinfo {author} {\bibfnamefont
  {E.}~\bibnamefont {Del~Barco}},\ }\bibfield  {title} {\enquote {\bibinfo
  {title} {Subterahertz spin pumping from an insulating antiferromagnet},}\
  }\href@noop {} {\bibfield  {journal} {\bibinfo  {journal} {Science}\ }\textbf
  {\bibinfo {volume} {368}},\ \bibinfo {pages} {160--165} (\bibinfo {year}
  {2020})}\BibitemShut {NoStop}%
\bibitem [{\citenamefont {Boventer}\ \emph {et~al.}(2021)\citenamefont
  {Boventer}, \citenamefont {Simensen}, \citenamefont {Anane}, \citenamefont
  {Kl{\"a}ui}, \citenamefont {Brataas},\ and\ \citenamefont
  {Lebrun}}]{boventer2021room}%
  \BibitemOpen
  \bibfield  {author} {\bibinfo {author} {\bibfnamefont {I.}~\bibnamefont
  {Boventer}}, \bibinfo {author} {\bibfnamefont {H.~T.}\ \bibnamefont
  {Simensen}}, \bibinfo {author} {\bibfnamefont {A.}~\bibnamefont {Anane}},
  \bibinfo {author} {\bibfnamefont {M.}~\bibnamefont {Kl{\"a}ui}}, \bibinfo
  {author} {\bibfnamefont {A.}~\bibnamefont {Brataas}},\ and\ \bibinfo {author}
  {\bibfnamefont {R.}~\bibnamefont {Lebrun}},\ }\bibfield  {title} {\enquote
  {\bibinfo {title} {Room-temperature antiferromagnetic resonance and inverse
  spin-{H}all voltage in canted antiferromagnets},}\ }\href@noop {} {\bibfield
  {journal} {\bibinfo  {journal} {Physical Review Letters}\ }\textbf {\bibinfo
  {volume} {126}},\ \bibinfo {pages} {187201} (\bibinfo {year}
  {2021})}\BibitemShut {NoStop}%
\bibitem [{\citenamefont {Puliafito}\ \emph {et~al.}(2019)\citenamefont
  {Puliafito}, \citenamefont {Khymyn}, \citenamefont {Carpentieri},
  \citenamefont {Azzerboni}, \citenamefont {Tiberkevich}, \citenamefont
  {Slavin},\ and\ \citenamefont {Finocchio}}]{puliafito2019micromagnetic}%
  \BibitemOpen
  \bibfield  {author} {\bibinfo {author} {\bibfnamefont {V.}~\bibnamefont
  {Puliafito}}, \bibinfo {author} {\bibfnamefont {R.}~\bibnamefont {Khymyn}},
  \bibinfo {author} {\bibfnamefont {M.}~\bibnamefont {Carpentieri}}, \bibinfo
  {author} {\bibfnamefont {B.}~\bibnamefont {Azzerboni}}, \bibinfo {author}
  {\bibfnamefont {V.}~\bibnamefont {Tiberkevich}}, \bibinfo {author}
  {\bibfnamefont {A.}~\bibnamefont {Slavin}},\ and\ \bibinfo {author}
  {\bibfnamefont {G.}~\bibnamefont {Finocchio}},\ }\bibfield  {title} {\enquote
  {\bibinfo {title} {Micromagnetic modeling of terahertz oscillations in an
  antiferromagnetic material driven by the spin {H}all effect},}\ }\href@noop
  {} {\bibfield  {journal} {\bibinfo  {journal} {Physical Review B}\ }\textbf
  {\bibinfo {volume} {99}},\ \bibinfo {pages} {024405} (\bibinfo {year}
  {2019})}\BibitemShut {NoStop}%
\bibitem [{\citenamefont {Shi}\ \emph {et~al.}(2020)\citenamefont {Shi},
  \citenamefont {Lopez-Dominguez}, \citenamefont {Garesci}, \citenamefont
  {Wang}, \citenamefont {Almasi}, \citenamefont {Grayson}, \citenamefont
  {Finocchio},\ and\ \citenamefont {Khalili~Amiri}}]{shi2020electrical}%
  \BibitemOpen
  \bibfield  {author} {\bibinfo {author} {\bibfnamefont {J.}~\bibnamefont
  {Shi}}, \bibinfo {author} {\bibfnamefont {V.}~\bibnamefont
  {Lopez-Dominguez}}, \bibinfo {author} {\bibfnamefont {F.}~\bibnamefont
  {Garesci}}, \bibinfo {author} {\bibfnamefont {C.}~\bibnamefont {Wang}},
  \bibinfo {author} {\bibfnamefont {H.}~\bibnamefont {Almasi}}, \bibinfo
  {author} {\bibfnamefont {M.}~\bibnamefont {Grayson}}, \bibinfo {author}
  {\bibfnamefont {G.}~\bibnamefont {Finocchio}},\ and\ \bibinfo {author}
  {\bibfnamefont {P.}~\bibnamefont {Khalili~Amiri}},\ }\bibfield  {title}
  {\enquote {\bibinfo {title} {Electrical manipulation of the magnetic order in
  antiferromagnetic {PtMn} pillars},}\ }\href@noop {} {\bibfield  {journal}
  {\bibinfo  {journal} {Nature Electronics}\ }\textbf {\bibinfo {volume} {3}},\
  \bibinfo {pages} {92--98} (\bibinfo {year} {2020})}\BibitemShut {NoStop}%
\bibitem [{\citenamefont {Ni}\ \emph {et~al.}(2021)\citenamefont {Ni},
  \citenamefont {Jin}, \citenamefont {Song}, \citenamefont {Zhou},
  \citenamefont {Chen}, \citenamefont {Song}, \citenamefont {Peng},
  \citenamefont {Zhang}, \citenamefont {Pan}, \citenamefont {Ma} \emph
  {et~al.}}]{ni2021temperature}%
  \BibitemOpen
  \bibfield  {author} {\bibinfo {author} {\bibfnamefont {Y.}~\bibnamefont
  {Ni}}, \bibinfo {author} {\bibfnamefont {Z.}~\bibnamefont {Jin}}, \bibinfo
  {author} {\bibfnamefont {B.}~\bibnamefont {Song}}, \bibinfo {author}
  {\bibfnamefont {X.}~\bibnamefont {Zhou}}, \bibinfo {author} {\bibfnamefont
  {H.}~\bibnamefont {Chen}}, \bibinfo {author} {\bibfnamefont {C.}~\bibnamefont
  {Song}}, \bibinfo {author} {\bibfnamefont {Y.}~\bibnamefont {Peng}}, \bibinfo
  {author} {\bibfnamefont {C.}~\bibnamefont {Zhang}}, \bibinfo {author}
  {\bibfnamefont {F.}~\bibnamefont {Pan}}, \bibinfo {author} {\bibfnamefont
  {G.}~\bibnamefont {Ma}}, \emph {et~al.},\ }\bibfield  {title} {\enquote
  {\bibinfo {title} {Temperature-dependent terahertz emission from {Co/Mn2Au}
  spintronic bilayers},}\ }\href@noop {} {\bibfield  {journal} {\bibinfo
  {journal} {Physica Status Solidi (RRL)--Rapid Research Letters}\ }\textbf
  {\bibinfo {volume} {15}},\ \bibinfo {pages} {2100290} (\bibinfo {year}
  {2021})}\BibitemShut {NoStop}%
\bibitem [{\citenamefont {Baldrati}\ \emph {et~al.}(2019)\citenamefont
  {Baldrati}, \citenamefont {Gomonay}, \citenamefont {Ross}, \citenamefont
  {Filianina}, \citenamefont {Lebrun}, \citenamefont {Ramos}, \citenamefont
  {Leveille}, \citenamefont {Fuhrmann}, \citenamefont {Forrest}, \citenamefont
  {Maccherozzi} \emph {et~al.}}]{baldrati2019mechanism}%
  \BibitemOpen
  \bibfield  {author} {\bibinfo {author} {\bibfnamefont {L.}~\bibnamefont
  {Baldrati}}, \bibinfo {author} {\bibfnamefont {O.}~\bibnamefont {Gomonay}},
  \bibinfo {author} {\bibfnamefont {A.}~\bibnamefont {Ross}}, \bibinfo {author}
  {\bibfnamefont {M.}~\bibnamefont {Filianina}}, \bibinfo {author}
  {\bibfnamefont {R.}~\bibnamefont {Lebrun}}, \bibinfo {author} {\bibfnamefont
  {R.}~\bibnamefont {Ramos}}, \bibinfo {author} {\bibfnamefont
  {C.}~\bibnamefont {Leveille}}, \bibinfo {author} {\bibfnamefont
  {F.}~\bibnamefont {Fuhrmann}}, \bibinfo {author} {\bibfnamefont
  {T.}~\bibnamefont {Forrest}}, \bibinfo {author} {\bibfnamefont
  {F.}~\bibnamefont {Maccherozzi}}, \emph {et~al.},\ }\bibfield  {title}
  {\enquote {\bibinfo {title} {Mechanism of {N}{\'e}el order switching in
  antiferromagnetic thin films revealed by magnetotransport and direct
  imaging},}\ }\href@noop {} {\bibfield  {journal} {\bibinfo  {journal}
  {Physical Review Letters}\ }\textbf {\bibinfo {volume} {123}},\ \bibinfo
  {pages} {177201} (\bibinfo {year} {2019})}\BibitemShut {NoStop}%
\bibitem [{\citenamefont {Chen}\ \emph {et~al.}(2019)\citenamefont {Chen},
  \citenamefont {Zhou}, \citenamefont {Cheng}, \citenamefont {Song},
  \citenamefont {Zhang}, \citenamefont {Wu}, \citenamefont {Ba}, \citenamefont
  {Li}, \citenamefont {Sun}, \citenamefont {You} \emph
  {et~al.}}]{chen2019electric}%
  \BibitemOpen
  \bibfield  {author} {\bibinfo {author} {\bibfnamefont {X.}~\bibnamefont
  {Chen}}, \bibinfo {author} {\bibfnamefont {X.}~\bibnamefont {Zhou}}, \bibinfo
  {author} {\bibfnamefont {R.}~\bibnamefont {Cheng}}, \bibinfo {author}
  {\bibfnamefont {C.}~\bibnamefont {Song}}, \bibinfo {author} {\bibfnamefont
  {J.}~\bibnamefont {Zhang}}, \bibinfo {author} {\bibfnamefont
  {Y.}~\bibnamefont {Wu}}, \bibinfo {author} {\bibfnamefont {Y.}~\bibnamefont
  {Ba}}, \bibinfo {author} {\bibfnamefont {H.}~\bibnamefont {Li}}, \bibinfo
  {author} {\bibfnamefont {Y.}~\bibnamefont {Sun}}, \bibinfo {author}
  {\bibfnamefont {Y.}~\bibnamefont {You}}, \emph {et~al.},\ }\bibfield  {title}
  {\enquote {\bibinfo {title} {Electric field control of {N}{\'e}el spin--orbit
  torque in an antiferromagnet},}\ }\href@noop {} {\bibfield  {journal}
  {\bibinfo  {journal} {Nature Materials}\ }\textbf {\bibinfo {volume} {18}},\
  \bibinfo {pages} {931--935} (\bibinfo {year} {2019})}\BibitemShut {NoStop}%
\bibitem [{\citenamefont {Gray}\ \emph {et~al.}(2019)\citenamefont {Gray},
  \citenamefont {Moriyama}, \citenamefont {Sivadas}, \citenamefont {Stiehl},
  \citenamefont {Heron}, \citenamefont {Need}, \citenamefont {Kirby},
  \citenamefont {Low}, \citenamefont {Nowack}, \citenamefont {Schlom} \emph
  {et~al.}}]{gray2019spin}%
  \BibitemOpen
  \bibfield  {author} {\bibinfo {author} {\bibfnamefont {I.}~\bibnamefont
  {Gray}}, \bibinfo {author} {\bibfnamefont {T.}~\bibnamefont {Moriyama}},
  \bibinfo {author} {\bibfnamefont {N.}~\bibnamefont {Sivadas}}, \bibinfo
  {author} {\bibfnamefont {G.~M.}\ \bibnamefont {Stiehl}}, \bibinfo {author}
  {\bibfnamefont {J.~T.}\ \bibnamefont {Heron}}, \bibinfo {author}
  {\bibfnamefont {R.}~\bibnamefont {Need}}, \bibinfo {author} {\bibfnamefont
  {B.~J.}\ \bibnamefont {Kirby}}, \bibinfo {author} {\bibfnamefont {D.~H.}\
  \bibnamefont {Low}}, \bibinfo {author} {\bibfnamefont {K.~C.}\ \bibnamefont
  {Nowack}}, \bibinfo {author} {\bibfnamefont {D.~G.}\ \bibnamefont {Schlom}},
  \emph {et~al.},\ }\bibfield  {title} {\enquote {\bibinfo {title} {Spin
  {S}eebeck imaging of spin-torque switching in antiferromagnetic {Pt/NiO}
  heterostructures},}\ }\href@noop {} {\bibfield  {journal} {\bibinfo
  {journal} {Physical Review X}\ }\textbf {\bibinfo {volume} {9}},\ \bibinfo
  {pages} {041016} (\bibinfo {year} {2019})}\BibitemShut {NoStop}%
\bibitem [{\citenamefont {Squire}\ \emph {et~al.}(2012)\citenamefont {Squire},
  \citenamefont {Berg}, \citenamefont {Bloom}, \citenamefont {Du~Lac},
  \citenamefont {Ghosh},\ and\ \citenamefont
  {Spitzer}}]{squire2012fundamental}%
  \BibitemOpen
  \bibfield  {author} {\bibinfo {author} {\bibfnamefont {L.}~\bibnamefont
  {Squire}}, \bibinfo {author} {\bibfnamefont {D.}~\bibnamefont {Berg}},
  \bibinfo {author} {\bibfnamefont {F.~E.}\ \bibnamefont {Bloom}}, \bibinfo
  {author} {\bibfnamefont {S.}~\bibnamefont {Du~Lac}}, \bibinfo {author}
  {\bibfnamefont {A.}~\bibnamefont {Ghosh}},\ and\ \bibinfo {author}
  {\bibfnamefont {N.~C.}\ \bibnamefont {Spitzer}},\ }\href@noop {} {\emph
  {\bibinfo {title} {Fundamental Neuroscience}}}\ (\bibinfo  {publisher}
  {Academic press},\ \bibinfo {year} {2012})\BibitemShut {NoStop}%
\bibitem [{\citenamefont {Levakova}\ \emph {et~al.}(2015)\citenamefont
  {Levakova}, \citenamefont {Tamborrino}, \citenamefont {Ditlevsen},\ and\
  \citenamefont {Lansky}}]{levakova2015review}%
  \BibitemOpen
  \bibfield  {author} {\bibinfo {author} {\bibfnamefont {M.}~\bibnamefont
  {Levakova}}, \bibinfo {author} {\bibfnamefont {M.}~\bibnamefont
  {Tamborrino}}, \bibinfo {author} {\bibfnamefont {S.}~\bibnamefont
  {Ditlevsen}},\ and\ \bibinfo {author} {\bibfnamefont {P.}~\bibnamefont
  {Lansky}},\ }\bibfield  {title} {\enquote {\bibinfo {title} {A review of the
  methods for neuronal response latency estimation},}\ }\href@noop {}
  {\bibfield  {journal} {\bibinfo  {journal} {Biosystems}\ }\textbf {\bibinfo
  {volume} {136}},\ \bibinfo {pages} {23--34} (\bibinfo {year}
  {2015})}\BibitemShut {NoStop}%
\bibitem [{\citenamefont {Overton}\ and\ \citenamefont
  {Clark}(1997)}]{overton1997burst}%
  \BibitemOpen
  \bibfield  {author} {\bibinfo {author} {\bibfnamefont {P.}~\bibnamefont
  {Overton}}\ and\ \bibinfo {author} {\bibfnamefont {D.}~\bibnamefont
  {Clark}},\ }\bibfield  {title} {\enquote {\bibinfo {title} {Burst firing in
  midbrain dopaminergic neurons},}\ }\href@noop {} {\bibfield  {journal}
  {\bibinfo  {journal} {Brain Research Reviews}\ }\textbf {\bibinfo {volume}
  {25}},\ \bibinfo {pages} {312--334} (\bibinfo {year} {1997})}\BibitemShut
  {NoStop}%
\bibitem [{\citenamefont {Krahe}\ and\ \citenamefont
  {Gabbiani}(2004)}]{krahe2004burst}%
  \BibitemOpen
  \bibfield  {author} {\bibinfo {author} {\bibfnamefont {R.}~\bibnamefont
  {Krahe}}\ and\ \bibinfo {author} {\bibfnamefont {F.}~\bibnamefont
  {Gabbiani}},\ }\bibfield  {title} {\enquote {\bibinfo {title} {Burst firing
  in sensory systems},}\ }\href@noop {} {\bibfield  {journal} {\bibinfo
  {journal} {Nature Reviews Neuroscience}\ }\textbf {\bibinfo {volume} {5}},\
  \bibinfo {pages} {13--23} (\bibinfo {year} {2004})}\BibitemShut {NoStop}%
\bibitem [{\citenamefont {Izhikevich}(2004)}]{izhikevich2004model}%
  \BibitemOpen
  \bibfield  {author} {\bibinfo {author} {\bibfnamefont {E.~M.}\ \bibnamefont
  {Izhikevich}},\ }\bibfield  {title} {\enquote {\bibinfo {title} {Which model
  to use for cortical spiking neurons?}}\ }\href@noop {} {\bibfield  {journal}
  {\bibinfo  {journal} {IEEE Transactions on Neural Networks}\ }\textbf
  {\bibinfo {volume} {15}},\ \bibinfo {pages} {1063--1070} (\bibinfo {year}
  {2004})}\BibitemShut {NoStop}%
\bibitem [{\citenamefont {Gerstner}\ \emph {et~al.}(2014)\citenamefont
  {Gerstner}, \citenamefont {Kistler}, \citenamefont {Naud},\ and\
  \citenamefont {Paninski}}]{gerstner2014neuronal}%
  \BibitemOpen
  \bibfield  {author} {\bibinfo {author} {\bibfnamefont {W.}~\bibnamefont
  {Gerstner}}, \bibinfo {author} {\bibfnamefont {W.~M.}\ \bibnamefont
  {Kistler}}, \bibinfo {author} {\bibfnamefont {R.}~\bibnamefont {Naud}},\ and\
  \bibinfo {author} {\bibfnamefont {L.}~\bibnamefont {Paninski}},\ }\href@noop
  {} {\emph {\bibinfo {title} {Neuronal Dynamics: From Single Neurons to
  Networks and Models of Cognition}}}\ (\bibinfo  {publisher} {Cambridge
  University Press},\ \bibinfo {year} {2014})\BibitemShut {NoStop}%
\bibitem [{\citenamefont {Aoyagi}\ \emph {et~al.}(2001)\citenamefont {Aoyagi},
  \citenamefont {Terada}, \citenamefont {Kang}, \citenamefont {Kaneko},\ and\
  \citenamefont {Fukai}}]{aoyagi2001bursting}%
  \BibitemOpen
  \bibfield  {author} {\bibinfo {author} {\bibfnamefont {T.}~\bibnamefont
  {Aoyagi}}, \bibinfo {author} {\bibfnamefont {N.}~\bibnamefont {Terada}},
  \bibinfo {author} {\bibfnamefont {Y.}~\bibnamefont {Kang}}, \bibinfo {author}
  {\bibfnamefont {T.}~\bibnamefont {Kaneko}},\ and\ \bibinfo {author}
  {\bibfnamefont {T.}~\bibnamefont {Fukai}},\ }\bibfield  {title} {\enquote
  {\bibinfo {title} {A bursting mechanism of chattering neurons based on
  {Ca}$^{2+}$-dependent cationic currents},}\ }\href@noop {} {\bibfield
  {journal} {\bibinfo  {journal} {Neurocomputing}\ }\textbf {\bibinfo {volume}
  {38}},\ \bibinfo {pages} {93--98} (\bibinfo {year} {2001})}\BibitemShut
  {NoStop}%
\bibitem [{\citenamefont {Silverthorn}(2015)}]{silverthorn2015human}%
  \BibitemOpen
  \bibfield  {author} {\bibinfo {author} {\bibfnamefont {D.~U.}\ \bibnamefont
  {Silverthorn}},\ }\href@noop {} {\emph {\bibinfo {title} {Human
  Physiology}}}\ (\bibinfo  {publisher} {Jones \& Bartlett Publishers},\
  \bibinfo {year} {2015})\BibitemShut {NoStop}%
\bibitem [{\citenamefont {Purves}\ \emph {et~al.}(2001)\citenamefont {Purves},
  \citenamefont {Augustine}, \citenamefont {Fitzpatrick}, \citenamefont {Katz},
  \citenamefont {LaMantia}, \citenamefont {McNamara},\ and\ \citenamefont
  {Williams}}]{Purves_Williams_2001}%
  \BibitemOpen
  \bibfield  {author} {\bibinfo {author} {\bibfnamefont {D.}~\bibnamefont
  {Purves}}, \bibinfo {author} {\bibfnamefont {G.~J.}\ \bibnamefont
  {Augustine}}, \bibinfo {author} {\bibfnamefont {D.}~\bibnamefont
  {Fitzpatrick}}, \bibinfo {author} {\bibfnamefont {L.~C.}\ \bibnamefont
  {Katz}}, \bibinfo {author} {\bibfnamefont {A.-S.}\ \bibnamefont {LaMantia}},
  \bibinfo {author} {\bibfnamefont {J.~O.}\ \bibnamefont {McNamara}},\ and\
  \bibinfo {author} {\bibfnamefont {S.~M.}\ \bibnamefont {Williams}},\
  }\href@noop {} {\emph {\bibinfo {title} {Neuroscience}}},\ \bibinfo {edition}
  {2nd}\ ed.\ (\bibinfo  {publisher} {Sinauer Associates},\ \bibinfo {address}
  {Sunderland, Mass},\ \bibinfo {year} {2001})\BibitemShut {NoStop}%
\bibitem [{\citenamefont {Tocci}, \citenamefont {Widmer},\ and\ \citenamefont
  {Moss}(2007)}]{Tocci_Widmer_Moss_2007}%
  \BibitemOpen
  \bibfield  {author} {\bibinfo {author} {\bibfnamefont {R.~J.}\ \bibnamefont
  {Tocci}}, \bibinfo {author} {\bibfnamefont {N.~S.}\ \bibnamefont {Widmer}},\
  and\ \bibinfo {author} {\bibfnamefont {G.~L.}\ \bibnamefont {Moss}},\
  }\href@noop {} {\emph {\bibinfo {title} {Digital Systems: Principles and
  Applications}}},\ \bibinfo {edition} {10th}\ ed.\ (\bibinfo  {publisher}
  {Pearson/Prentice Hall},\ \bibinfo {year} {2007})\BibitemShut {NoStop}%
\bibitem [{\citenamefont {Jo}\ \emph {et~al.}(2010)\citenamefont {Jo},
  \citenamefont {Chang}, \citenamefont {Ebong}, \citenamefont {Bhadviya},
  \citenamefont {Mazumder},\ and\ \citenamefont {Lu}}]{jo2010nanoscale}%
  \BibitemOpen
  \bibfield  {author} {\bibinfo {author} {\bibfnamefont {S.~H.}\ \bibnamefont
  {Jo}}, \bibinfo {author} {\bibfnamefont {T.}~\bibnamefont {Chang}}, \bibinfo
  {author} {\bibfnamefont {I.}~\bibnamefont {Ebong}}, \bibinfo {author}
  {\bibfnamefont {B.~B.}\ \bibnamefont {Bhadviya}}, \bibinfo {author}
  {\bibfnamefont {P.}~\bibnamefont {Mazumder}},\ and\ \bibinfo {author}
  {\bibfnamefont {W.}~\bibnamefont {Lu}},\ }\bibfield  {title} {\enquote
  {\bibinfo {title} {Nanoscale memristor device as synapse in neuromorphic
  systems},}\ }\href@noop {} {\bibfield  {journal} {\bibinfo  {journal} {Nano
  Letters}\ }\textbf {\bibinfo {volume} {10}},\ \bibinfo {pages} {1297--1301}
  (\bibinfo {year} {2010})}\BibitemShut {NoStop}%
\bibitem [{\citenamefont {Lu}\ \emph {et~al.}(2019)\citenamefont {Lu},
  \citenamefont {Shen}, \citenamefont {Shang},\ and\ \citenamefont
  {Sun}}]{lu2019artificial}%
  \BibitemOpen
  \bibfield  {author} {\bibinfo {author} {\bibfnamefont {P.-P.}\ \bibnamefont
  {Lu}}, \bibinfo {author} {\bibfnamefont {J.-X.}\ \bibnamefont {Shen}},
  \bibinfo {author} {\bibfnamefont {D.-S.}\ \bibnamefont {Shang}},\ and\
  \bibinfo {author} {\bibfnamefont {Y.}~\bibnamefont {Sun}},\ }\bibfield
  {title} {\enquote {\bibinfo {title} {Artificial synaptic device based on a
  multiferroic heterostructure},}\ }\href@noop {} {\bibfield  {journal}
  {\bibinfo  {journal} {Journal of Physics D: Applied Physics}\ }\textbf
  {\bibinfo {volume} {52}},\ \bibinfo {pages} {465303} (\bibinfo {year}
  {2019})}\BibitemShut {NoStop}%
\bibitem [{\citenamefont {Schneider}\ \emph {et~al.}(2018)\citenamefont
  {Schneider}, \citenamefont {Donnelly}, \citenamefont {Russek}, \citenamefont
  {Baek}, \citenamefont {Pufall}, \citenamefont {Hopkins}, \citenamefont
  {Dresselhaus}, \citenamefont {Benz},\ and\ \citenamefont
  {Rippard}}]{Schneider_Donnelly_Russek_Baek_Pufall_Hopkins_Dresselhaus_Benz_Rippard_2018}%
  \BibitemOpen
  \bibfield  {author} {\bibinfo {author} {\bibfnamefont {M.~L.}\ \bibnamefont
  {Schneider}}, \bibinfo {author} {\bibfnamefont {C.~A.}\ \bibnamefont
  {Donnelly}}, \bibinfo {author} {\bibfnamefont {S.~E.}\ \bibnamefont
  {Russek}}, \bibinfo {author} {\bibfnamefont {B.}~\bibnamefont {Baek}},
  \bibinfo {author} {\bibfnamefont {M.~R.}\ \bibnamefont {Pufall}}, \bibinfo
  {author} {\bibfnamefont {P.~F.}\ \bibnamefont {Hopkins}}, \bibinfo {author}
  {\bibfnamefont {P.~D.}\ \bibnamefont {Dresselhaus}}, \bibinfo {author}
  {\bibfnamefont {S.~P.}\ \bibnamefont {Benz}},\ and\ \bibinfo {author}
  {\bibfnamefont {W.~H.}\ \bibnamefont {Rippard}},\ }\bibfield  {title}
  {\enquote {\bibinfo {title} {Ultralow power artificial synapses using
  nanotextured magnetic {Josephson} junctions},}\ }\href
  {https://doi.org/10.1126/sciadv.1701329} {\bibfield  {journal} {\bibinfo
  {journal} {Science Advances}\ }\textbf {\bibinfo {volume} {4}},\ \bibinfo
  {pages} {e1701329} (\bibinfo {year} {2018})}\BibitemShut {NoStop}%
\bibitem [{\citenamefont {Bartolozzi}\ and\ \citenamefont
  {Indiveri}(2007)}]{bartolozzi2007synaptic}%
  \BibitemOpen
  \bibfield  {author} {\bibinfo {author} {\bibfnamefont {C.}~\bibnamefont
  {Bartolozzi}}\ and\ \bibinfo {author} {\bibfnamefont {G.}~\bibnamefont
  {Indiveri}},\ }\bibfield  {title} {\enquote {\bibinfo {title} {Synaptic
  dynamics in analog {VLSI}},}\ }\href@noop {} {\bibfield  {journal} {\bibinfo
  {journal} {Neural Computation}\ }\textbf {\bibinfo {volume} {19}},\ \bibinfo
  {pages} {2581--2603} (\bibinfo {year} {2007})}\BibitemShut {NoStop}%
\bibitem [{\citenamefont {Leroux}\ \emph
  {et~al.}(2021{\natexlab{a}})\citenamefont {Leroux}, \citenamefont {Mizrahi},
  \citenamefont {Markovi{\'c}}, \citenamefont {Sanz-Hern{\'a}ndez},
  \citenamefont {Trastoy}, \citenamefont {Bortolotti}, \citenamefont {Martins},
  \citenamefont {Jenkins}, \citenamefont {Ferreira},\ and\ \citenamefont
  {Grollier}}]{leroux2021hardware}%
  \BibitemOpen
  \bibfield  {author} {\bibinfo {author} {\bibfnamefont {N.}~\bibnamefont
  {Leroux}}, \bibinfo {author} {\bibfnamefont {A.}~\bibnamefont {Mizrahi}},
  \bibinfo {author} {\bibfnamefont {D.}~\bibnamefont {Markovi{\'c}}}, \bibinfo
  {author} {\bibfnamefont {D.}~\bibnamefont {Sanz-Hern{\'a}ndez}}, \bibinfo
  {author} {\bibfnamefont {J.}~\bibnamefont {Trastoy}}, \bibinfo {author}
  {\bibfnamefont {P.}~\bibnamefont {Bortolotti}}, \bibinfo {author}
  {\bibfnamefont {L.}~\bibnamefont {Martins}}, \bibinfo {author} {\bibfnamefont
  {A.}~\bibnamefont {Jenkins}}, \bibinfo {author} {\bibfnamefont
  {R.}~\bibnamefont {Ferreira}},\ and\ \bibinfo {author} {\bibfnamefont
  {J.}~\bibnamefont {Grollier}},\ }\bibfield  {title} {\enquote {\bibinfo
  {title} {Hardware realization of the multiply and accumulate operation on
  radio-frequency signals with magnetic tunnel junctions},}\ }\href@noop {}
  {\bibfield  {journal} {\bibinfo  {journal} {Neuromorphic Computing and
  Engineering}\ }\textbf {\bibinfo {volume} {1}},\ \bibinfo {pages} {011001}
  (\bibinfo {year} {2021}{\natexlab{a}})}\BibitemShut {NoStop}%
\bibitem [{\citenamefont {Leroux}\ \emph
  {et~al.}(2021{\natexlab{b}})\citenamefont {Leroux}, \citenamefont
  {Markovi{\'c}}, \citenamefont {Martin}, \citenamefont {Petrisor},
  \citenamefont {Querlioz}, \citenamefont {Mizrahi},\ and\ \citenamefont
  {Grollier}}]{leroux2021radio}%
  \BibitemOpen
  \bibfield  {author} {\bibinfo {author} {\bibfnamefont {N.}~\bibnamefont
  {Leroux}}, \bibinfo {author} {\bibfnamefont {D.}~\bibnamefont
  {Markovi{\'c}}}, \bibinfo {author} {\bibfnamefont {E.}~\bibnamefont
  {Martin}}, \bibinfo {author} {\bibfnamefont {T.}~\bibnamefont {Petrisor}},
  \bibinfo {author} {\bibfnamefont {D.}~\bibnamefont {Querlioz}}, \bibinfo
  {author} {\bibfnamefont {A.}~\bibnamefont {Mizrahi}},\ and\ \bibinfo {author}
  {\bibfnamefont {J.}~\bibnamefont {Grollier}},\ }\bibfield  {title} {\enquote
  {\bibinfo {title} {Radio-frequency multiply-and-accumulate operations with
  spintronic synapses},}\ }\href@noop {} {\bibfield  {journal} {\bibinfo
  {journal} {Physical Review Applied}\ }\textbf {\bibinfo {volume} {15}},\
  \bibinfo {pages} {034067} (\bibinfo {year} {2021}{\natexlab{b}})}\BibitemShut
  {NoStop}%
\bibitem [{\citenamefont {Akinola}\ \emph {et~al.}(2019)\citenamefont
  {Akinola}, \citenamefont {Hu}, \citenamefont {Bennett}, \citenamefont
  {Marinella}, \citenamefont {Friedman},\ and\ \citenamefont
  {Incorvia}}]{akinola2019three}%
  \BibitemOpen
  \bibfield  {author} {\bibinfo {author} {\bibfnamefont {O.}~\bibnamefont
  {Akinola}}, \bibinfo {author} {\bibfnamefont {X.}~\bibnamefont {Hu}},
  \bibinfo {author} {\bibfnamefont {C.~H.}\ \bibnamefont {Bennett}}, \bibinfo
  {author} {\bibfnamefont {M.}~\bibnamefont {Marinella}}, \bibinfo {author}
  {\bibfnamefont {J.~S.}\ \bibnamefont {Friedman}},\ and\ \bibinfo {author}
  {\bibfnamefont {J.~A.~C.}\ \bibnamefont {Incorvia}},\ }\bibfield  {title}
  {\enquote {\bibinfo {title} {Three-terminal magnetic tunnel junction synapse
  circuits showing spike-timing-dependent plasticity},}\ }\href@noop {}
  {\bibfield  {journal} {\bibinfo  {journal} {Journal of Physics D: Applied
  Physics}\ }\textbf {\bibinfo {volume} {52}},\ \bibinfo {pages} {49LT01}
  (\bibinfo {year} {2019})}\BibitemShut {NoStop}%
\bibitem [{\citenamefont {Narasimman}\ \emph {et~al.}(2016)\citenamefont
  {Narasimman}, \citenamefont {Roy}, \citenamefont {Fong}, \citenamefont {Roy},
  \citenamefont {Chang},\ and\ \citenamefont {Basu}}]{narasimman2016low}%
  \BibitemOpen
  \bibfield  {author} {\bibinfo {author} {\bibfnamefont {G.}~\bibnamefont
  {Narasimman}}, \bibinfo {author} {\bibfnamefont {S.}~\bibnamefont {Roy}},
  \bibinfo {author} {\bibfnamefont {X.}~\bibnamefont {Fong}}, \bibinfo {author}
  {\bibfnamefont {K.}~\bibnamefont {Roy}}, \bibinfo {author} {\bibfnamefont
  {C.-H.}\ \bibnamefont {Chang}},\ and\ \bibinfo {author} {\bibfnamefont
  {A.}~\bibnamefont {Basu}},\ }\bibfield  {title} {\enquote {\bibinfo {title}
  {A low-voltage, low power {STDP} synapse implementation using domain-wall
  magnets for spiking neural networks},}\ }in\ \href@noop {} {\emph {\bibinfo
  {booktitle} {2016 IEEE International Symposium on Circuits and Systems
  (ISCAS)}}}\ (\bibinfo {organization} {IEEE},\ \bibinfo {year} {2016})\ pp.\
  \bibinfo {pages} {914--917}\BibitemShut {NoStop}%
\bibitem [{\citenamefont {Bhowmik}\ \emph {et~al.}(2019)\citenamefont
  {Bhowmik}, \citenamefont {Saxena}, \citenamefont {Dankar}, \citenamefont
  {Verma}, \citenamefont {Kaushik}, \citenamefont {Chatterjee},\ and\
  \citenamefont {Singh}}]{bhowmik2019chip}%
  \BibitemOpen
  \bibfield  {author} {\bibinfo {author} {\bibfnamefont {D.}~\bibnamefont
  {Bhowmik}}, \bibinfo {author} {\bibfnamefont {U.}~\bibnamefont {Saxena}},
  \bibinfo {author} {\bibfnamefont {A.}~\bibnamefont {Dankar}}, \bibinfo
  {author} {\bibfnamefont {A.}~\bibnamefont {Verma}}, \bibinfo {author}
  {\bibfnamefont {D.}~\bibnamefont {Kaushik}}, \bibinfo {author} {\bibfnamefont
  {S.}~\bibnamefont {Chatterjee}},\ and\ \bibinfo {author} {\bibfnamefont
  {U.}~\bibnamefont {Singh}},\ }\bibfield  {title} {\enquote {\bibinfo {title}
  {On-chip learning for domain wall synapse based fully connected neural
  network},}\ }\href@noop {} {\bibfield  {journal} {\bibinfo  {journal}
  {Journal of Magnetism and Magnetic Materials}\ }\textbf {\bibinfo {volume}
  {489}},\ \bibinfo {pages} {165434} (\bibinfo {year} {2019})}\BibitemShut
  {NoStop}%
\bibitem [{\citenamefont {Song}\ \emph {et~al.}(2020)\citenamefont {Song},
  \citenamefont {Jeong}, \citenamefont {Pan}, \citenamefont {Zhang},
  \citenamefont {Xia}, \citenamefont {Cha}, \citenamefont {Park}, \citenamefont
  {Kim}, \citenamefont {Finizio}, \citenamefont {Raabe} \emph
  {et~al.}}]{song2020skyrmion}%
  \BibitemOpen
  \bibfield  {author} {\bibinfo {author} {\bibfnamefont {K.~M.}\ \bibnamefont
  {Song}}, \bibinfo {author} {\bibfnamefont {J.-S.}\ \bibnamefont {Jeong}},
  \bibinfo {author} {\bibfnamefont {B.}~\bibnamefont {Pan}}, \bibinfo {author}
  {\bibfnamefont {X.}~\bibnamefont {Zhang}}, \bibinfo {author} {\bibfnamefont
  {J.}~\bibnamefont {Xia}}, \bibinfo {author} {\bibfnamefont {S.}~\bibnamefont
  {Cha}}, \bibinfo {author} {\bibfnamefont {T.-E.}\ \bibnamefont {Park}},
  \bibinfo {author} {\bibfnamefont {K.}~\bibnamefont {Kim}}, \bibinfo {author}
  {\bibfnamefont {S.}~\bibnamefont {Finizio}}, \bibinfo {author} {\bibfnamefont
  {J.}~\bibnamefont {Raabe}}, \emph {et~al.},\ }\bibfield  {title} {\enquote
  {\bibinfo {title} {Skyrmion-based artificial synapses for neuromorphic
  computing},}\ }\href@noop {} {\bibfield  {journal} {\bibinfo  {journal}
  {Nature Electronics}\ }\textbf {\bibinfo {volume} {3}},\ \bibinfo {pages}
  {148--155} (\bibinfo {year} {2020})}\BibitemShut {NoStop}%
\bibitem [{\citenamefont {Huang}\ \emph {et~al.}(2017)\citenamefont {Huang},
  \citenamefont {Kang}, \citenamefont {Zhang}, \citenamefont {Zhou},\ and\
  \citenamefont {Zhao}}]{huang2017magnetic}%
  \BibitemOpen
  \bibfield  {author} {\bibinfo {author} {\bibfnamefont {Y.}~\bibnamefont
  {Huang}}, \bibinfo {author} {\bibfnamefont {W.}~\bibnamefont {Kang}},
  \bibinfo {author} {\bibfnamefont {X.}~\bibnamefont {Zhang}}, \bibinfo
  {author} {\bibfnamefont {Y.}~\bibnamefont {Zhou}},\ and\ \bibinfo {author}
  {\bibfnamefont {W.}~\bibnamefont {Zhao}},\ }\bibfield  {title} {\enquote
  {\bibinfo {title} {Magnetic skyrmion-based synaptic devices},}\ }\href@noop
  {} {\bibfield  {journal} {\bibinfo  {journal} {Nanotechnology}\ }\textbf
  {\bibinfo {volume} {28}},\ \bibinfo {pages} {08LT02} (\bibinfo {year}
  {2017})}\BibitemShut {NoStop}%
\bibitem [{\citenamefont {Das}\ \emph {et~al.}(2022)\citenamefont {Das},
  \citenamefont {Cen}, \citenamefont {Wang},\ and\ \citenamefont
  {Fong}}]{das2022bilayer}%
  \BibitemOpen
  \bibfield  {author} {\bibinfo {author} {\bibfnamefont {D.}~\bibnamefont
  {Das}}, \bibinfo {author} {\bibfnamefont {Y.}~\bibnamefont {Cen}}, \bibinfo
  {author} {\bibfnamefont {J.}~\bibnamefont {Wang}},\ and\ \bibinfo {author}
  {\bibfnamefont {X.}~\bibnamefont {Fong}},\ }\bibfield  {title} {\enquote
  {\bibinfo {title} {Bilayer-skyrmion based design of neuron and synapse for
  spiking neural network},}\ }\href@noop {} {\bibfield  {journal} {\bibinfo
  {journal} {arXiv preprint arXiv:2203.02171}\ } (\bibinfo {year}
  {2022})}\BibitemShut {NoStop}%
\bibitem [{\citenamefont {Kasabov}(2018)}]{Kasabov_2018}%
  \BibitemOpen
  \bibfield  {author} {\bibinfo {author} {\bibfnamefont {N.~K.}\ \bibnamefont
  {Kasabov}},\ }\href@noop {} {\emph {\bibinfo {title} {Time-Space, Spiking
  Neural Networks and Brain-Inspired Artificial Intelligence}}}\ (\bibinfo
  {publisher} {Springer Berlin Heidelberg},\ \bibinfo {year}
  {2018})\BibitemShut {NoStop}%
\bibitem [{\citenamefont {Maekawa}\ \emph {et~al.}(2017)\citenamefont
  {Maekawa}, \citenamefont {Valenzuela}, \citenamefont {Kimura},\ and\
  \citenamefont {Saitoh}}]{Maekawa_Valenzuela_Saitoh_2017}%
  \BibitemOpen
  \bibfield  {author} {\bibinfo {author} {\bibfnamefont {S.}~\bibnamefont
  {Maekawa}}, \bibinfo {author} {\bibfnamefont {S.~O.}\ \bibnamefont
  {Valenzuela}}, \bibinfo {author} {\bibfnamefont {T.}~\bibnamefont {Kimura}},\
  and\ \bibinfo {author} {\bibfnamefont {E.}~\bibnamefont {Saitoh}},\
  }\href@noop {} {\emph {\bibinfo {title} {Spin Current}}}\ (\bibinfo
  {publisher} {Oxford University Press},\ \bibinfo {year} {2017})\BibitemShut
  {NoStop}%
\end{thebibliography}%

\end{document}